\documentclass[aps,prx,amsmath,amssymb,floatfix,twocolumn]{revtex4-1}
\usepackage{parskip} 
\usepackage{tikz} 
\usepackage{subfigure}
\usepackage{caption}
\usepackage{amsmath}
\usepackage{bbold}
\usepackage{slashed}
\usepackage{amssymb}
\usepackage{braket}
\usepackage{array}
\usepackage[colorlinks]{hyperref}
\usepackage{float}
\usepackage[margin=1in]{geometry}

\usepackage{natbib}

\begin{document}
\preprint{APS/123-QED}

\title{Solitons and real-space screening of bulk topology of quantum materials}
\author{Alexander C. Tyner$^{1}$ and Pallab Goswami$^{1,2}$}
\email{pallab.goswami@northwestern.edu}
\affiliation{$^{1}$ Graduate Program in Applied Physics, Northwestern University, Evanston, Illinois, 60208, USA}
\affiliation{$^{2}$ Department of Physics and Astronomy, Northwestern University, Evanston, Illinois, 60208, USA}

\date{\today}

\begin{abstract} 
Recent years have seen multiple high-throughput studies reveal an immense number of topological materials through use of symmetry indicators. Despite this success, three-dimensional topological insulators (TIs) admitting a band-gap larger than Bi$_{2}$Se$_{3}$ and two-dimensional TIs admitting a band gap larger than $\beta$-bismuthene, two of the originally proposed TIs, remain extremely rare. Simultaneously, a significant effort has been made to understand and identify topological phases ``invisible" to symmetry indicators. Such phases offer a unique opportunity to expand the search for a large band-gap TI, however their identification requires sophisticated probes of bulk topology. Magnetic flux tubes or vortices have emerged as one such probe in two-dimensions when inserted into the bulk. In this work, we develop an automated workflow to perform vortex insertion and apply it to a current database of high-quality, experimentally realized, two-dimensional insulators. The results reveal multiple novel two-dimensional topological insulators supporting large bands gaps, including the 1H-MX$_{2}$ (M=Mo,W) and (X=S,Se,Te) family of transition metal dichalcogenides. Our work has broad implications for current theoretical and experimental efforts to employ these materials in superconducting and Moire systems.

\end{abstract}

\maketitle
\par 
\emph{Introduction:}\label{intro}
Experimental control and interest in two-dimensional time-reversal ($\mathcal{T}$) invariant insulators (quantum spin-Hall insulators) has advanced rapidly in recent years, prompting the creation of multiple two-dimensional material databases\cite{mounet2018two,zhou20192dmatpedia,rasmussen2015computational,haastrup2018computational}. These databases have been subsequently scanned, utilizing momentum-space diagnostics, for novel topological phases. Despite the impressive scope and success of these high-throughput studies, a two-dimensional quantum spin-Hall (QSH) insulator supporting a band-gap larger than that of $\beta$-bismuthene, proposed more than two decades ago as a QSH system\cite{MurakamiQSHBi,WadaBi}, remains elusive. As a result, interest in topological phases ``invisible" to symmetry indicators has grown. This presents a tremendous challenge as identification of these phases involves considerable computational complexity and can not be achieved using the momentum-space tools available within the current set of community software packages. Real-space probes of topology offer a unique opportunity to overcome these challenges. In two-dimensions, insertion of a magnetic flux tube (vortex) into an insulator has been demonstrated to be a reliable real-space probe of bulk topology beyond symmetry indicators. Importantly, this process can be integrated with \emph{ab initio} derived models. In this work, we utilize vortex insertion to scan 141 high-quality and experimentally realized, two-dimensional, $\mathcal{T}$-invariant insulators for topology invisible to symmetry indicators. 
\par 
The current literature standard for diagnosis of non-trivial topology in a $\mathcal{T}$-preserving system relies on assignment of the Fu-Kane $\mathbb{Z}_{2}$ index, $\nu_{0}$\cite{Kane2005,FuKaneMele2007,FuKane,Roy2009}. When inversion symmetry ($\mathcal{P}$), is independently conserved, $\nu_{0}$ is efficiently computed through multiplication of the inversion eigenvalues for all occupied bands at each time-reversal invariant momenta (TRIM) point in the Brillouin zone. In the absence of inversion-symmetry, the $\mathbb{Z}_{2}$ index is computed through calculation of the Wilson loop, or Wannier center charges (WCCs). In both cases, the $\mathbb{Z}_{2}$ classification is often supplemented through an analysis of the surface spectra, with the presence of gapless states constituting bulk-boundary correspondence in first-order topological insulators.
\par 
 While efficient in its computation, the $\mathbb{Z}_{2}$ index is unable to distinguish insulators supporting a non-zero, even-integer, ground state spin-Chern number ($\mathcal{C}_{s,G}$) from trivial insulators with a vanishing invariant. As a consequence of being trivial under the $\mathbb{Z}_{2}$ index, materials supporting an even-integer spin Chern number are elusive, but they are in principle allowed. This remains true in the absence of U(1) spin-conservation and/or inversion symmetry, where it is possible even to trivialize the Wilson loop of a two-dimensional insulator supporting an odd integer spin-Chern number\cite{Sheng2006,Prodan2009,tyner2020topology,tyner2021quantized}. Each of these phases falls under the category of topological materials ``invisible" to symmetry indicators. 
\begin{figure*}
    \centering
    \includegraphics[width=14cm]{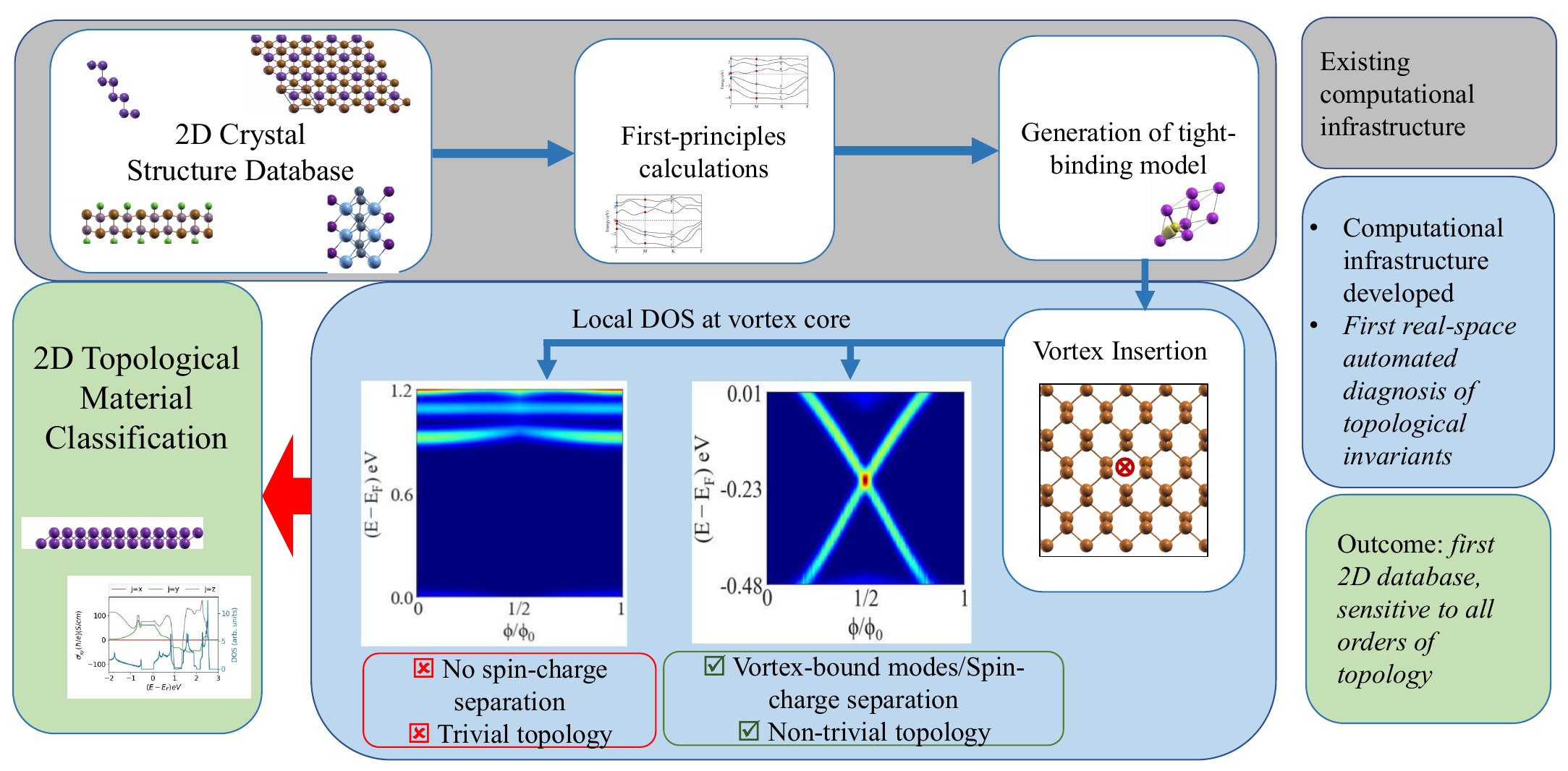}
    \caption{Diagram detailing work flow for automated identification of bulk topological invariant in 2D materials. Flow ends with plots of local DOS at the inserted vortex core. Examples given are for antimonene (labeled trivial) and bismuthene (labeled non-trivial) in vicinity of the Fermi energy\cite{tynerbismuthene}. Bismuthene is non-trivial as we can clearly observe the electromagnetic vortex binding a Kramers pair in the bulk-gap.}
    \label{fig:CompFlow}
\end{figure*}

\par 
Recent works by Bai et. al\cite{Bai2022Doubled} and Wang et. al\cite{bansilspin}, have demonstrated that $\alpha$-bismuthene and $\alpha$-antimonene realize a ground state with $|\mathcal{C}_{s,G}|=2$. This was accomplished through an analysis of the spectrum of WCCs calculated along a primary crystalline axis as a function of the transverse momenta\cite{yu2011equivalent,Soluyanov2011,alexandradinata2014wilson,bouhon2019wilson,bradlyn2019disconnected,Z2pack}. In each case, an examination of the surface states was completed. The results constituted bulk-boundary correspondence as the WCC and surface state spectra displayed two gapless locations\cite{Taherinejad2014}. However, it has been demonstrated that assignment of a quantized spin-Chern number in the absence of U(1) spin-conservation symmetry does not dictate the existence of gapless edge states or WCC spectra\cite{Sheng2006,Prodan2009,tyner2020topology}. 
\par 
Due to the significant computational challenges associated with momentum-space calculation of the spin-Chern number in \emph{ab initio} models of QSH insulators lacking spin-conservation symmetry\cite{tyner2021quantized,Lin2022Spin}, we are unaware of any works that have unambiguously identified a two-dimensional QSH insulator supporting gapped edge states and a non-zero, even-integer spin-Chern number. Nevertheless, alternative signatures of bulk-topology such as spin-Hall conductivity\cite{costa2021discovery} and corner localized states\cite{sodequist2022abundance} have been used in material scans to identify potential candidates. 
\par
Recently, Tyner et. al\cite{tynerbismuthene} demonstrated that the ground state topological invariant in two-dimensions can be computed in realistic, \emph{ab initio} data, through insertion of a magnetic flux tube or vortex, a concept first proposed by Qi and Zhang\cite{QiSpinCharge} and Ran et. al\cite{SpinChargeVishwanath}. Upon insertion of a magnetic flux tube of strength $\phi=\phi_{0}/2$, where $\phi_{0}=hc/e$ is the flux quanta, in two-dimensional QSH insulators defined by ground state invariant $\mathcal{C}_{s,G}=N$ with $N\in \mathbb{Z}$, the flux tube binds $2N$, mid gap states. Notably, it has been shown that this method goes beyond the $\mathbb{Z}_{2}$ formulation, classifying the magnitude of the ground state bulk-invariant for arbitrary natural numbers, even in the absence of a U(1) spin-conservation symmetry\cite{slager2012,MESAROS2013977,Wang_2010,schindler2022topological}. 
\par 
Motivated by this important technical development, we reexamine the database of high-quality, non-magnetic, two-dimensional insulators developed by Mounet et. al\cite{mounet2018two} through the method of vortex insertion. The results of an automated computational screening of 141 materials reveals 19 high-quality topological insulators, of which only one, $\beta$-bismuthene, has been previously identified as topological via the $\mathbb{Z}_{2}$ index. Within those identified to be topological is ZnCl$_{2}$ which supports a calculated band gap of 4.5 eV, far beyond any current record for a predicted two-dimensional topological insulator, to our knowledge. 
\par 
For clarity of presentation, we concentrate on the 1H-MX$_{2}$ phase of transition metal dichalcogenides (TMDs), with M=(Mo, W) and X=(Te, Se, S) in the main body. Our work concludes that these materials are generalized spin-Hall insulators with ground state topological invariant of magnitude $|\mathcal{C}_{s,G}|=2$. The identification of non-trivial topology in these materials is of particular importance given the current experimental interest in their use for superconducting devices\cite{ye2012superconducting,taniguchi2012electric,shi2015superconductivity}, where insertion of magnetic vortices becomes a natural and physically realizable process, as well as Moire systems. Further results of the screening, outside the MX$_{2}$ family, are presented in appendix E. 
\begin{figure}
\centering
\subfigure[]{\includegraphics[scale=0.3]{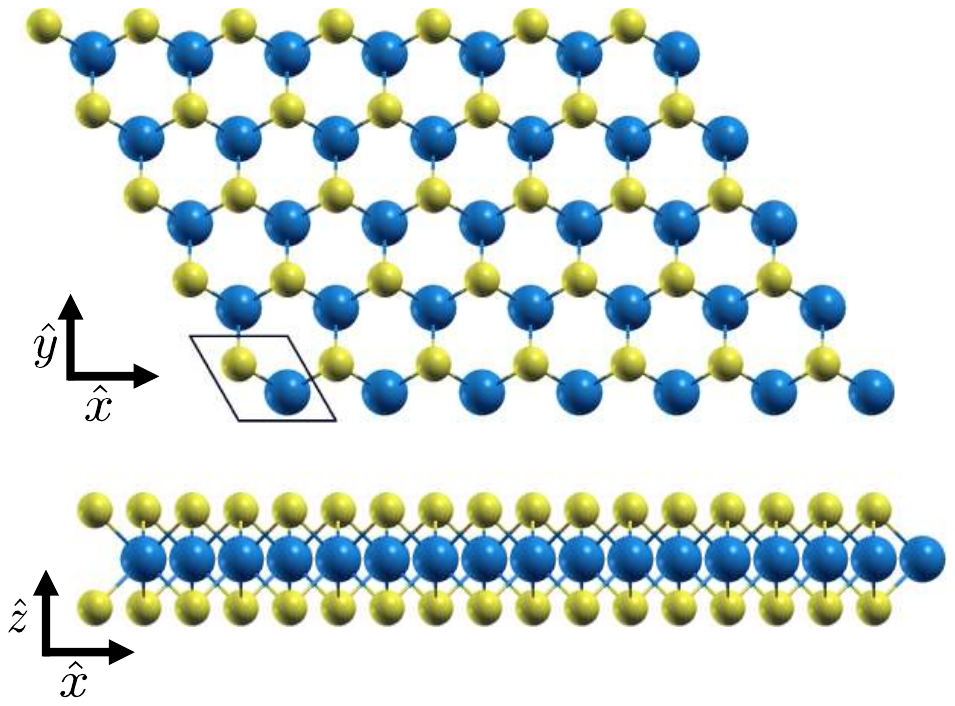}
\label{fig:MX2Structure}}
\subfigure[]{\includegraphics[scale=0.3]{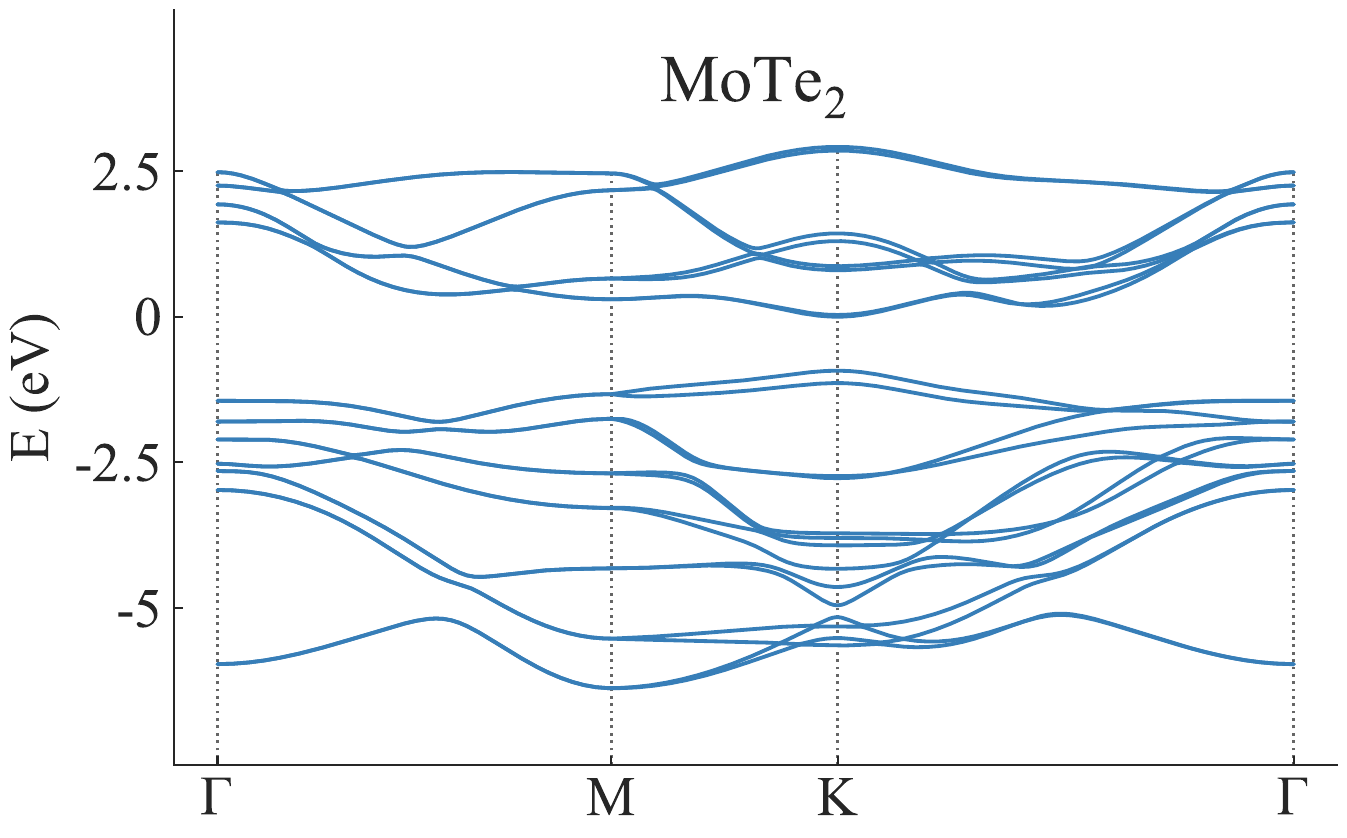}
\label{fig:MoTe2Bands}}
\subfigure[]{
    \begin{tabular}{|m{2.0cm}|m{2.0cm}|}
        \hline
          & $\Delta E_{DFT}$  \\
        \hline
        MoTe$_{2}$ & 1.1 eV \\
        \hline
        MoS$_{2}$  & 1.6 eV   \\
        \hline
        MoSe$_{2}$  & 1.5 eV   \\
        \hline
         WSe$_{2}$ & 1.6 eV  \\
        \hline
         WS$_{2}$ & 1.8 eV    \\
        \hline
         WTe$_{2}$ & 1.1 eV    \\
        \hline
    \end{tabular}
    \label{tab:deltaE}
}
\caption{(a) Structure of 1H-MX$_{2}$ phase for two-dimensional transition metal dichalcogenides. Blue atoms indicate transition metal (M), and yellow atoms indicate chalcogen atoms. (b) Bulk band structure along high-symmetry path for 1H-MoTe$_{2}$. (c) Calculated bulk band-gap for materials of interest. }
\end{figure}

\begin{figure*}
\centering
\subfigure[]{
\includegraphics[scale=0.28]{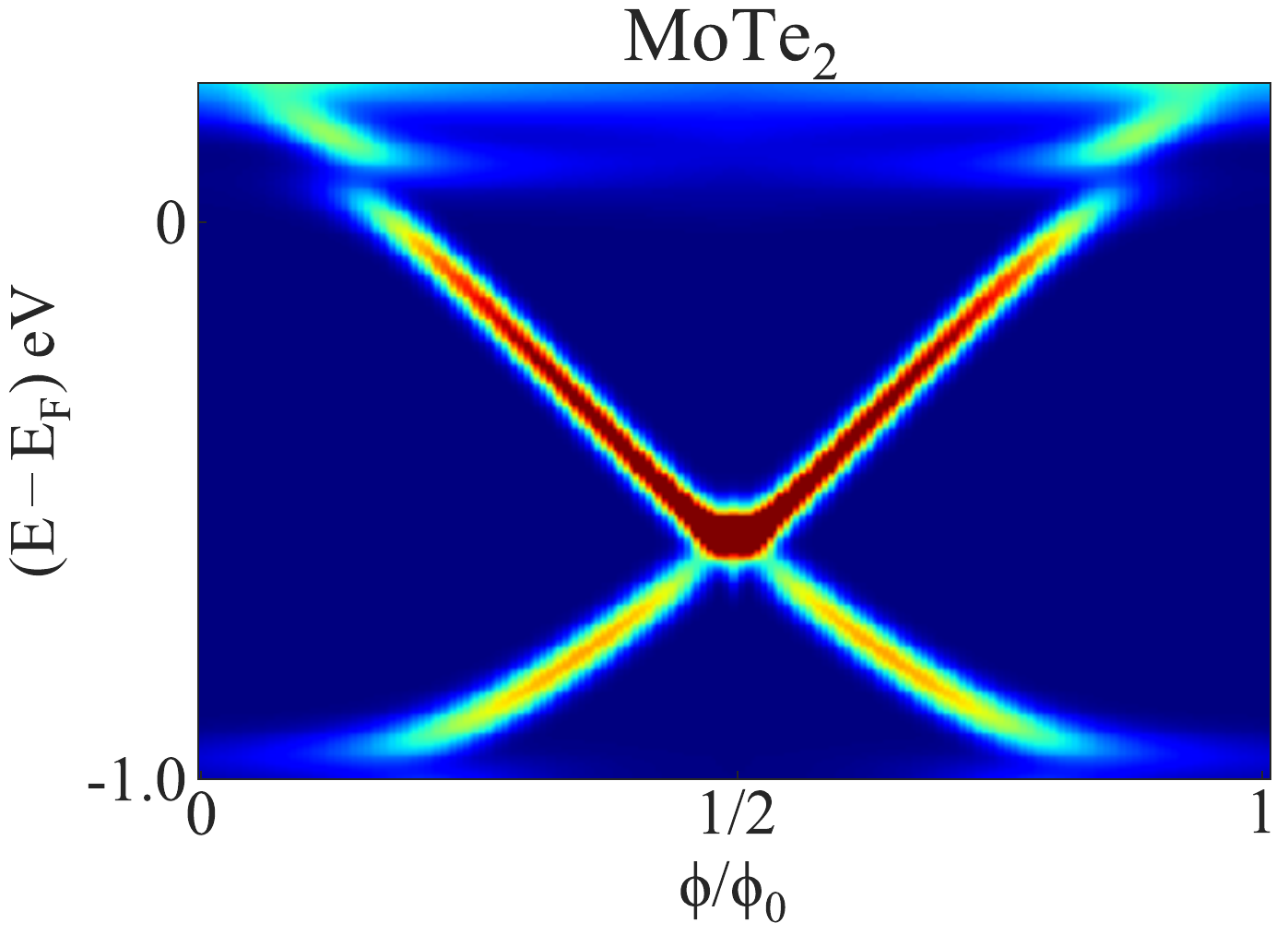}
\label{fig:}}
\subfigure[]{
\includegraphics[scale=0.28]{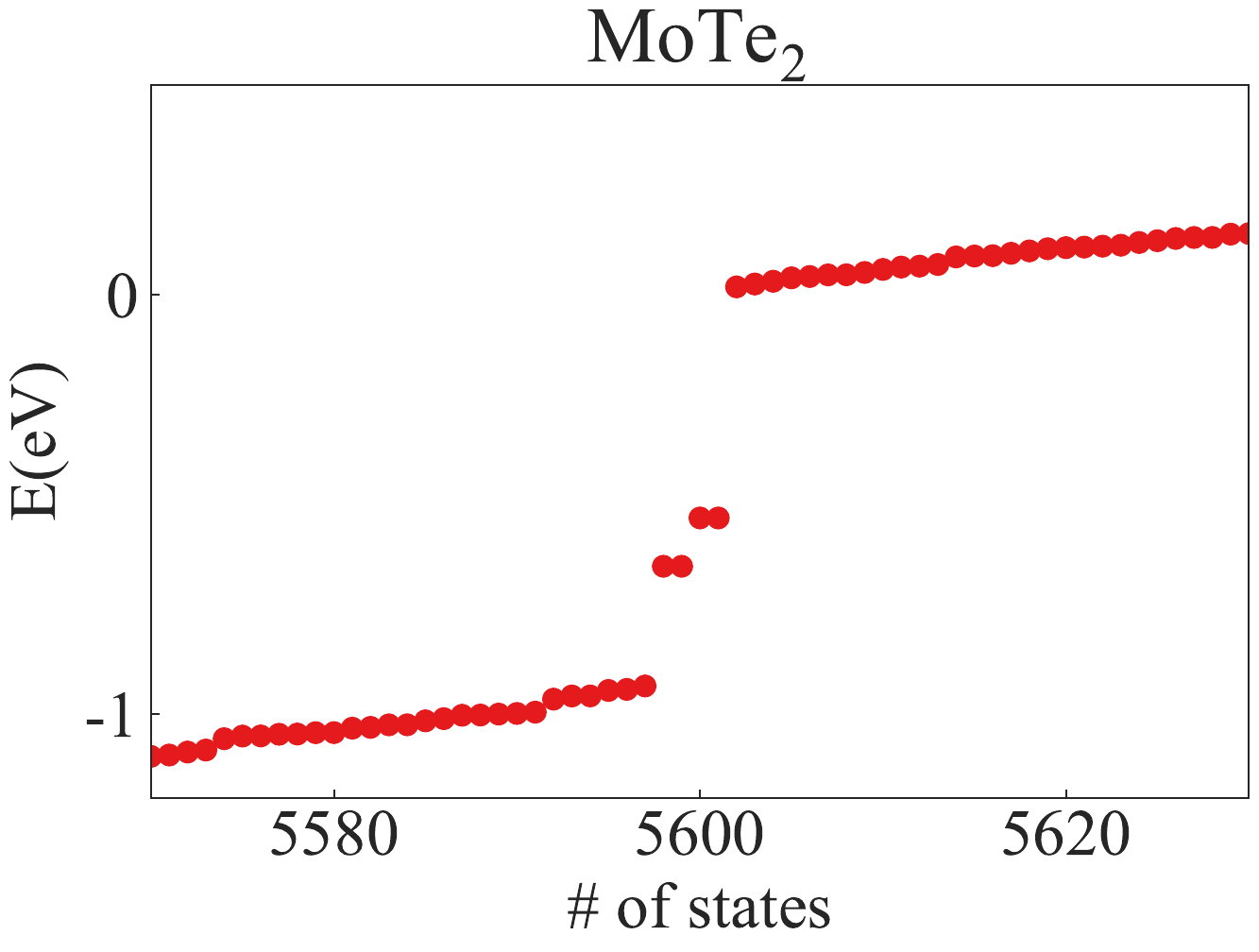}
\label{fig:}}
\subfigure[]{
\includegraphics[scale=0.28]{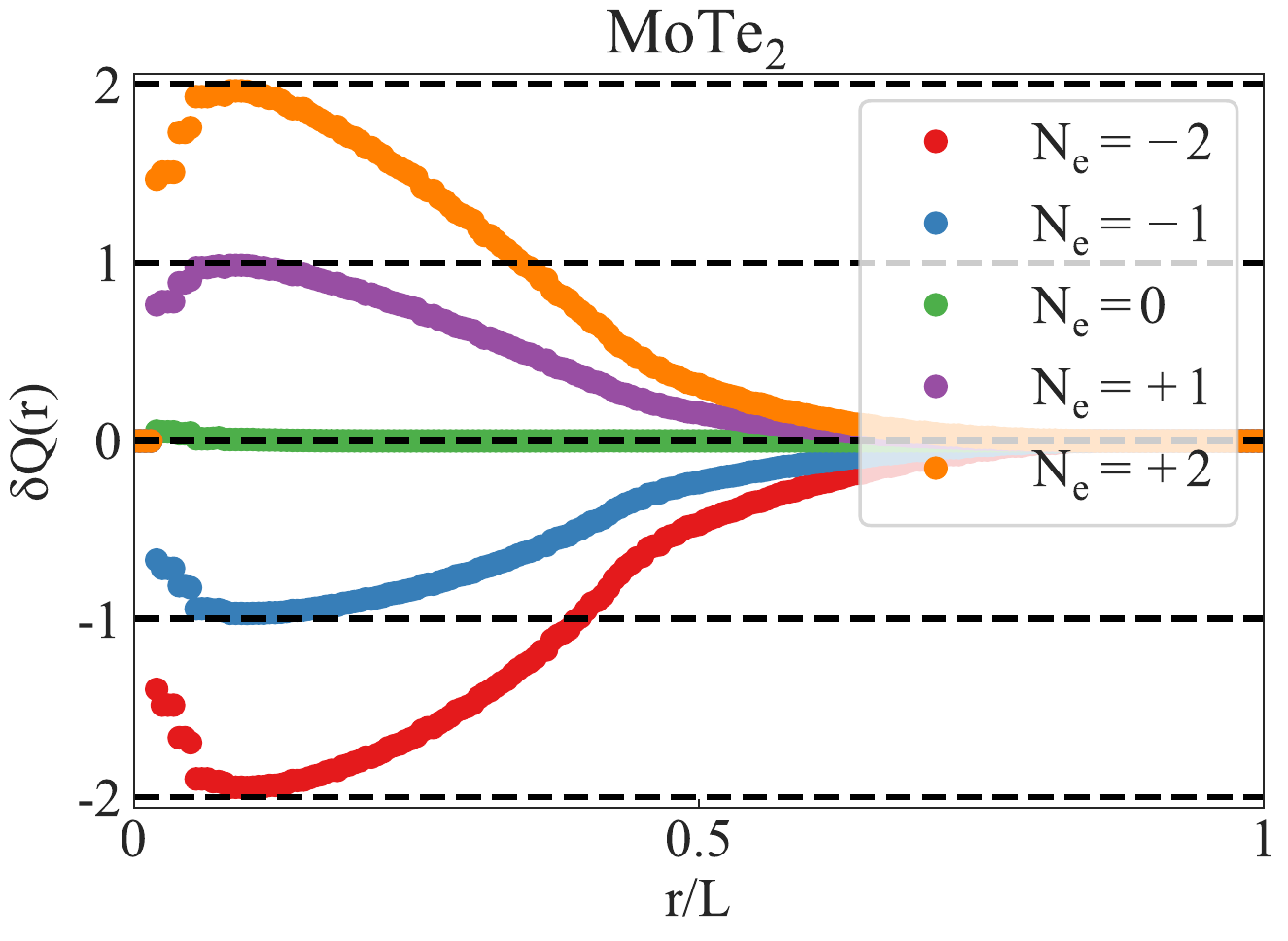}
\label{fig:}}
\subfigure[]{
\includegraphics[scale=0.28]{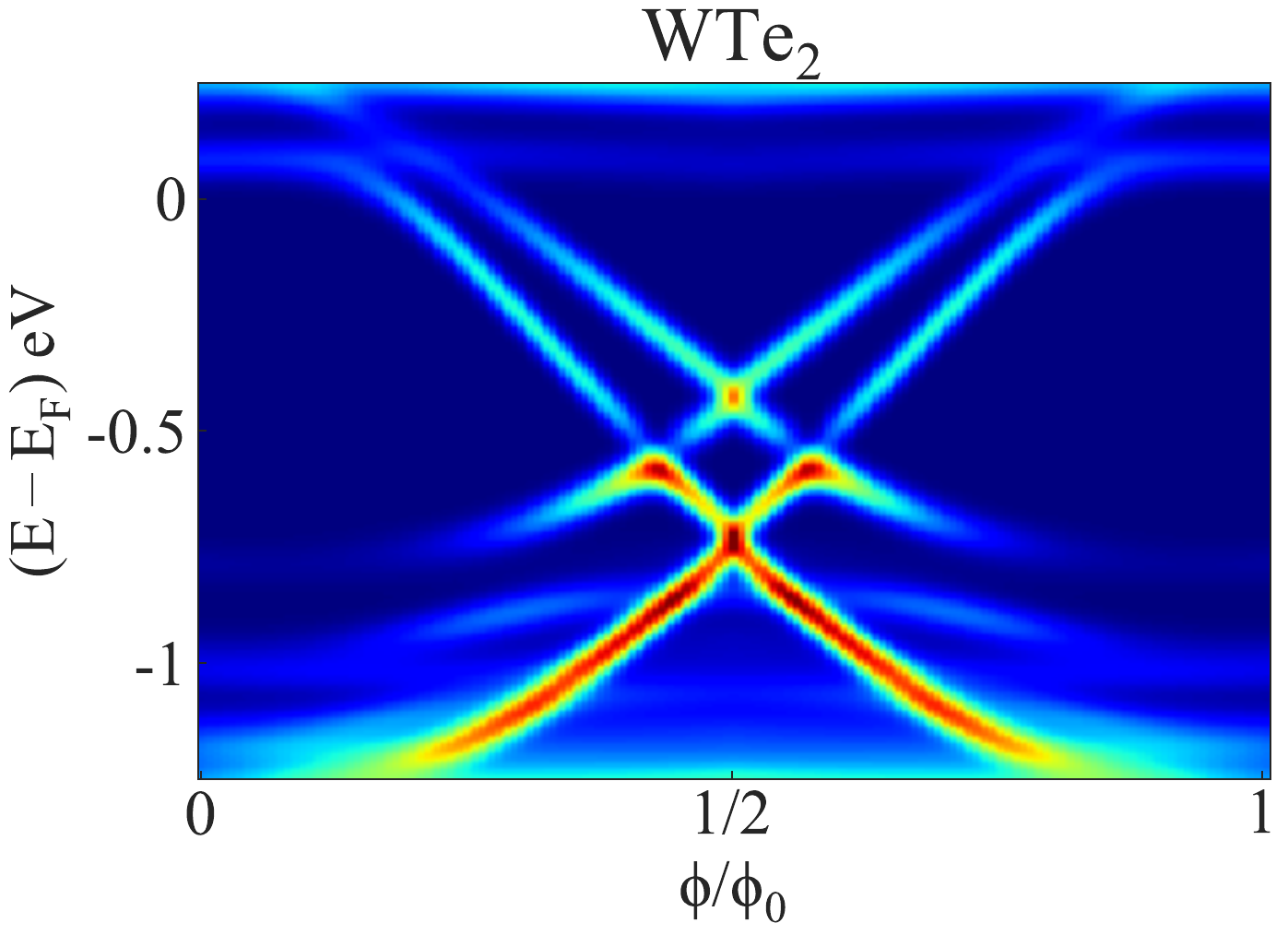}
\label{fig:}}
\subfigure[]{
\includegraphics[scale=0.28]{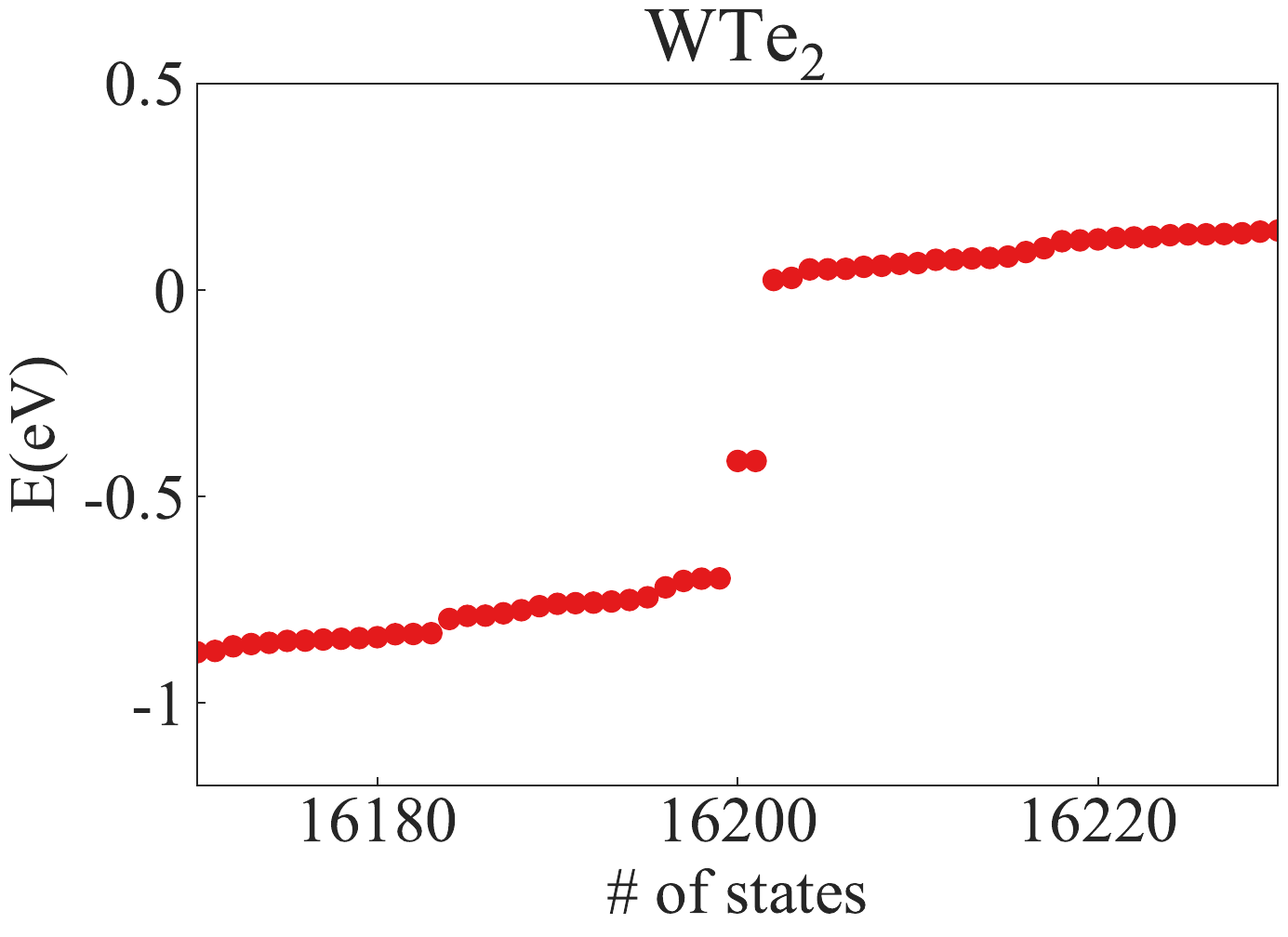}
\label{fig:}}
\subfigure[]{
\includegraphics[scale=0.28]{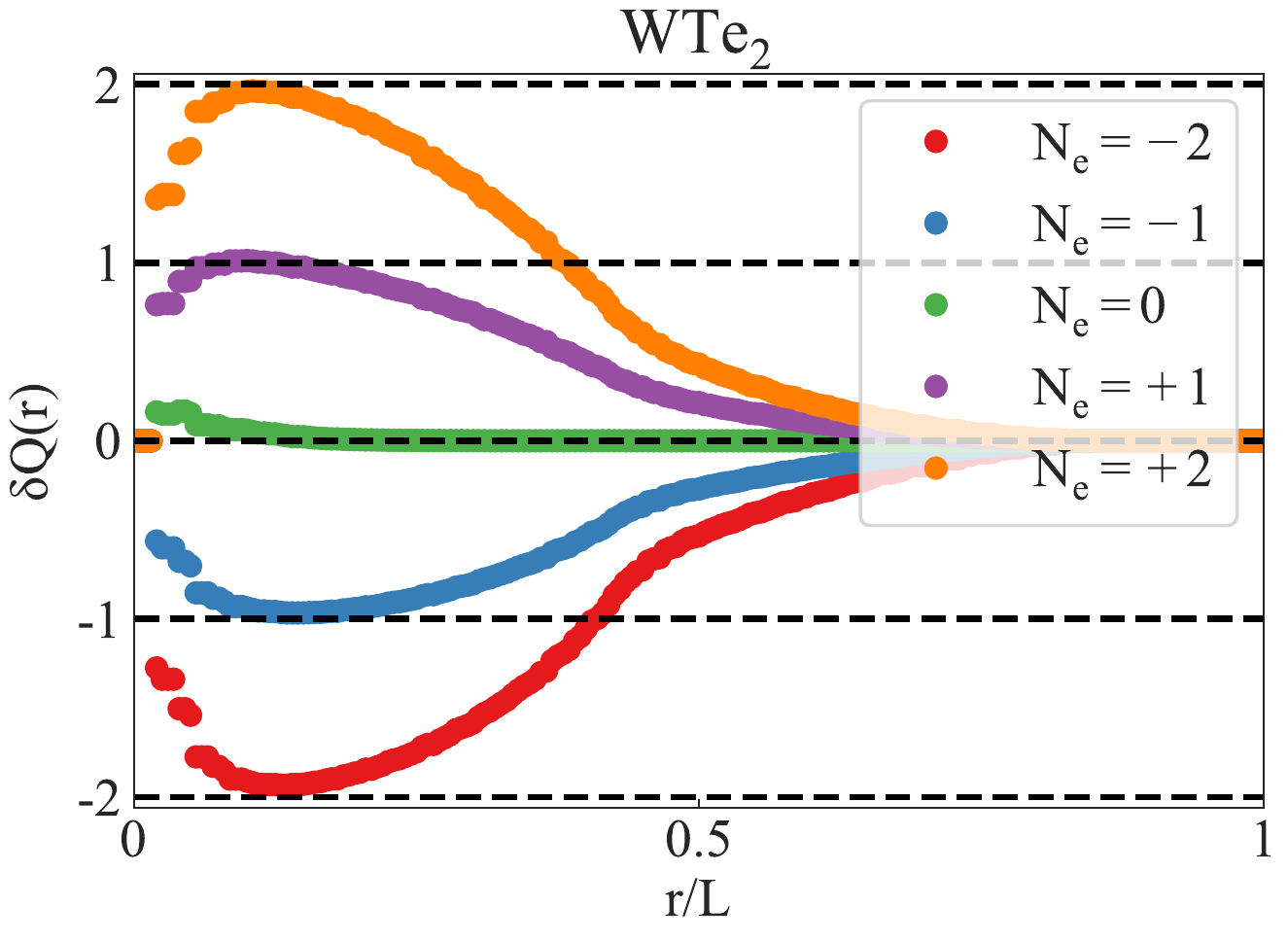}
\label{fig:}}
\caption{Local density of states on vortex for (a) 1H-MoTe$_{2}$ and (d) 1H-WTe$_{2}$, TMDs admitting ground state spin Chern number $|\mathcal{C}_{s,G}|=2$, as a function of flux strength. State vs. energy upon insertion of $\pi$-flux vortex in a system of 20 $\times$ 20 unit cells for (b) 1H-MoTe$_{2}$ and (e) 1H-WTe$_{2}$. Induced charge on the vortex as a function of doping away from half-filling of vortex bound modes (VBMs) by $N_{e}$ states. Results display robust nature of spin-charge separation and relation, $\delta Q= |\mathcal{C}_{s,G}|\times e$, even in the absence of additional symmetries forcing a complete degeneracy of VBMs at $\pi$-flux for (c) 1H-MoTe$_{2}$ and (f) 1H-WTe$_{2}$.}
\label{fig:VortexWTe2MoTe2}
\end{figure*}

\par 
\emph{Computational screening:}
In this work, all first principles calculations based on density-functional theory (DFT) are carried out using the Quantum Espresso software package \cite{QE-2009,QE-2017,QE-2020}. Exchange-correlation potentials use the Perdew-Burke-Ernzerhof (PBE) parametrization of the generalized gradient approximation (GGA) \cite{Perdew1996}. The Wannier90 and Z2pack software packages were utilized in calculation of all topological invariants \cite{Pizzi2020,Z2pack,Soluyanov2011}. We consider all non-magnetic two-dimensional insulators detailed by Mounet et. al\cite{mounet2018two}. Utilizing the listed optimal lattice parameters and atomic positions in all calculations. A schematic of the automated workflow is shown in Fig. \eqref{fig:CompFlow}.  
\par
In order to facilitate automated analysis of the bulk topology, Wannier tight-binding (WTB) models are constructed through use of the SCDM method introduced by Vitale et. al\cite{vitale2020automated}. Manipulation of Wannier tight-binding model for vortex insertion is done with a custom python program which will be made publicly available upon being developed into a stand-alone package. After an initial screening utilizing the automated Wannier tight-binding model, materials classified as topological are reexamined. The criteria for topological classification is based on the concept of spin-charge separation. 
\par
QSH insulators were first proposed by Qi and Zhang\cite{QiSpinCharge} and Ran et. al\cite{SpinChargeVishwanath}, as platforms for spin-charge separation. The mechanism of spin-charge separation was shown to be insertion of a magnetic flux tube (vortex). In the original works, it was shown that a $\phi=hc/(2e)$ ($\pi$-flux) tube binds $2N$, degenerate states in a non-trivial $\mathbb{Z}_{2}$ insulator with odd integer spin Chern number $\mathcal{C}_{s,G}=N$. This concept was then extended to the situation of arbitrary spin-Chern numbers by Tyner et. al\cite{tynerbismuthene}. Spin charge separation can be observed by tuning filling fraction of the mid-gap vortex bound modes (VBMs). If the VBMs are half-filled, the vortex acquires induced spin but no induced charge. If we dope by $N_{e} \in [-N,+N]$ electrons away from half-filling, occupying all VBMs, the vortex acquires induced charge $\delta Q= N_{e}\times e$. If this condition is satisfied, the spin Chern number can be directly calculated by fixing $N_{e}=N$ such that $\delta Q= |\mathcal{C}_{s,G}|\times e$. \emph{This is the criteria for topological classification employed in this work.} Due to the computational burden associated with computation of induced spin-densities we reserve it for a proof-of-principle calculation in a tight-binding model presented in Appendix A, which captures the topological nature of the materials examined in our work.

\par 
It was further shown by Qi et. al\cite{QiSpinCharge}, that it is possible to confirm that a two-dimensional insulator supports spin-charge separation without explicit calculation of induced charge. This is accomplished by tuning magnitude of the flux, $\phi$, from 0 to the flux quanta $\phi_{0}=hc/e$. As the flux is tuned, if the vortex acts as a spin-pump, pumping N states from the valence(conduction) subspace to the conduction(valence) subspace it is confirmed that the insulator supports spin-charge separation and the bulk invariant can be determined by observing the spectral flow of VBMs across the bulk gap. 
\par 
For two-dimensional topological insulators trivial under $\mathbb{Z}_{2}$ classification, additional symmetries such as Kramers degeneracy throughout the Brillouin zone ($\mathcal{P}\mathcal{T}$) and particle-hole symmetry are required to realize full degeneracy of VBMs at $\pi$-flux, yielding a fully connected spectral flow. In the absence of additional symmetries it is possible to break the degeneracy of vortex bound modes at $\pi$-flux. This situation does not disallow the assignment of a ground state bulk invariant\cite{Sheng2006,Prodan2009} or spin-pumping. Rather, verification that the non-degenerate VBMs are correlated and support spin-charge separation as a function of filling and spin-pumping as a function of flux, is required through direct calculation. For exemplification of this process and verification that these systems remain platforms for spin-charge separation in a simplified tight-binding model, please consult appendix A.
\par
To allow for more fine-grained analysis of those materials in the 1H-MX$_{2}$ family, a second WTB model is produced exactly replicating only the Kramers-degenerate bands nearest to the Fermi energy using carefully selected orbitals rather than an automated selection. As these bands are significantly separated energetically from all other bands, the TB model reproduces the DFT data precisely. In construction of both TB models, a 40 x 40 x 1 Monkhorst-Pack grid of k-points is utilized as well as a plane wave cutoff of 100 Ry. Spin-orbit coupling is accounted for in all calculations. To computationally simulate the flux tube, all hopping elements $H_{ij}$, connecting lattice sites $\mathbf{r}_{i}$  and $\mathbf{r}_{j}$ are modified to $H_{ij} e^{i\phi_{ij}}$, where we define the Peierls factor,  
\begin{equation}
    \phi_{ij}=\frac{\phi}{\phi_{0}}\int_{\mathbf{r}_{i}}^{\mathbf{r}_{j}}\frac{\hat{z}\times \mathbf{r}}{\mathbf{r}^2}\cdot d\mathbf{l}. 
\end{equation}
\par 
While a our sample size of 141 materials is considerably smaller than recent high-throughout studies of topological materials\cite{tang2019efficient,zhang2019catalogue,vergniory2019complete,tang2019comprehensive,xu2020high}, the materials identified are experimentally realizable and support significant band gaps. This is in contrast to other high-throughput scans which, despite displaying tremendous technical progress, have not identified a topological insulator supporting a larger band-gap than previously known topological insulators\cite{lantagne2019topology}. 

\par 
\emph{1H-MX$_{2}$ Material Family:}
Transition metal dichalcogenides (TMDs) have emerged as a premier class of topological materials. In three-dimensions, $\gamma$-WTe$_{2}$ and $\gamma$-MoTe$_{2}$ have been identified as type-II Weyl semimetals\cite{soluyanov2015type,Wang2016Obs,li2017evidence,Wang2016MoTe2,Tamai2016,deng2016experimental,jiang2017signature} while $\beta$-MoTe$_{2}$ was proposed to be a higher-order topological insulator (HOTI)\cite{Wang2019Higher}. In two-dimensions, Qian et. al\cite{qian2014quantum} famously demonstrated that the 1T' phase of 2D TMDs in the MX$_{2}$ family with M=(molybdenum, tungsten) and X=(selenium, tellurium, or sulfur) are unit strength spin-Hall insulators. As a consequence, TMDs have received considerable experimental attention in the fields of Moire systems\cite{andrei2021marvels,Naik2018,Wu2018Hubbard,Zang2022,zhang2017interlayer,wang2020correlated, wang2015electronic}, topological superconductivity\cite{Castro2001,ye2012superconducting,taniguchi2012electric,shi2015superconductivity,Yuan2014TopSup,Chu2014,Zhou2016,hsu2017topological,CXLiu2017,li2021observation}, and beyond. 
\par
The 1H phase of MX$_{2}$ TMDs is constructed from a Bernal ABA stacking of two hexagonal layers of chalcogen atoms (X) with an intermediate hexagonal layer of a transition-metal (M)\cite{wang2012electronics,manzeli20172d}. A schematic of the structure can be seen in Fig. \eqref{fig:MX2Structure}, displaying the three-fold rotational symmetry. The 1H-MX$_{2}$ family has been reported to support band-gaps in the range of 1.1-1.8 eV. The calculated band-gaps are shown in Tab. \eqref{tab:deltaE}. Due to these values, giant in the context of topological insulators, they offer significant experimental control and are considered prime candidates for technological applications. The bulk electronic band structure along the high-symmetry path, $\Gamma - M -K -\Gamma$ is shown in Fig. \eqref{fig:MoTe2Bands} for MoTe$_{2}$, band structures for the remaining materials in this family are available in appendix C. As we are primarily concerned with bulk topological classification the surface spectra is relegated to appendix B, however we note that the surface spectra is gapped in each case.

\par 
To best visualize the effect of vortex insertion, we plot the local density of states at the vortex core in Fig. \eqref{fig:VortexWTe2MoTe2} as function of flux strength for MoTe$_{2}$ and WTe$_{2}$, reserving an identical analysis of the remaining TMDs for the supplementary material. In addition, we display the results of an exact-diagonalization calculation, fixing $\phi=\phi_{0}/2$, in the vicinity of the bulk-gap, demonstrating that there are four states bound to the flux tube. 
\par
To prove the existence of a double spin Chern number, we calculate charge induced on the vortex when all bound modes are occupied. This is done for $N_{e}$ occupied states by taking the difference in area charge densities, $\delta \sigma(\mathbf{r}_{i}, N_{e})=\sigma(\mathbf{r}_{i}, N_{e})_{1}-\sigma(\mathbf{r}_{i}, N_{e})_{0}$, where $\sigma(\mathbf{r}_{i}, N_{e})_{1}(\sigma(\mathbf{r}_{i}, N_{e})_{0})$ is the area charge density in the presence (absence) of the $\pi$-flux vortex. The total induced charge within a circle of radius $r$, centered at the flux tube, then follows as $\delta Q(r, N_{e})=\sum_{|\mathbf{r}_{i}|<r}\delta \sigma (\mathbf{r}_{i},N_{e})$. The results in Fig. \eqref{fig:VortexWTe2MoTe2}, provide a final and definite proof of the doubled ground-state topological invariant. 
\par 
\emph{Summary and impact:}
This work provides confirmation that multiple, experimentally relevant, high-quality two-dimensional materials admit non-trivial bulk topological classification invisible to symmetry indicators and the literature standard momentum space techniques. While thousands of topological materials have been identified in recent years, the materials presented in this work are unique in having an established track record for being experimentally controlled and \emph{supporting a significant direct band-gap.} As such, our results motivate a reexamination of experimental data.  
\par 
The presence of a non-zero ground state topological invariant has far-reaching experimental consequences. In the absence of spin-conservation symmetry an insulator identified with a ground-state bulk invariant, $\mathcal{C}_{s,G}=N$, need not realize a quantized value of spin-Hall conductivity $\sigma^{j=x,y,z}_{xy}=2Ne^{2}/h$. Nevertheless, an enhanced spin-Hall conductivity is expected and calculated for the 1H-MX$_{2}$ family in appendix C utilizing the WannierBerri software package\cite{tsirkin2021high}. Furthermore, in the presence of gapped surface states, it is possible for such phases to exhibit corner-localized modes. This has been demonstrated to be the case for 1H-MX$_{2}$ TMDs\cite{sodequist2022abundance}. 
\par 
Multiple studies have also reported on the presence of superconductivity in 1H-MX$_{2}$ systems\cite{ye2012superconducting,taniguchi2012electric,shi2015superconductivity}. While the question of whether the topology of the normal phase survives in the superconducting phase remains open, vortex insertion becomes a natural and realizable phenomenon in the superconducting phase. This opens a unique opportunity to access vortex-bound Majorana states.  
\par 
Finally, our work is unique in proposing a high-throughput search for topological materials through calculation of genuine bulk invariants rather than symmetry indices. While more computationally demanding, it suggests that many high-quality topological materials invisible to symmetry indicators exist, motivating a wider search. 
\par 

\bibliography{ref.bib}

\begin{thebibliography}{0}%
\makeatletter
\providecommand \@ifxundefined [1]{%
 \@ifx{#1\undefined}
}%
\providecommand \@ifnum [1]{%
 \ifnum #1\expandafter \@firstoftwo
 \else \expandafter \@secondoftwo
 \fi
}%
\providecommand \@ifx [1]{%
 \ifx #1\expandafter \@firstoftwo
 \else \expandafter \@secondoftwo
 \fi
}%
\providecommand \natexlab [1]{#1}%
\providecommand \enquote  [1]{``#1''}%
\providecommand \bibnamefont  [1]{#1}%
\providecommand \bibfnamefont [1]{#1}%
\providecommand \citenamefont [1]{#1}%
\providecommand \href@noop [0]{\@secondoftwo}%
\providecommand \href [0]{\begingroup \@sanitize@url \@href}%
\providecommand \@href[1]{\@@startlink{#1}\@@href}%
\providecommand \@@href[1]{\endgroup#1\@@endlink}%
\providecommand \@sanitize@url [0]{\catcode `\\12\catcode `\$12\catcode
  `\&12\catcode `\#12\catcode `\^12\catcode `\_12\catcode `\%12\relax}%
\providecommand \@@startlink[1]{}%
\providecommand \@@endlink[0]{}%
\providecommand \url  [0]{\begingroup\@sanitize@url \@url }%
\providecommand \@url [1]{\endgroup\@href {#1}{\urlprefix }}%
\providecommand \urlprefix  [0]{URL }%
\providecommand \Eprint [0]{\href }%
\providecommand \doibase [0]{http://dx.doi.org/}%
\providecommand \selectlanguage [0]{\@gobble}%
\providecommand \bibinfo  [0]{\@secondoftwo}%
\providecommand \bibfield  [0]{\@secondoftwo}%
\providecommand \translation [1]{[#1]}%
\providecommand \BibitemOpen [0]{}%
\providecommand \bibitemStop [0]{}%
\providecommand \bibitemNoStop [0]{.\EOS\space}%
\providecommand \EOS [0]{\spacefactor3000\relax}%
\providecommand \BibitemShut  [1]{\csname bibitem#1\endcsname}%
\let\auto@bib@innerbib\@empty
\end{thebibliography}%


\begin{thebibliography}{80}%
\makeatletter
\providecommand \@ifxundefined [1]{%
 \@ifx{#1\undefined}
}%
\providecommand \@ifnum [1]{%
 \ifnum #1\expandafter \@firstoftwo
 \else \expandafter \@secondoftwo
 \fi
}%
\providecommand \@ifx [1]{%
 \ifx #1\expandafter \@firstoftwo
 \else \expandafter \@secondoftwo
 \fi
}%
\providecommand \natexlab [1]{#1}%
\providecommand \enquote  [1]{``#1''}%
\providecommand \bibnamefont  [1]{#1}%
\providecommand \bibfnamefont [1]{#1}%
\providecommand \citenamefont [1]{#1}%
\providecommand \href@noop [0]{\@secondoftwo}%
\providecommand \href [0]{\begingroup \@sanitize@url \@href}%
\providecommand \@href[1]{\@@startlink{#1}\@@href}%
\providecommand \@@href[1]{\endgroup#1\@@endlink}%
\providecommand \@sanitize@url [0]{\catcode `\\12\catcode `\$12\catcode
  `\&12\catcode `\#12\catcode `\^12\catcode `\_12\catcode `\%12\relax}%
\providecommand \@@startlink[1]{}%
\providecommand \@@endlink[0]{}%
\providecommand \url  [0]{\begingroup\@sanitize@url \@url }%
\providecommand \@url [1]{\endgroup\@href {#1}{\urlprefix }}%
\providecommand \urlprefix  [0]{URL }%
\providecommand \Eprint [0]{\href }%
\providecommand \doibase [0]{http://dx.doi.org/}%
\providecommand \selectlanguage [0]{\@gobble}%
\providecommand \bibinfo  [0]{\@secondoftwo}%
\providecommand \bibfield  [0]{\@secondoftwo}%
\providecommand \translation [1]{[#1]}%
\providecommand \BibitemOpen [0]{}%
\providecommand \bibitemStop [0]{}%
\providecommand \bibitemNoStop [0]{.\EOS\space}%
\providecommand \EOS [0]{\spacefactor3000\relax}%
\providecommand \BibitemShut  [1]{\csname bibitem#1\endcsname}%
\let\auto@bib@innerbib\@empty
\bibitem [{\citenamefont {Mounet}\ \emph {et~al.}(2018)\citenamefont {Mounet},
  \citenamefont {Gibertini}, \citenamefont {Schwaller}, \citenamefont {Campi},
  \citenamefont {Merkys}, \citenamefont {Marrazzo}, \citenamefont {Sohier},
  \citenamefont {Castelli}, \citenamefont {Cepellotti}, \citenamefont {Pizzi}
  \emph {et~al.}}]{mounet2018two}%
  \BibitemOpen
  \bibfield  {author} {\bibinfo {author} {\bibfnamefont {N.}~\bibnamefont
  {Mounet}}, \bibinfo {author} {\bibfnamefont {M.}~\bibnamefont {Gibertini}},
  \bibinfo {author} {\bibfnamefont {P.}~\bibnamefont {Schwaller}}, \bibinfo
  {author} {\bibfnamefont {D.}~\bibnamefont {Campi}}, \bibinfo {author}
  {\bibfnamefont {A.}~\bibnamefont {Merkys}}, \bibinfo {author} {\bibfnamefont
  {A.}~\bibnamefont {Marrazzo}}, \bibinfo {author} {\bibfnamefont
  {T.}~\bibnamefont {Sohier}}, \bibinfo {author} {\bibfnamefont {I.~E.}\
  \bibnamefont {Castelli}}, \bibinfo {author} {\bibfnamefont {A.}~\bibnamefont
  {Cepellotti}}, \bibinfo {author} {\bibfnamefont {G.}~\bibnamefont {Pizzi}},
  \emph {et~al.},\ }\href {\doibase 10.1038/s41565-017-0035-5} {\bibfield
  {journal} {\bibinfo  {journal} {Nat. nanotechnol.}\ }\textbf {\bibinfo
  {volume} {13}},\ \bibinfo {pages} {246} (\bibinfo {year} {2018})}\BibitemShut
  {NoStop}%
\bibitem [{\citenamefont {Zhou}\ \emph {et~al.}(2019)\citenamefont {Zhou},
  \citenamefont {Shen}, \citenamefont {Costa}, \citenamefont {Persson},
  \citenamefont {Ong}, \citenamefont {Huck}, \citenamefont {Lu}, \citenamefont
  {Ma}, \citenamefont {Chen}, \citenamefont {Tang} \emph
  {et~al.}}]{zhou20192dmatpedia}%
  \BibitemOpen
  \bibfield  {author} {\bibinfo {author} {\bibfnamefont {J.}~\bibnamefont
  {Zhou}}, \bibinfo {author} {\bibfnamefont {L.}~\bibnamefont {Shen}}, \bibinfo
  {author} {\bibfnamefont {M.~D.}\ \bibnamefont {Costa}}, \bibinfo {author}
  {\bibfnamefont {K.~A.}\ \bibnamefont {Persson}}, \bibinfo {author}
  {\bibfnamefont {S.~P.}\ \bibnamefont {Ong}}, \bibinfo {author} {\bibfnamefont
  {P.}~\bibnamefont {Huck}}, \bibinfo {author} {\bibfnamefont {Y.}~\bibnamefont
  {Lu}}, \bibinfo {author} {\bibfnamefont {X.}~\bibnamefont {Ma}}, \bibinfo
  {author} {\bibfnamefont {Y.}~\bibnamefont {Chen}}, \bibinfo {author}
  {\bibfnamefont {H.}~\bibnamefont {Tang}},  \emph {et~al.},\ }\href {\doibase
  10.1038/s41597-019-0097-3} {\bibfield  {journal} {\bibinfo  {journal} {Sci.
  Data}\ }\textbf {\bibinfo {volume} {6}},\ \bibinfo {pages} {86} (\bibinfo
  {year} {2019})}\BibitemShut {NoStop}%
\bibitem [{\citenamefont {Rasmussen}\ and\ \citenamefont
  {Thygesen}(2015)}]{rasmussen2015computational}%
  \BibitemOpen
  \bibfield  {author} {\bibinfo {author} {\bibfnamefont {F.~A.}\ \bibnamefont
  {Rasmussen}}\ and\ \bibinfo {author} {\bibfnamefont {K.~S.}\ \bibnamefont
  {Thygesen}},\ }\href {\doibase 10.1021/acs.jpcc.5b02950} {\bibfield
  {journal} {\bibinfo  {journal} {J. Phys. Chem. C}\ }\textbf {\bibinfo
  {volume} {119}},\ \bibinfo {pages} {13169} (\bibinfo {year}
  {2015})}\BibitemShut {NoStop}%
\bibitem [{\citenamefont {Haastrup}\ \emph {et~al.}(2018)\citenamefont
  {Haastrup}, \citenamefont {Strange}, \citenamefont {Pandey}, \citenamefont
  {Deilmann}, \citenamefont {Schmidt}, \citenamefont {Hinsche}, \citenamefont
  {Gjerding}, \citenamefont {Torelli}, \citenamefont {Larsen}, \citenamefont
  {Riis-Jensen} \emph {et~al.}}]{haastrup2018computational}%
  \BibitemOpen
  \bibfield  {author} {\bibinfo {author} {\bibfnamefont {S.}~\bibnamefont
  {Haastrup}}, \bibinfo {author} {\bibfnamefont {M.}~\bibnamefont {Strange}},
  \bibinfo {author} {\bibfnamefont {M.}~\bibnamefont {Pandey}}, \bibinfo
  {author} {\bibfnamefont {T.}~\bibnamefont {Deilmann}}, \bibinfo {author}
  {\bibfnamefont {P.~S.}\ \bibnamefont {Schmidt}}, \bibinfo {author}
  {\bibfnamefont {N.~F.}\ \bibnamefont {Hinsche}}, \bibinfo {author}
  {\bibfnamefont {M.~N.}\ \bibnamefont {Gjerding}}, \bibinfo {author}
  {\bibfnamefont {D.}~\bibnamefont {Torelli}}, \bibinfo {author} {\bibfnamefont
  {P.~M.}\ \bibnamefont {Larsen}}, \bibinfo {author} {\bibfnamefont {A.~C.}\
  \bibnamefont {Riis-Jensen}},  \emph {et~al.},\ }\href {\doibase
  10.1088/2053-1583/aacfc1} {\bibfield  {journal} {\bibinfo  {journal} {2D
  Mat.}\ }\textbf {\bibinfo {volume} {5}},\ \bibinfo {pages} {042002} (\bibinfo
  {year} {2018})}\BibitemShut {NoStop}%
\bibitem [{\citenamefont {Murakami}(2006)}]{MurakamiQSHBi}%
  \BibitemOpen
  \bibfield  {author} {\bibinfo {author} {\bibfnamefont {S.}~\bibnamefont
  {Murakami}},\ }\href {\doibase 10.1103/PhysRevLett.97.236805} {\bibfield
  {journal} {\bibinfo  {journal} {Phys. Rev. Lett.}\ }\textbf {\bibinfo
  {volume} {97}},\ \bibinfo {pages} {236805} (\bibinfo {year}
  {2006})}\BibitemShut {NoStop}%
\bibitem [{\citenamefont {Wada}\ \emph {et~al.}(2011)\citenamefont {Wada},
  \citenamefont {Murakami}, \citenamefont {Freimuth},\ and\ \citenamefont
  {Bihlmayer}}]{WadaBi}%
  \BibitemOpen
  \bibfield  {author} {\bibinfo {author} {\bibfnamefont {M.}~\bibnamefont
  {Wada}}, \bibinfo {author} {\bibfnamefont {S.}~\bibnamefont {Murakami}},
  \bibinfo {author} {\bibfnamefont {F.}~\bibnamefont {Freimuth}}, \ and\
  \bibinfo {author} {\bibfnamefont {G.}~\bibnamefont {Bihlmayer}},\ }\href
  {\doibase 10.1103/PhysRevB.83.121310} {\bibfield  {journal} {\bibinfo
  {journal} {Phys. Rev. B}\ }\textbf {\bibinfo {volume} {83}},\ \bibinfo
  {pages} {121310} (\bibinfo {year} {2011})}\BibitemShut {NoStop}%
\bibitem [{\citenamefont {Kane}\ and\ \citenamefont {Mele}(2005)}]{Kane2005}%
  \BibitemOpen
  \bibfield  {author} {\bibinfo {author} {\bibfnamefont {C.~L.}\ \bibnamefont
  {Kane}}\ and\ \bibinfo {author} {\bibfnamefont {E.~J.}\ \bibnamefont
  {Mele}},\ }\href {\doibase 10.1103/PhysRevLett.95.146802} {\bibfield
  {journal} {\bibinfo  {journal} {Phys. Rev. Lett.}\ }\textbf {\bibinfo
  {volume} {95}},\ \bibinfo {pages} {146802} (\bibinfo {year}
  {2005})}\BibitemShut {NoStop}%
\bibitem [{\citenamefont {Fu}\ \emph {et~al.}(2007)\citenamefont {Fu},
  \citenamefont {Kane},\ and\ \citenamefont {Mele}}]{FuKaneMele2007}%
  \BibitemOpen
  \bibfield  {author} {\bibinfo {author} {\bibfnamefont {L.}~\bibnamefont
  {Fu}}, \bibinfo {author} {\bibfnamefont {C.~L.}\ \bibnamefont {Kane}}, \ and\
  \bibinfo {author} {\bibfnamefont {E.~J.}\ \bibnamefont {Mele}},\ }\href
  {\doibase 10.1103/PhysRevLett.98.106803} {\bibfield  {journal} {\bibinfo
  {journal} {Phys. Rev. Lett.}\ }\textbf {\bibinfo {volume} {98}},\ \bibinfo
  {pages} {106803} (\bibinfo {year} {2007})}\BibitemShut {NoStop}%
\bibitem [{\citenamefont {Fu}\ and\ \citenamefont {Kane}(2007)}]{FuKane}%
  \BibitemOpen
  \bibfield  {author} {\bibinfo {author} {\bibfnamefont {L.}~\bibnamefont
  {Fu}}\ and\ \bibinfo {author} {\bibfnamefont {C.~L.}\ \bibnamefont {Kane}},\
  }\href {\doibase 10.1103/PhysRevB.76.045302} {\bibfield  {journal} {\bibinfo
  {journal} {Phys. Rev. B}\ }\textbf {\bibinfo {volume} {76}},\ \bibinfo
  {pages} {045302} (\bibinfo {year} {2007})}\BibitemShut {NoStop}%
\bibitem [{\citenamefont {Roy}(2009)}]{Roy2009}%
  \BibitemOpen
  \bibfield  {author} {\bibinfo {author} {\bibfnamefont {R.}~\bibnamefont
  {Roy}},\ }\href {\doibase 10.1103/PhysRevB.79.195321} {\bibfield  {journal}
  {\bibinfo  {journal} {Phys. Rev. B}\ }\textbf {\bibinfo {volume} {79}},\
  \bibinfo {pages} {195321} (\bibinfo {year} {2009})}\BibitemShut {NoStop}%
\bibitem [{\citenamefont {Sheng}\ \emph {et~al.}(2006)\citenamefont {Sheng},
  \citenamefont {Weng}, \citenamefont {Sheng},\ and\ \citenamefont
  {Haldane}}]{Sheng2006}%
  \BibitemOpen
  \bibfield  {author} {\bibinfo {author} {\bibfnamefont {D.~N.}\ \bibnamefont
  {Sheng}}, \bibinfo {author} {\bibfnamefont {Z.~Y.}\ \bibnamefont {Weng}},
  \bibinfo {author} {\bibfnamefont {L.}~\bibnamefont {Sheng}}, \ and\ \bibinfo
  {author} {\bibfnamefont {F.~D.~M.}\ \bibnamefont {Haldane}},\ }\href
  {\doibase 10.1103/PhysRevLett.97.036808} {\bibfield  {journal} {\bibinfo
  {journal} {Phys. Rev. Lett.}\ }\textbf {\bibinfo {volume} {97}},\ \bibinfo
  {pages} {036808} (\bibinfo {year} {2006})}\BibitemShut {NoStop}%
\bibitem [{\citenamefont {Prodan}(2009)}]{Prodan2009}%
  \BibitemOpen
  \bibfield  {author} {\bibinfo {author} {\bibfnamefont {E.}~\bibnamefont
  {Prodan}},\ }\href {\doibase 10.1103/PhysRevB.80.125327} {\bibfield
  {journal} {\bibinfo  {journal} {Phys. Rev. B}\ }\textbf {\bibinfo {volume}
  {80}},\ \bibinfo {pages} {125327} (\bibinfo {year} {2009})}\BibitemShut
  {NoStop}%
\bibitem [{\citenamefont {Tyner}\ \emph {et~al.}(2020)\citenamefont {Tyner},
  \citenamefont {Sur}, \citenamefont {Puggioni}, \citenamefont {Rondinelli},\
  and\ \citenamefont {Goswami}}]{tyner2020topology}%
  \BibitemOpen
  \bibfield  {author} {\bibinfo {author} {\bibfnamefont {A.~C.}\ \bibnamefont
  {Tyner}}, \bibinfo {author} {\bibfnamefont {S.}~\bibnamefont {Sur}}, \bibinfo
  {author} {\bibfnamefont {D.}~\bibnamefont {Puggioni}}, \bibinfo {author}
  {\bibfnamefont {J.~M.}\ \bibnamefont {Rondinelli}}, \ and\ \bibinfo {author}
  {\bibfnamefont {P.}~\bibnamefont {Goswami}},\ }\href
  {https://doi.org/10.48550/arXiv.2012.12906} {\bibfield  {journal} {\bibinfo
  {journal} {arXiv:2012.12906}\ } (\bibinfo {year} {2020})}\BibitemShut
  {NoStop}%
\bibitem [{\citenamefont {Tyner}\ \emph {et~al.}(2021)\citenamefont {Tyner},
  \citenamefont {Sur}, \citenamefont {Zhou}, \citenamefont {Puggioni},
  \citenamefont {Darancet}, \citenamefont {Rondinelli},\ and\ \citenamefont
  {Goswami}}]{tyner2021quantized}%
  \BibitemOpen
  \bibfield  {author} {\bibinfo {author} {\bibfnamefont {A.~C.}\ \bibnamefont
  {Tyner}}, \bibinfo {author} {\bibfnamefont {S.}~\bibnamefont {Sur}}, \bibinfo
  {author} {\bibfnamefont {Q.}~\bibnamefont {Zhou}}, \bibinfo {author}
  {\bibfnamefont {D.}~\bibnamefont {Puggioni}}, \bibinfo {author}
  {\bibfnamefont {P.}~\bibnamefont {Darancet}}, \bibinfo {author}
  {\bibfnamefont {J.~M.}\ \bibnamefont {Rondinelli}}, \ and\ \bibinfo {author}
  {\bibfnamefont {P.}~\bibnamefont {Goswami}},\ }\href
  {https://doi.org/10.48550/arXiv.2102.06207} {\bibfield  {journal} {\bibinfo
  {journal} {arXiv:2102.06207}\ } (\bibinfo {year} {2021})}\BibitemShut
  {NoStop}%
\bibitem [{\citenamefont {Tyner}\ and\ \citenamefont
  {Goswami}(2022)}]{tynerbismuthene}%
  \BibitemOpen
  \bibfield  {author} {\bibinfo {author} {\bibfnamefont {A.~C.}\ \bibnamefont
  {Tyner}}\ and\ \bibinfo {author} {\bibfnamefont {P.}~\bibnamefont
  {Goswami}},\ }\href {\doibase 10.48550/arXiv.2209.13582} {\  (\bibinfo {year}
  {2022}),\ 10.48550/arXiv.2209.13582}\BibitemShut {NoStop}%
\bibitem [{\citenamefont {Bai}\ \emph {et~al.}(2022)\citenamefont {Bai},
  \citenamefont {Cai}, \citenamefont {Mao}, \citenamefont {Li}, \citenamefont
  {Dai}, \citenamefont {Huang},\ and\ \citenamefont {Niu}}]{Bai2022Doubled}%
  \BibitemOpen
  \bibfield  {author} {\bibinfo {author} {\bibfnamefont {Y.}~\bibnamefont
  {Bai}}, \bibinfo {author} {\bibfnamefont {L.}~\bibnamefont {Cai}}, \bibinfo
  {author} {\bibfnamefont {N.}~\bibnamefont {Mao}}, \bibinfo {author}
  {\bibfnamefont {R.}~\bibnamefont {Li}}, \bibinfo {author} {\bibfnamefont
  {Y.}~\bibnamefont {Dai}}, \bibinfo {author} {\bibfnamefont {B.}~\bibnamefont
  {Huang}}, \ and\ \bibinfo {author} {\bibfnamefont {C.}~\bibnamefont {Niu}},\
  }\href {\doibase 10.1103/PhysRevB.105.195142} {\bibfield  {journal} {\bibinfo
   {journal} {Phys. Rev. B}\ }\textbf {\bibinfo {volume} {105}},\ \bibinfo
  {pages} {195142} (\bibinfo {year} {2022})}\BibitemShut {NoStop}%
\bibitem [{\citenamefont {Wang}\ \emph {et~al.}(2022)\citenamefont {Wang},
  \citenamefont {Zhou}, \citenamefont {Lin}, \citenamefont {Lin},\ and\
  \citenamefont {Bansil}}]{bansilspin}%
  \BibitemOpen
  \bibfield  {author} {\bibinfo {author} {\bibfnamefont {B.}~\bibnamefont
  {Wang}}, \bibinfo {author} {\bibfnamefont {X.}~\bibnamefont {Zhou}}, \bibinfo
  {author} {\bibfnamefont {Y.-C.}\ \bibnamefont {Lin}}, \bibinfo {author}
  {\bibfnamefont {H.}~\bibnamefont {Lin}}, \ and\ \bibinfo {author}
  {\bibfnamefont {A.}~\bibnamefont {Bansil}},\ }\href
  {https://doi.org/10.48550/arXiv.2202.04162} {\bibfield  {journal} {\bibinfo
  {journal} {arXiv:2202.04162}\ } (\bibinfo {year} {2022})}\BibitemShut
  {NoStop}%
\bibitem [{\citenamefont {Yu}\ \emph {et~al.}(2011)\citenamefont {Yu},
  \citenamefont {Qi}, \citenamefont {Bernevig}, \citenamefont {Fang},\ and\
  \citenamefont {Dai}}]{yu2011equivalent}%
  \BibitemOpen
  \bibfield  {author} {\bibinfo {author} {\bibfnamefont {R.}~\bibnamefont
  {Yu}}, \bibinfo {author} {\bibfnamefont {X.~L.}\ \bibnamefont {Qi}}, \bibinfo
  {author} {\bibfnamefont {A.}~\bibnamefont {Bernevig}}, \bibinfo {author}
  {\bibfnamefont {Z.}~\bibnamefont {Fang}}, \ and\ \bibinfo {author}
  {\bibfnamefont {X.}~\bibnamefont {Dai}},\ }\href {\doibase
  10.1103/PhysRevB.84.075119} {\bibfield  {journal} {\bibinfo  {journal} {Phys.
  Rev. B}\ }\textbf {\bibinfo {volume} {84}},\ \bibinfo {pages} {075119}
  (\bibinfo {year} {2011})}\BibitemShut {NoStop}%
\bibitem [{\citenamefont {Soluyanov}\ and\ \citenamefont
  {Vanderbilt}(2011)}]{Soluyanov2011}%
  \BibitemOpen
  \bibfield  {author} {\bibinfo {author} {\bibfnamefont {A.~A.}\ \bibnamefont
  {Soluyanov}}\ and\ \bibinfo {author} {\bibfnamefont {D.}~\bibnamefont
  {Vanderbilt}},\ }\href {\doibase 10.1103/PhysRevB.83.235401} {\bibfield
  {journal} {\bibinfo  {journal} {Phys. Rev. B}\ }\textbf {\bibinfo {volume}
  {83}},\ \bibinfo {pages} {235401} (\bibinfo {year} {2011})}\BibitemShut
  {NoStop}%
\bibitem [{\citenamefont {Alexandradinata}\ \emph {et~al.}(2014)\citenamefont
  {Alexandradinata}, \citenamefont {Dai},\ and\ \citenamefont
  {Bernevig}}]{alexandradinata2014wilson}%
  \BibitemOpen
  \bibfield  {author} {\bibinfo {author} {\bibfnamefont {A.}~\bibnamefont
  {Alexandradinata}}, \bibinfo {author} {\bibfnamefont {X.}~\bibnamefont
  {Dai}}, \ and\ \bibinfo {author} {\bibfnamefont {B.~A.}\ \bibnamefont
  {Bernevig}},\ }\href {\doibase 10.1103/PhysRevB.89.155114} {\bibfield
  {journal} {\bibinfo  {journal} {Phys. Rev. B}\ }\textbf {\bibinfo {volume}
  {89}},\ \bibinfo {pages} {155114} (\bibinfo {year} {2014})}\BibitemShut
  {NoStop}%
\bibitem [{\citenamefont {Bouhon}\ \emph {et~al.}(2019)\citenamefont {Bouhon},
  \citenamefont {Black-Schaffer},\ and\ \citenamefont
  {Slager}}]{bouhon2019wilson}%
  \BibitemOpen
  \bibfield  {author} {\bibinfo {author} {\bibfnamefont {A.}~\bibnamefont
  {Bouhon}}, \bibinfo {author} {\bibfnamefont {A.~M.}\ \bibnamefont
  {Black-Schaffer}}, \ and\ \bibinfo {author} {\bibfnamefont {R.-J.}\
  \bibnamefont {Slager}},\ }\href {\doibase 10.1103/PhysRevB.100.195135}
  {\bibfield  {journal} {\bibinfo  {journal} {Phys. Rev. B}\ }\textbf {\bibinfo
  {volume} {100}},\ \bibinfo {pages} {195135} (\bibinfo {year}
  {2019})}\BibitemShut {NoStop}%
\bibitem [{\citenamefont {Bradlyn}\ \emph {et~al.}(2019)\citenamefont
  {Bradlyn}, \citenamefont {Wang}, \citenamefont {Cano},\ and\ \citenamefont
  {Bernevig}}]{bradlyn2019disconnected}%
  \BibitemOpen
  \bibfield  {author} {\bibinfo {author} {\bibfnamefont {B.}~\bibnamefont
  {Bradlyn}}, \bibinfo {author} {\bibfnamefont {Z.}~\bibnamefont {Wang}},
  \bibinfo {author} {\bibfnamefont {J.}~\bibnamefont {Cano}}, \ and\ \bibinfo
  {author} {\bibfnamefont {B.~A.}\ \bibnamefont {Bernevig}},\ }\href {\doibase
  10.1103/PhysRevB.99.045140} {\bibfield  {journal} {\bibinfo  {journal} {Phys.
  Rev. B}\ }\textbf {\bibinfo {volume} {99}},\ \bibinfo {pages} {045140}
  (\bibinfo {year} {2019})}\BibitemShut {NoStop}%
\bibitem [{\citenamefont {Gresch}\ \emph {et~al.}(2017)\citenamefont {Gresch},
  \citenamefont {Aut\`es}, \citenamefont {Yazyev}, \citenamefont {Troyer},
  \citenamefont {Vanderbilt}, \citenamefont {Bernevig},\ and\ \citenamefont
  {Soluyanov}}]{Z2pack}%
  \BibitemOpen
  \bibfield  {author} {\bibinfo {author} {\bibfnamefont {D.}~\bibnamefont
  {Gresch}}, \bibinfo {author} {\bibfnamefont {G.}~\bibnamefont {Aut\`es}},
  \bibinfo {author} {\bibfnamefont {O.~V.}\ \bibnamefont {Yazyev}}, \bibinfo
  {author} {\bibfnamefont {M.}~\bibnamefont {Troyer}}, \bibinfo {author}
  {\bibfnamefont {D.}~\bibnamefont {Vanderbilt}}, \bibinfo {author}
  {\bibfnamefont {B.~A.}\ \bibnamefont {Bernevig}}, \ and\ \bibinfo {author}
  {\bibfnamefont {A.~A.}\ \bibnamefont {Soluyanov}},\ }\href {\doibase
  10.1103/PhysRevB.95.075146} {\bibfield  {journal} {\bibinfo  {journal} {Phys.
  Rev. B}\ }\textbf {\bibinfo {volume} {95}},\ \bibinfo {pages} {075146}
  (\bibinfo {year} {2017})}\BibitemShut {NoStop}%
\bibitem [{\citenamefont {Taherinejad}\ \emph {et~al.}(2014)\citenamefont
  {Taherinejad}, \citenamefont {Garrity},\ and\ \citenamefont
  {Vanderbilt}}]{Taherinejad2014}%
  \BibitemOpen
  \bibfield  {author} {\bibinfo {author} {\bibfnamefont {M.}~\bibnamefont
  {Taherinejad}}, \bibinfo {author} {\bibfnamefont {K.~F.}\ \bibnamefont
  {Garrity}}, \ and\ \bibinfo {author} {\bibfnamefont {D.}~\bibnamefont
  {Vanderbilt}},\ }\href {\doibase 10.1103/PhysRevB.89.115102} {\bibfield
  {journal} {\bibinfo  {journal} {Phys. Rev. B}\ }\textbf {\bibinfo {volume}
  {89}},\ \bibinfo {pages} {1} (\bibinfo {year} {2014})},\ \Eprint
  {http://arxiv.org/abs/1312.6940} {1312.6940} \BibitemShut {NoStop}%
\bibitem [{\citenamefont {Lin}\ \emph {et~al.}(2022)\citenamefont {Lin},
  \citenamefont {Palumbo}, \citenamefont {Guo}, \citenamefont {Blackburn},
  \citenamefont {Shoemaker}, \citenamefont {Mahmood}, \citenamefont {Wang},
  \citenamefont {Fiete}, \citenamefont {Wieder},\ and\ \citenamefont
  {Bradlyn}}]{Lin2022Spin}%
  \BibitemOpen
  \bibfield  {author} {\bibinfo {author} {\bibfnamefont {K.-S.}\ \bibnamefont
  {Lin}}, \bibinfo {author} {\bibfnamefont {G.}~\bibnamefont {Palumbo}},
  \bibinfo {author} {\bibfnamefont {Z.}~\bibnamefont {Guo}}, \bibinfo {author}
  {\bibfnamefont {J.}~\bibnamefont {Blackburn}}, \bibinfo {author}
  {\bibfnamefont {D.}~\bibnamefont {Shoemaker}}, \bibinfo {author}
  {\bibfnamefont {F.}~\bibnamefont {Mahmood}}, \bibinfo {author} {\bibfnamefont
  {Z.}~\bibnamefont {Wang}}, \bibinfo {author} {\bibfnamefont {G.}~\bibnamefont
  {Fiete}}, \bibinfo {author} {\bibfnamefont {B.}~\bibnamefont {Wieder}}, \
  and\ \bibinfo {author} {\bibfnamefont {B.}~\bibnamefont {Bradlyn}},\ }\href
  {https://doi.org/10.48550/arXiv:2207.10099} {\bibfield  {journal} {\bibinfo
  {journal} {arXiv:2207.10099}\ } (\bibinfo {year} {2022})}\BibitemShut
  {NoStop}%
\bibitem [{\citenamefont {Costa}\ \emph {et~al.}(2021)\citenamefont {Costa},
  \citenamefont {Schleder}, \citenamefont {Mera~Acosta}, \citenamefont
  {Padilha}, \citenamefont {Cerasoli}, \citenamefont {Buongiorno~Nardelli},\
  and\ \citenamefont {Fazzio}}]{costa2021discovery}%
  \BibitemOpen
  \bibfield  {author} {\bibinfo {author} {\bibfnamefont {M.}~\bibnamefont
  {Costa}}, \bibinfo {author} {\bibfnamefont {G.~R.}\ \bibnamefont {Schleder}},
  \bibinfo {author} {\bibfnamefont {C.}~\bibnamefont {Mera~Acosta}}, \bibinfo
  {author} {\bibfnamefont {A.~C.}\ \bibnamefont {Padilha}}, \bibinfo {author}
  {\bibfnamefont {F.}~\bibnamefont {Cerasoli}}, \bibinfo {author}
  {\bibfnamefont {M.}~\bibnamefont {Buongiorno~Nardelli}}, \ and\ \bibinfo
  {author} {\bibfnamefont {A.}~\bibnamefont {Fazzio}},\ }\href {\doibase
  10.1038/s41524-021-00518-4} {\bibfield  {journal} {\bibinfo  {journal} {npj
  Comp. Mat.}\ }\textbf {\bibinfo {volume} {7}},\ \bibinfo {pages} {49}
  (\bibinfo {year} {2021})}\BibitemShut {NoStop}%
\bibitem [{\citenamefont {S{\o}dequist}\ \emph {et~al.}(2022)\citenamefont
  {S{\o}dequist}, \citenamefont {Petralanda},\ and\ \citenamefont
  {Olsen}}]{sodequist2022abundance}%
  \BibitemOpen
  \bibfield  {author} {\bibinfo {author} {\bibfnamefont {J.}~\bibnamefont
  {S{\o}dequist}}, \bibinfo {author} {\bibfnamefont {U.}~\bibnamefont
  {Petralanda}}, \ and\ \bibinfo {author} {\bibfnamefont {T.}~\bibnamefont
  {Olsen}},\ }\href {\doibase 10.1088/2053-1583/ac9fe2} {\bibfield  {journal}
  {\bibinfo  {journal} {2D Mat.}\ }\textbf {\bibinfo {volume} {10}},\ \bibinfo
  {pages} {015009} (\bibinfo {year} {2022})}\BibitemShut {NoStop}%
\bibitem [{\citenamefont {Qi}\ and\ \citenamefont
  {Zhang}(2008)}]{QiSpinCharge}%
  \BibitemOpen
  \bibfield  {author} {\bibinfo {author} {\bibfnamefont {X.-L.}\ \bibnamefont
  {Qi}}\ and\ \bibinfo {author} {\bibfnamefont {S.-C.}\ \bibnamefont {Zhang}},\
  }\href {\doibase 10.1103/PhysRevLett.101.086802} {\bibfield  {journal}
  {\bibinfo  {journal} {Phys. Rev. Lett.}\ }\textbf {\bibinfo {volume} {101}},\
  \bibinfo {pages} {086802} (\bibinfo {year} {2008})}\BibitemShut {NoStop}%
\bibitem [{\citenamefont {Ran}\ \emph {et~al.}(2008)\citenamefont {Ran},
  \citenamefont {Vishwanath},\ and\ \citenamefont
  {Lee}}]{SpinChargeVishwanath}%
  \BibitemOpen
  \bibfield  {author} {\bibinfo {author} {\bibfnamefont {Y.}~\bibnamefont
  {Ran}}, \bibinfo {author} {\bibfnamefont {A.}~\bibnamefont {Vishwanath}}, \
  and\ \bibinfo {author} {\bibfnamefont {D.-H.}\ \bibnamefont {Lee}},\ }\href
  {\doibase 10.1103/PhysRevLett.101.086801} {\bibfield  {journal} {\bibinfo
  {journal} {Phys. Rev. Lett.}\ }\textbf {\bibinfo {volume} {101}},\ \bibinfo
  {pages} {086801} (\bibinfo {year} {2008})}\BibitemShut {NoStop}%
\bibitem [{\citenamefont {Juri\ifmmode \check{c}\else
  \v{c}\fi{}i\ifmmode~\acute{c}\else \'{c}\fi{}}\ \emph
  {et~al.}(2012)\citenamefont {Juri\ifmmode \check{c}\else
  \v{c}\fi{}i\ifmmode~\acute{c}\else \'{c}\fi{}}, \citenamefont {Mesaros},
  \citenamefont {Slager},\ and\ \citenamefont {Zaanen}}]{slager2012}%
  \BibitemOpen
  \bibfield  {author} {\bibinfo {author} {\bibfnamefont {V.}~\bibnamefont
  {Juri\ifmmode \check{c}\else \v{c}\fi{}i\ifmmode~\acute{c}\else \'{c}\fi{}}},
  \bibinfo {author} {\bibfnamefont {A.}~\bibnamefont {Mesaros}}, \bibinfo
  {author} {\bibfnamefont {R.-J.}\ \bibnamefont {Slager}}, \ and\ \bibinfo
  {author} {\bibfnamefont {J.}~\bibnamefont {Zaanen}},\ }\href {\doibase
  10.1103/PhysRevLett.108.106403} {\bibfield  {journal} {\bibinfo  {journal}
  {Phys. Rev. Lett.}\ }\textbf {\bibinfo {volume} {108}},\ \bibinfo {pages}
  {106403} (\bibinfo {year} {2012})}\BibitemShut {NoStop}%
\bibitem [{\citenamefont {Mesaros}\ \emph {et~al.}(2013)\citenamefont
  {Mesaros}, \citenamefont {Slager}, \citenamefont {Zaanen},\ and\
  \citenamefont {Juri{\v{c}}i{\'c}}}]{MESAROS2013977}%
  \BibitemOpen
  \bibfield  {author} {\bibinfo {author} {\bibfnamefont {A.}~\bibnamefont
  {Mesaros}}, \bibinfo {author} {\bibfnamefont {R.-J.}\ \bibnamefont {Slager}},
  \bibinfo {author} {\bibfnamefont {J.}~\bibnamefont {Zaanen}}, \ and\ \bibinfo
  {author} {\bibfnamefont {V.}~\bibnamefont {Juri{\v{c}}i{\'c}}},\ }\href
  {\doibase 10.1016/j.nuclphysb.2012.10.022} {\bibfield  {journal} {\bibinfo
  {journal} {Nuc. Phys. B}\ }\textbf {\bibinfo {volume} {867}},\ \bibinfo
  {pages} {977} (\bibinfo {year} {2013})}\BibitemShut {NoStop}%
\bibitem [{\citenamefont {Wang}\ and\ \citenamefont {Zhang}(2010)}]{Wang_2010}%
  \BibitemOpen
  \bibfield  {author} {\bibinfo {author} {\bibfnamefont {Z.}~\bibnamefont
  {Wang}}\ and\ \bibinfo {author} {\bibfnamefont {P.}~\bibnamefont {Zhang}},\
  }\href {\doibase 10.1088/1367-2630/12/4/043055} {\bibfield  {journal}
  {\bibinfo  {journal} {New Journal of Physics}\ }\textbf {\bibinfo {volume}
  {12}},\ \bibinfo {pages} {043055} (\bibinfo {year} {2010})}\BibitemShut
  {NoStop}%
\bibitem [{\citenamefont {Schindler}\ \emph {et~al.}(2022)\citenamefont
  {Schindler}, \citenamefont {Tsirkin}, \citenamefont {Neupert}, \citenamefont
  {Andrei~Bernevig},\ and\ \citenamefont {Wieder}}]{schindler2022topological}%
  \BibitemOpen
  \bibfield  {author} {\bibinfo {author} {\bibfnamefont {F.}~\bibnamefont
  {Schindler}}, \bibinfo {author} {\bibfnamefont {S.~S.}\ \bibnamefont
  {Tsirkin}}, \bibinfo {author} {\bibfnamefont {T.}~\bibnamefont {Neupert}},
  \bibinfo {author} {\bibfnamefont {B.}~\bibnamefont {Andrei~Bernevig}}, \ and\
  \bibinfo {author} {\bibfnamefont {B.~J.}\ \bibnamefont {Wieder}},\ }\href
  {\doibase 10.1038/s41467-022-33471-x} {\bibfield  {journal} {\bibinfo
  {journal} {Nat. commun.}\ }\textbf {\bibinfo {volume} {13}},\ \bibinfo
  {pages} {5791} (\bibinfo {year} {2022})}\BibitemShut {NoStop}%
\bibitem [{\citenamefont {Ye}\ \emph {et~al.}(2012)\citenamefont {Ye},
  \citenamefont {Zhang}, \citenamefont {Akashi}, \citenamefont {Bahramy},
  \citenamefont {Arita},\ and\ \citenamefont {Iwasa}}]{ye2012superconducting}%
  \BibitemOpen
  \bibfield  {author} {\bibinfo {author} {\bibfnamefont {J.}~\bibnamefont
  {Ye}}, \bibinfo {author} {\bibfnamefont {Y.~J.}\ \bibnamefont {Zhang}},
  \bibinfo {author} {\bibfnamefont {R.}~\bibnamefont {Akashi}}, \bibinfo
  {author} {\bibfnamefont {M.~S.}\ \bibnamefont {Bahramy}}, \bibinfo {author}
  {\bibfnamefont {R.}~\bibnamefont {Arita}}, \ and\ \bibinfo {author}
  {\bibfnamefont {Y.}~\bibnamefont {Iwasa}},\ }\href {\doibase
  10.1126/science.1228006} {\bibfield  {journal} {\bibinfo  {journal}
  {Science}\ }\textbf {\bibinfo {volume} {338}},\ \bibinfo {pages} {1193}
  (\bibinfo {year} {2012})}\BibitemShut {NoStop}%
\bibitem [{\citenamefont {Taniguchi}\ \emph {et~al.}(2012)\citenamefont
  {Taniguchi}, \citenamefont {Matsumoto}, \citenamefont {Shimotani},\ and\
  \citenamefont {Takagi}}]{taniguchi2012electric}%
  \BibitemOpen
  \bibfield  {author} {\bibinfo {author} {\bibfnamefont {K.}~\bibnamefont
  {Taniguchi}}, \bibinfo {author} {\bibfnamefont {A.}~\bibnamefont
  {Matsumoto}}, \bibinfo {author} {\bibfnamefont {H.}~\bibnamefont
  {Shimotani}}, \ and\ \bibinfo {author} {\bibfnamefont {H.}~\bibnamefont
  {Takagi}},\ }\href {\doibase 10.1063/1.4740268} {\bibfield  {journal}
  {\bibinfo  {journal} {App. Phys. Lett.}\ }\textbf {\bibinfo {volume} {101}},\
  \bibinfo {pages} {042603} (\bibinfo {year} {2012})}\BibitemShut {NoStop}%
\bibitem [{\citenamefont {Shi}\ \emph {et~al.}(2015)\citenamefont {Shi},
  \citenamefont {Ye}, \citenamefont {Zhang}, \citenamefont {Suzuki},
  \citenamefont {Yoshida}, \citenamefont {Miyazaki}, \citenamefont {Inoue},
  \citenamefont {Saito},\ and\ \citenamefont
  {Iwasa}}]{shi2015superconductivity}%
  \BibitemOpen
  \bibfield  {author} {\bibinfo {author} {\bibfnamefont {W.}~\bibnamefont
  {Shi}}, \bibinfo {author} {\bibfnamefont {J.}~\bibnamefont {Ye}}, \bibinfo
  {author} {\bibfnamefont {Y.}~\bibnamefont {Zhang}}, \bibinfo {author}
  {\bibfnamefont {R.}~\bibnamefont {Suzuki}}, \bibinfo {author} {\bibfnamefont
  {M.}~\bibnamefont {Yoshida}}, \bibinfo {author} {\bibfnamefont
  {J.}~\bibnamefont {Miyazaki}}, \bibinfo {author} {\bibfnamefont
  {N.}~\bibnamefont {Inoue}}, \bibinfo {author} {\bibfnamefont
  {Y.}~\bibnamefont {Saito}}, \ and\ \bibinfo {author} {\bibfnamefont
  {Y.}~\bibnamefont {Iwasa}},\ }\href {\doibase 10.1038/srep12534} {\bibfield
  {journal} {\bibinfo  {journal} {Sci. rep.}\ }\textbf {\bibinfo {volume}
  {5}},\ \bibinfo {pages} {1} (\bibinfo {year} {2015})}\BibitemShut {NoStop}%
\bibitem [{\citenamefont {Giannozzi}\ \emph {et~al.}(2009)\citenamefont
  {Giannozzi}, \citenamefont {Baroni}, \citenamefont {Bonini}, \citenamefont
  {Calandra}, \citenamefont {Car}, \citenamefont {Cavazzoni}, \citenamefont
  {Ceresoli}, \citenamefont {Chiarotti}, \citenamefont {Cococcioni},
  \citenamefont {Dabo}, \citenamefont {{Dal Corso}}, \citenamefont
  {de~Gironcoli}, \citenamefont {Fabris}, \citenamefont {Fratesi},
  \citenamefont {Gebauer}, \citenamefont {Gerstmann}, \citenamefont
  {Gougoussis}, \citenamefont {Kokalj}, \citenamefont {Lazzeri}, \citenamefont
  {Martin-Samos}, \citenamefont {Marzari}, \citenamefont {Mauri}, \citenamefont
  {Mazzarello}, \citenamefont {Paolini}, \citenamefont {Pasquarello},
  \citenamefont {Paulatto}, \citenamefont {Sbraccia}, \citenamefont {Scandolo},
  \citenamefont {Sclauzero}, \citenamefont {Seitsonen}, \citenamefont
  {Smogunov}, \citenamefont {Umari},\ and\ \citenamefont
  {Wentzcovitch}}]{QE-2009}%
  \BibitemOpen
  \bibfield  {author} {\bibinfo {author} {\bibfnamefont {P.}~\bibnamefont
  {Giannozzi}}, \bibinfo {author} {\bibfnamefont {S.}~\bibnamefont {Baroni}},
  \bibinfo {author} {\bibfnamefont {N.}~\bibnamefont {Bonini}}, \bibinfo
  {author} {\bibfnamefont {M.}~\bibnamefont {Calandra}}, \bibinfo {author}
  {\bibfnamefont {R.}~\bibnamefont {Car}}, \bibinfo {author} {\bibfnamefont
  {C.}~\bibnamefont {Cavazzoni}}, \bibinfo {author} {\bibfnamefont
  {D.}~\bibnamefont {Ceresoli}}, \bibinfo {author} {\bibfnamefont {G.~L.}\
  \bibnamefont {Chiarotti}}, \bibinfo {author} {\bibfnamefont {M.}~\bibnamefont
  {Cococcioni}}, \bibinfo {author} {\bibfnamefont {I.}~\bibnamefont {Dabo}},
  \bibinfo {author} {\bibfnamefont {A.}~\bibnamefont {{Dal Corso}}}, \bibinfo
  {author} {\bibfnamefont {S.}~\bibnamefont {de~Gironcoli}}, \bibinfo {author}
  {\bibfnamefont {S.}~\bibnamefont {Fabris}}, \bibinfo {author} {\bibfnamefont
  {G.}~\bibnamefont {Fratesi}}, \bibinfo {author} {\bibfnamefont
  {R.}~\bibnamefont {Gebauer}}, \bibinfo {author} {\bibfnamefont
  {U.}~\bibnamefont {Gerstmann}}, \bibinfo {author} {\bibfnamefont
  {C.}~\bibnamefont {Gougoussis}}, \bibinfo {author} {\bibfnamefont
  {A.}~\bibnamefont {Kokalj}}, \bibinfo {author} {\bibfnamefont
  {M.}~\bibnamefont {Lazzeri}}, \bibinfo {author} {\bibfnamefont
  {L.}~\bibnamefont {Martin-Samos}}, \bibinfo {author} {\bibfnamefont
  {N.}~\bibnamefont {Marzari}}, \bibinfo {author} {\bibfnamefont
  {F.}~\bibnamefont {Mauri}}, \bibinfo {author} {\bibfnamefont
  {R.}~\bibnamefont {Mazzarello}}, \bibinfo {author} {\bibfnamefont
  {S.}~\bibnamefont {Paolini}}, \bibinfo {author} {\bibfnamefont
  {A.}~\bibnamefont {Pasquarello}}, \bibinfo {author} {\bibfnamefont
  {L.}~\bibnamefont {Paulatto}}, \bibinfo {author} {\bibfnamefont
  {C.}~\bibnamefont {Sbraccia}}, \bibinfo {author} {\bibfnamefont
  {S.}~\bibnamefont {Scandolo}}, \bibinfo {author} {\bibfnamefont
  {G.}~\bibnamefont {Sclauzero}}, \bibinfo {author} {\bibfnamefont {A.~P.}\
  \bibnamefont {Seitsonen}}, \bibinfo {author} {\bibfnamefont {A.}~\bibnamefont
  {Smogunov}}, \bibinfo {author} {\bibfnamefont {P.}~\bibnamefont {Umari}}, \
  and\ \bibinfo {author} {\bibfnamefont {R.~M.}\ \bibnamefont {Wentzcovitch}},\
  }\href {http://www.quantum-espresso.org} {\bibfield  {journal} {\bibinfo
  {journal} {J. Phys. Condens. Matter}\ }\textbf {\bibinfo {volume} {21}},\
  \bibinfo {pages} {395502 (19pp)} (\bibinfo {year} {2009})}\BibitemShut
  {NoStop}%
\bibitem [{\citenamefont {Giannozzi}\ \emph {et~al.}(2017)\citenamefont
  {Giannozzi}, \citenamefont {Andreussi}, \citenamefont {Brumme}, \citenamefont
  {Bunau}, \citenamefont {Nardelli}, \citenamefont {Calandra}, \citenamefont
  {Car}, \citenamefont {Cavazzoni}, \citenamefont {Ceresoli}, \citenamefont
  {Cococcioni}, \citenamefont {Colonna}, \citenamefont {Carnimeo},
  \citenamefont {Corso}, \citenamefont {de~Gironcoli}, \citenamefont {Delugas},
  \citenamefont {Jr}, \citenamefont {Ferretti}, \citenamefont {Floris},
  \citenamefont {Fratesi}, \citenamefont {Fugallo}, \citenamefont {Gebauer},
  \citenamefont {Gerstmann}, \citenamefont {Giustino}, \citenamefont {Gorni},
  \citenamefont {Jia}, \citenamefont {Kawamura}, \citenamefont {Ko},
  \citenamefont {Kokalj}, \citenamefont {Küçükbenli}, \citenamefont
  {Lazzeri}, \citenamefont {Marsili}, \citenamefont {Marzari}, \citenamefont
  {Mauri}, \citenamefont {Nguyen}, \citenamefont {Nguyen}, \citenamefont {de-la
  Roza}, \citenamefont {Paulatto}, \citenamefont {Poncé}, \citenamefont
  {Rocca}, \citenamefont {Sabatini}, \citenamefont {Santra}, \citenamefont
  {Schlipf}, \citenamefont {Seitsonen}, \citenamefont {Smogunov}, \citenamefont
  {Timrov}, \citenamefont {Thonhauser}, \citenamefont {Umari}, \citenamefont
  {Vast}, \citenamefont {Wu},\ and\ \citenamefont {Baroni}}]{QE-2017}%
  \BibitemOpen
  \bibfield  {author} {\bibinfo {author} {\bibfnamefont {P.}~\bibnamefont
  {Giannozzi}}, \bibinfo {author} {\bibfnamefont {O.}~\bibnamefont
  {Andreussi}}, \bibinfo {author} {\bibfnamefont {T.}~\bibnamefont {Brumme}},
  \bibinfo {author} {\bibfnamefont {O.}~\bibnamefont {Bunau}}, \bibinfo
  {author} {\bibfnamefont {M.~B.}\ \bibnamefont {Nardelli}}, \bibinfo {author}
  {\bibfnamefont {M.}~\bibnamefont {Calandra}}, \bibinfo {author}
  {\bibfnamefont {R.}~\bibnamefont {Car}}, \bibinfo {author} {\bibfnamefont
  {C.}~\bibnamefont {Cavazzoni}}, \bibinfo {author} {\bibfnamefont
  {D.}~\bibnamefont {Ceresoli}}, \bibinfo {author} {\bibfnamefont
  {M.}~\bibnamefont {Cococcioni}}, \bibinfo {author} {\bibfnamefont
  {N.}~\bibnamefont {Colonna}}, \bibinfo {author} {\bibfnamefont
  {I.}~\bibnamefont {Carnimeo}}, \bibinfo {author} {\bibfnamefont {A.~D.}\
  \bibnamefont {Corso}}, \bibinfo {author} {\bibfnamefont {S.}~\bibnamefont
  {de~Gironcoli}}, \bibinfo {author} {\bibfnamefont {P.}~\bibnamefont
  {Delugas}}, \bibinfo {author} {\bibfnamefont {R.~A.~D.}\ \bibnamefont {Jr}},
  \bibinfo {author} {\bibfnamefont {A.}~\bibnamefont {Ferretti}}, \bibinfo
  {author} {\bibfnamefont {A.}~\bibnamefont {Floris}}, \bibinfo {author}
  {\bibfnamefont {G.}~\bibnamefont {Fratesi}}, \bibinfo {author} {\bibfnamefont
  {G.}~\bibnamefont {Fugallo}}, \bibinfo {author} {\bibfnamefont
  {R.}~\bibnamefont {Gebauer}}, \bibinfo {author} {\bibfnamefont
  {U.}~\bibnamefont {Gerstmann}}, \bibinfo {author} {\bibfnamefont
  {F.}~\bibnamefont {Giustino}}, \bibinfo {author} {\bibfnamefont
  {T.}~\bibnamefont {Gorni}}, \bibinfo {author} {\bibfnamefont
  {J.}~\bibnamefont {Jia}}, \bibinfo {author} {\bibfnamefont {M.}~\bibnamefont
  {Kawamura}}, \bibinfo {author} {\bibfnamefont {H.-Y.}\ \bibnamefont {Ko}},
  \bibinfo {author} {\bibfnamefont {A.}~\bibnamefont {Kokalj}}, \bibinfo
  {author} {\bibfnamefont {E.}~\bibnamefont {Küçükbenli}}, \bibinfo {author}
  {\bibfnamefont {M.}~\bibnamefont {Lazzeri}}, \bibinfo {author} {\bibfnamefont
  {M.}~\bibnamefont {Marsili}}, \bibinfo {author} {\bibfnamefont
  {N.}~\bibnamefont {Marzari}}, \bibinfo {author} {\bibfnamefont
  {F.}~\bibnamefont {Mauri}}, \bibinfo {author} {\bibfnamefont {N.~L.}\
  \bibnamefont {Nguyen}}, \bibinfo {author} {\bibfnamefont {H.-V.}\
  \bibnamefont {Nguyen}}, \bibinfo {author} {\bibfnamefont {A.~O.}\
  \bibnamefont {de-la Roza}}, \bibinfo {author} {\bibfnamefont
  {L.}~\bibnamefont {Paulatto}}, \bibinfo {author} {\bibfnamefont
  {S.}~\bibnamefont {Poncé}}, \bibinfo {author} {\bibfnamefont
  {D.}~\bibnamefont {Rocca}}, \bibinfo {author} {\bibfnamefont
  {R.}~\bibnamefont {Sabatini}}, \bibinfo {author} {\bibfnamefont
  {B.}~\bibnamefont {Santra}}, \bibinfo {author} {\bibfnamefont
  {M.}~\bibnamefont {Schlipf}}, \bibinfo {author} {\bibfnamefont {A.~P.}\
  \bibnamefont {Seitsonen}}, \bibinfo {author} {\bibfnamefont {A.}~\bibnamefont
  {Smogunov}}, \bibinfo {author} {\bibfnamefont {I.}~\bibnamefont {Timrov}},
  \bibinfo {author} {\bibfnamefont {T.}~\bibnamefont {Thonhauser}}, \bibinfo
  {author} {\bibfnamefont {P.}~\bibnamefont {Umari}}, \bibinfo {author}
  {\bibfnamefont {N.}~\bibnamefont {Vast}}, \bibinfo {author} {\bibfnamefont
  {X.}~\bibnamefont {Wu}}, \ and\ \bibinfo {author} {\bibfnamefont
  {S.}~\bibnamefont {Baroni}},\ }\href
  {http://stacks.iop.org/0953-8984/29/i=46/a=465901} {\bibfield  {journal}
  {\bibinfo  {journal} {J. Phys. Condens. Matter}\ }\textbf {\bibinfo {volume}
  {29}},\ \bibinfo {pages} {465901} (\bibinfo {year} {2017})}\BibitemShut
  {NoStop}%
\bibitem [{\citenamefont {Giannozzi}\ \emph {et~al.}(2020)\citenamefont
  {Giannozzi}, \citenamefont {Baseggio}, \citenamefont {Bonfà}, \citenamefont
  {Brunato}, \citenamefont {Car}, \citenamefont {Carnimeo}, \citenamefont
  {Cavazzoni}, \citenamefont {de~Gironcoli}, \citenamefont {Delugas},
  \citenamefont {Ferrari~Ruffino}, \citenamefont {Ferretti}, \citenamefont
  {Marzari}, \citenamefont {Timrov}, \citenamefont {Urru},\ and\ \citenamefont
  {Baroni}}]{QE-2020}%
  \BibitemOpen
  \bibfield  {author} {\bibinfo {author} {\bibfnamefont {P.}~\bibnamefont
  {Giannozzi}}, \bibinfo {author} {\bibfnamefont {O.}~\bibnamefont {Baseggio}},
  \bibinfo {author} {\bibfnamefont {P.}~\bibnamefont {Bonfà}}, \bibinfo
  {author} {\bibfnamefont {D.}~\bibnamefont {Brunato}}, \bibinfo {author}
  {\bibfnamefont {R.}~\bibnamefont {Car}}, \bibinfo {author} {\bibfnamefont
  {I.}~\bibnamefont {Carnimeo}}, \bibinfo {author} {\bibfnamefont
  {C.}~\bibnamefont {Cavazzoni}}, \bibinfo {author} {\bibfnamefont
  {S.}~\bibnamefont {de~Gironcoli}}, \bibinfo {author} {\bibfnamefont
  {P.}~\bibnamefont {Delugas}}, \bibinfo {author} {\bibfnamefont
  {F.}~\bibnamefont {Ferrari~Ruffino}}, \bibinfo {author} {\bibfnamefont
  {A.}~\bibnamefont {Ferretti}}, \bibinfo {author} {\bibfnamefont
  {N.}~\bibnamefont {Marzari}}, \bibinfo {author} {\bibfnamefont
  {I.}~\bibnamefont {Timrov}}, \bibinfo {author} {\bibfnamefont
  {A.}~\bibnamefont {Urru}}, \ and\ \bibinfo {author} {\bibfnamefont
  {S.}~\bibnamefont {Baroni}},\ }\href {\doibase 10.1063/5.0005082} {\bibfield
  {journal} {\bibinfo  {journal} {J. Chem. Phys.}\ }\textbf {\bibinfo {volume}
  {152}},\ \bibinfo {pages} {154105} (\bibinfo {year} {2020})}\BibitemShut
  {NoStop}%
\bibitem [{\citenamefont {Perdew}\ \emph {et~al.}(1997)\citenamefont {Perdew},
  \citenamefont {Burke},\ and\ \citenamefont {Ernzerhof}}]{Perdew1996}%
  \BibitemOpen
  \bibfield  {author} {\bibinfo {author} {\bibfnamefont {J.~P.}\ \bibnamefont
  {Perdew}}, \bibinfo {author} {\bibfnamefont {K.}~\bibnamefont {Burke}}, \
  and\ \bibinfo {author} {\bibfnamefont {M.}~\bibnamefont {Ernzerhof}},\ }\href
  {\doibase 10.1103/PhysRevLett.78.1396} {\bibfield  {journal} {\bibinfo
  {journal} {Phys. Rev. Lett.}\ }\textbf {\bibinfo {volume} {78}},\ \bibinfo
  {pages} {1396} (\bibinfo {year} {1997})}\BibitemShut {NoStop}%
\bibitem [{\citenamefont {Pizzi}\ \emph {et~al.}(2020)\citenamefont {Pizzi},
  \citenamefont {Vitale}, \citenamefont {Arita}, \citenamefont {Blugel},
  \citenamefont {Freimuth}, \citenamefont {G{\'{e}}ranton}, \citenamefont
  {Gibertini}, \citenamefont {Gresch}, \citenamefont {Johnson}, \citenamefont
  {Koretsune}, \citenamefont {Iba{\~{n}}ez-Azpiroz}, \citenamefont {Lee},
  \citenamefont {Lihm}, \citenamefont {Marchand}, \citenamefont {Marrazzo},
  \citenamefont {Mokrousov}, \citenamefont {Mustafa}, \citenamefont {Nohara},
  \citenamefont {Nomura}, \citenamefont {Paulatto}, \citenamefont
  {Ponc{\'{e}}}, \citenamefont {Ponweiser}, \citenamefont {Qiao}, \citenamefont
  {Thole}, \citenamefont {Tsirkin}, \citenamefont {Wierzbowska}, \citenamefont
  {Marzari}, \citenamefont {Vanderbilt}, \citenamefont {Souza}, \citenamefont
  {Mostofi},\ and\ \citenamefont {Yates}}]{Pizzi2020}%
  \BibitemOpen
  \bibfield  {author} {\bibinfo {author} {\bibfnamefont {G.}~\bibnamefont
  {Pizzi}}, \bibinfo {author} {\bibfnamefont {V.}~\bibnamefont {Vitale}},
  \bibinfo {author} {\bibfnamefont {R.}~\bibnamefont {Arita}}, \bibinfo
  {author} {\bibfnamefont {S.}~\bibnamefont {Blugel}}, \bibinfo {author}
  {\bibfnamefont {F.}~\bibnamefont {Freimuth}}, \bibinfo {author}
  {\bibfnamefont {G.}~\bibnamefont {G{\'{e}}ranton}}, \bibinfo {author}
  {\bibfnamefont {M.}~\bibnamefont {Gibertini}}, \bibinfo {author}
  {\bibfnamefont {D.}~\bibnamefont {Gresch}}, \bibinfo {author} {\bibfnamefont
  {C.}~\bibnamefont {Johnson}}, \bibinfo {author} {\bibfnamefont
  {T.}~\bibnamefont {Koretsune}}, \bibinfo {author} {\bibfnamefont
  {J.}~\bibnamefont {Iba{\~{n}}ez-Azpiroz}}, \bibinfo {author} {\bibfnamefont
  {H.}~\bibnamefont {Lee}}, \bibinfo {author} {\bibfnamefont {J.-M.}\
  \bibnamefont {Lihm}}, \bibinfo {author} {\bibfnamefont {D.}~\bibnamefont
  {Marchand}}, \bibinfo {author} {\bibfnamefont {A.}~\bibnamefont {Marrazzo}},
  \bibinfo {author} {\bibfnamefont {Y.}~\bibnamefont {Mokrousov}}, \bibinfo
  {author} {\bibfnamefont {J.~I.}\ \bibnamefont {Mustafa}}, \bibinfo {author}
  {\bibfnamefont {Y.}~\bibnamefont {Nohara}}, \bibinfo {author} {\bibfnamefont
  {Y.}~\bibnamefont {Nomura}}, \bibinfo {author} {\bibfnamefont
  {L.}~\bibnamefont {Paulatto}}, \bibinfo {author} {\bibfnamefont
  {S.}~\bibnamefont {Ponc{\'{e}}}}, \bibinfo {author} {\bibfnamefont
  {T.}~\bibnamefont {Ponweiser}}, \bibinfo {author} {\bibfnamefont
  {J.}~\bibnamefont {Qiao}}, \bibinfo {author} {\bibfnamefont {F.}~\bibnamefont
  {Thole}}, \bibinfo {author} {\bibfnamefont {S.~S.}\ \bibnamefont {Tsirkin}},
  \bibinfo {author} {\bibfnamefont {M.}~\bibnamefont {Wierzbowska}}, \bibinfo
  {author} {\bibfnamefont {N.}~\bibnamefont {Marzari}}, \bibinfo {author}
  {\bibfnamefont {D.}~\bibnamefont {Vanderbilt}}, \bibinfo {author}
  {\bibfnamefont {I.}~\bibnamefont {Souza}}, \bibinfo {author} {\bibfnamefont
  {A.~A.}\ \bibnamefont {Mostofi}}, \ and\ \bibinfo {author} {\bibfnamefont
  {J.~R.}\ \bibnamefont {Yates}},\ }\href {\doibase 10.1088/1361-648x/ab51ff}
  {\bibfield  {journal} {\bibinfo  {journal} {J. Phys. Condens. Matter}\
  }\textbf {\bibinfo {volume} {32}},\ \bibinfo {pages} {165902} (\bibinfo
  {year} {2020})}\BibitemShut {NoStop}%
\bibitem [{\citenamefont {Vitale}\ \emph {et~al.}(2020)\citenamefont {Vitale},
  \citenamefont {Pizzi}, \citenamefont {Marrazzo}, \citenamefont {Yates},
  \citenamefont {Marzari},\ and\ \citenamefont
  {Mostofi}}]{vitale2020automated}%
  \BibitemOpen
  \bibfield  {author} {\bibinfo {author} {\bibfnamefont {V.}~\bibnamefont
  {Vitale}}, \bibinfo {author} {\bibfnamefont {G.}~\bibnamefont {Pizzi}},
  \bibinfo {author} {\bibfnamefont {A.}~\bibnamefont {Marrazzo}}, \bibinfo
  {author} {\bibfnamefont {J.~R.}\ \bibnamefont {Yates}}, \bibinfo {author}
  {\bibfnamefont {N.}~\bibnamefont {Marzari}}, \ and\ \bibinfo {author}
  {\bibfnamefont {A.~A.}\ \bibnamefont {Mostofi}},\ }\href {\doibase
  10.1038/s41524-020-0312-y} {\bibfield  {journal} {\bibinfo  {journal} {npj
  Comput. Mater.}\ }\textbf {\bibinfo {volume} {6}},\ \bibinfo {pages} {1}
  (\bibinfo {year} {2020})}\BibitemShut {NoStop}%
\bibitem [{\citenamefont {Tang}\ \emph
  {et~al.}(2019{\natexlab{a}})\citenamefont {Tang}, \citenamefont {Po},
  \citenamefont {Vishwanath},\ and\ \citenamefont {Wan}}]{tang2019efficient}%
  \BibitemOpen
  \bibfield  {author} {\bibinfo {author} {\bibfnamefont {F.}~\bibnamefont
  {Tang}}, \bibinfo {author} {\bibfnamefont {H.~C.}\ \bibnamefont {Po}},
  \bibinfo {author} {\bibfnamefont {A.}~\bibnamefont {Vishwanath}}, \ and\
  \bibinfo {author} {\bibfnamefont {X.}~\bibnamefont {Wan}},\ }\href {\doibase
  10.1038/s41586-019-0937-5} {\bibfield  {journal} {\bibinfo  {journal} {Nat.
  Phys.}\ }\textbf {\bibinfo {volume} {15}},\ \bibinfo {pages} {470} (\bibinfo
  {year} {2019}{\natexlab{a}})}\BibitemShut {NoStop}%
\bibitem [{\citenamefont {Zhang}\ \emph {et~al.}(2019)\citenamefont {Zhang},
  \citenamefont {Jiang}, \citenamefont {Song}, \citenamefont {Huang},
  \citenamefont {He}, \citenamefont {Fang}, \citenamefont {Weng},\ and\
  \citenamefont {Fang}}]{zhang2019catalogue}%
  \BibitemOpen
  \bibfield  {author} {\bibinfo {author} {\bibfnamefont {T.}~\bibnamefont
  {Zhang}}, \bibinfo {author} {\bibfnamefont {Y.}~\bibnamefont {Jiang}},
  \bibinfo {author} {\bibfnamefont {Z.}~\bibnamefont {Song}}, \bibinfo {author}
  {\bibfnamefont {H.}~\bibnamefont {Huang}}, \bibinfo {author} {\bibfnamefont
  {Y.}~\bibnamefont {He}}, \bibinfo {author} {\bibfnamefont {Z.}~\bibnamefont
  {Fang}}, \bibinfo {author} {\bibfnamefont {H.}~\bibnamefont {Weng}}, \ and\
  \bibinfo {author} {\bibfnamefont {C.}~\bibnamefont {Fang}},\ }\href {\doibase
  10.1038/s41586-019-0944-6} {\bibfield  {journal} {\bibinfo  {journal}
  {Nature}\ }\textbf {\bibinfo {volume} {566}},\ \bibinfo {pages} {475}
  (\bibinfo {year} {2019})}\BibitemShut {NoStop}%
\bibitem [{\citenamefont {Vergniory}\ \emph {et~al.}(2019)\citenamefont
  {Vergniory}, \citenamefont {Elcoro}, \citenamefont {Felser}, \citenamefont
  {Regnault}, \citenamefont {Bernevig},\ and\ \citenamefont
  {Wang}}]{vergniory2019complete}%
  \BibitemOpen
  \bibfield  {author} {\bibinfo {author} {\bibfnamefont {M.}~\bibnamefont
  {Vergniory}}, \bibinfo {author} {\bibfnamefont {L.}~\bibnamefont {Elcoro}},
  \bibinfo {author} {\bibfnamefont {C.}~\bibnamefont {Felser}}, \bibinfo
  {author} {\bibfnamefont {N.}~\bibnamefont {Regnault}}, \bibinfo {author}
  {\bibfnamefont {B.~A.}\ \bibnamefont {Bernevig}}, \ and\ \bibinfo {author}
  {\bibfnamefont {Z.}~\bibnamefont {Wang}},\ }\href {\doibase
  10.1038/s41586-019-0954-4} {\bibfield  {journal} {\bibinfo  {journal}
  {Nature}\ }\textbf {\bibinfo {volume} {566}},\ \bibinfo {pages} {480}
  (\bibinfo {year} {2019})}\BibitemShut {NoStop}%
\bibitem [{\citenamefont {Tang}\ \emph
  {et~al.}(2019{\natexlab{b}})\citenamefont {Tang}, \citenamefont {Po},
  \citenamefont {Vishwanath},\ and\ \citenamefont
  {Wan}}]{tang2019comprehensive}%
  \BibitemOpen
  \bibfield  {author} {\bibinfo {author} {\bibfnamefont {F.}~\bibnamefont
  {Tang}}, \bibinfo {author} {\bibfnamefont {H.~C.}\ \bibnamefont {Po}},
  \bibinfo {author} {\bibfnamefont {A.}~\bibnamefont {Vishwanath}}, \ and\
  \bibinfo {author} {\bibfnamefont {X.}~\bibnamefont {Wan}},\ }\href {\doibase
  10.1038/s41586-019-0937-5} {\bibfield  {journal} {\bibinfo  {journal}
  {Nature}\ }\textbf {\bibinfo {volume} {566}},\ \bibinfo {pages} {486}
  (\bibinfo {year} {2019}{\natexlab{b}})}\BibitemShut {NoStop}%
\bibitem [{\citenamefont {Xu}\ \emph {et~al.}(2020)\citenamefont {Xu},
  \citenamefont {Elcoro}, \citenamefont {Song}, \citenamefont {Wieder},
  \citenamefont {Vergniory}, \citenamefont {Regnault}, \citenamefont {Chen},
  \citenamefont {Felser},\ and\ \citenamefont {Bernevig}}]{xu2020high}%
  \BibitemOpen
  \bibfield  {author} {\bibinfo {author} {\bibfnamefont {Y.}~\bibnamefont
  {Xu}}, \bibinfo {author} {\bibfnamefont {L.}~\bibnamefont {Elcoro}}, \bibinfo
  {author} {\bibfnamefont {Z.-D.}\ \bibnamefont {Song}}, \bibinfo {author}
  {\bibfnamefont {B.~J.}\ \bibnamefont {Wieder}}, \bibinfo {author}
  {\bibfnamefont {M.}~\bibnamefont {Vergniory}}, \bibinfo {author}
  {\bibfnamefont {N.}~\bibnamefont {Regnault}}, \bibinfo {author}
  {\bibfnamefont {Y.}~\bibnamefont {Chen}}, \bibinfo {author} {\bibfnamefont
  {C.}~\bibnamefont {Felser}}, \ and\ \bibinfo {author} {\bibfnamefont {B.~A.}\
  \bibnamefont {Bernevig}},\ }\href {\doibase 10.1038/s41586-020-2837-0}
  {\bibfield  {journal} {\bibinfo  {journal} {Nature}\ }\textbf {\bibinfo
  {volume} {586}},\ \bibinfo {pages} {702} (\bibinfo {year}
  {2020})}\BibitemShut {NoStop}%
\bibitem [{\citenamefont {Lantagne-Hurtubise}\ and\ \citenamefont
  {Franz}(2019)}]{lantagne2019topology}%
  \BibitemOpen
  \bibfield  {author} {\bibinfo {author} {\bibfnamefont {{\'E}.}~\bibnamefont
  {Lantagne-Hurtubise}}\ and\ \bibinfo {author} {\bibfnamefont
  {M.}~\bibnamefont {Franz}},\ }\href {\doibase 10.1038/s42254-019-0041-7}
  {\bibfield  {journal} {\bibinfo  {journal} {Nat. Rev. Phys.}\ }\textbf
  {\bibinfo {volume} {1}},\ \bibinfo {pages} {183} (\bibinfo {year}
  {2019})}\BibitemShut {NoStop}%
\bibitem [{\citenamefont {Soluyanov}\ \emph {et~al.}(2015)\citenamefont
  {Soluyanov}, \citenamefont {Gresch}, \citenamefont {Wang}, \citenamefont
  {Wu}, \citenamefont {Troyer}, \citenamefont {Dai},\ and\ \citenamefont
  {Bernevig}}]{soluyanov2015type}%
  \BibitemOpen
  \bibfield  {author} {\bibinfo {author} {\bibfnamefont {A.~A.}\ \bibnamefont
  {Soluyanov}}, \bibinfo {author} {\bibfnamefont {D.}~\bibnamefont {Gresch}},
  \bibinfo {author} {\bibfnamefont {Z.}~\bibnamefont {Wang}}, \bibinfo {author}
  {\bibfnamefont {Q.}~\bibnamefont {Wu}}, \bibinfo {author} {\bibfnamefont
  {M.}~\bibnamefont {Troyer}}, \bibinfo {author} {\bibfnamefont
  {X.}~\bibnamefont {Dai}}, \ and\ \bibinfo {author} {\bibfnamefont {B.~A.}\
  \bibnamefont {Bernevig}},\ }\href {\doibase 10.1038/nature15768} {\bibfield
  {journal} {\bibinfo  {journal} {Nature}\ }\textbf {\bibinfo {volume} {527}},\
  \bibinfo {pages} {495} (\bibinfo {year} {2015})}\BibitemShut {NoStop}%
\bibitem [{\citenamefont {Wang}\ \emph
  {et~al.}(2016{\natexlab{a}})\citenamefont {Wang}, \citenamefont {Zhang},
  \citenamefont {Huang}, \citenamefont {Nie}, \citenamefont {Liu},
  \citenamefont {Liang}, \citenamefont {Zhang}, \citenamefont {Shen},
  \citenamefont {Liu}, \citenamefont {Hu}, \citenamefont {Ding}, \citenamefont
  {Liu}, \citenamefont {Hu}, \citenamefont {He}, \citenamefont {Zhao},
  \citenamefont {Yu}, \citenamefont {Hu}, \citenamefont {Wei}, \citenamefont
  {Mao}, \citenamefont {Shi}, \citenamefont {Jia}, \citenamefont {Zhang},
  \citenamefont {Zhang}, \citenamefont {Yang}, \citenamefont {Wang},
  \citenamefont {Peng}, \citenamefont {Weng}, \citenamefont {Dai},
  \citenamefont {Fang}, \citenamefont {Xu}, \citenamefont {Chen},\ and\
  \citenamefont {Zhou}}]{Wang2016Obs}%
  \BibitemOpen
  \bibfield  {author} {\bibinfo {author} {\bibfnamefont {C.}~\bibnamefont
  {Wang}}, \bibinfo {author} {\bibfnamefont {Y.}~\bibnamefont {Zhang}},
  \bibinfo {author} {\bibfnamefont {J.}~\bibnamefont {Huang}}, \bibinfo
  {author} {\bibfnamefont {S.}~\bibnamefont {Nie}}, \bibinfo {author}
  {\bibfnamefont {G.}~\bibnamefont {Liu}}, \bibinfo {author} {\bibfnamefont
  {A.}~\bibnamefont {Liang}}, \bibinfo {author} {\bibfnamefont
  {Y.}~\bibnamefont {Zhang}}, \bibinfo {author} {\bibfnamefont
  {B.}~\bibnamefont {Shen}}, \bibinfo {author} {\bibfnamefont {J.}~\bibnamefont
  {Liu}}, \bibinfo {author} {\bibfnamefont {C.}~\bibnamefont {Hu}}, \bibinfo
  {author} {\bibfnamefont {Y.}~\bibnamefont {Ding}}, \bibinfo {author}
  {\bibfnamefont {D.}~\bibnamefont {Liu}}, \bibinfo {author} {\bibfnamefont
  {Y.}~\bibnamefont {Hu}}, \bibinfo {author} {\bibfnamefont {S.}~\bibnamefont
  {He}}, \bibinfo {author} {\bibfnamefont {L.}~\bibnamefont {Zhao}}, \bibinfo
  {author} {\bibfnamefont {L.}~\bibnamefont {Yu}}, \bibinfo {author}
  {\bibfnamefont {J.}~\bibnamefont {Hu}}, \bibinfo {author} {\bibfnamefont
  {J.}~\bibnamefont {Wei}}, \bibinfo {author} {\bibfnamefont {Z.}~\bibnamefont
  {Mao}}, \bibinfo {author} {\bibfnamefont {Y.}~\bibnamefont {Shi}}, \bibinfo
  {author} {\bibfnamefont {X.}~\bibnamefont {Jia}}, \bibinfo {author}
  {\bibfnamefont {F.}~\bibnamefont {Zhang}}, \bibinfo {author} {\bibfnamefont
  {S.}~\bibnamefont {Zhang}}, \bibinfo {author} {\bibfnamefont
  {F.}~\bibnamefont {Yang}}, \bibinfo {author} {\bibfnamefont {Z.}~\bibnamefont
  {Wang}}, \bibinfo {author} {\bibfnamefont {Q.}~\bibnamefont {Peng}}, \bibinfo
  {author} {\bibfnamefont {H.}~\bibnamefont {Weng}}, \bibinfo {author}
  {\bibfnamefont {X.}~\bibnamefont {Dai}}, \bibinfo {author} {\bibfnamefont
  {Z.}~\bibnamefont {Fang}}, \bibinfo {author} {\bibfnamefont {Z.}~\bibnamefont
  {Xu}}, \bibinfo {author} {\bibfnamefont {C.}~\bibnamefont {Chen}}, \ and\
  \bibinfo {author} {\bibfnamefont {X.~J.}\ \bibnamefont {Zhou}},\ }\href
  {\doibase 10.1103/PhysRevB.94.241119} {\bibfield  {journal} {\bibinfo
  {journal} {Phys. Rev. B}\ }\textbf {\bibinfo {volume} {94}},\ \bibinfo
  {pages} {241119} (\bibinfo {year} {2016}{\natexlab{a}})}\BibitemShut
  {NoStop}%
\bibitem [{\citenamefont {Li}\ \emph {et~al.}(2017)\citenamefont {Li},
  \citenamefont {Wen}, \citenamefont {He}, \citenamefont {Zhang}, \citenamefont
  {Xia}, \citenamefont {Yu}, \citenamefont {Yang}, \citenamefont {Zhu},
  \citenamefont {Alshareef},\ and\ \citenamefont {Zhang}}]{li2017evidence}%
  \BibitemOpen
  \bibfield  {author} {\bibinfo {author} {\bibfnamefont {P.}~\bibnamefont
  {Li}}, \bibinfo {author} {\bibfnamefont {Y.}~\bibnamefont {Wen}}, \bibinfo
  {author} {\bibfnamefont {X.}~\bibnamefont {He}}, \bibinfo {author}
  {\bibfnamefont {Q.}~\bibnamefont {Zhang}}, \bibinfo {author} {\bibfnamefont
  {C.}~\bibnamefont {Xia}}, \bibinfo {author} {\bibfnamefont {Z.-M.}\
  \bibnamefont {Yu}}, \bibinfo {author} {\bibfnamefont {S.~A.}\ \bibnamefont
  {Yang}}, \bibinfo {author} {\bibfnamefont {Z.}~\bibnamefont {Zhu}}, \bibinfo
  {author} {\bibfnamefont {H.~N.}\ \bibnamefont {Alshareef}}, \ and\ \bibinfo
  {author} {\bibfnamefont {X.-X.}\ \bibnamefont {Zhang}},\ }\href {\doibase
  10.1038/s41467-017-02237-1} {\bibfield  {journal} {\bibinfo  {journal}
  {Nature communications}\ }\textbf {\bibinfo {volume} {8}},\ \bibinfo {pages}
  {2150} (\bibinfo {year} {2017})}\BibitemShut {NoStop}%
\bibitem [{\citenamefont {Wang}\ \emph
  {et~al.}(2016{\natexlab{b}})\citenamefont {Wang}, \citenamefont {Gresch},
  \citenamefont {Soluyanov}, \citenamefont {Xie}, \citenamefont {Kushwaha},
  \citenamefont {Dai}, \citenamefont {Troyer}, \citenamefont {Cava},\ and\
  \citenamefont {Bernevig}}]{Wang2016MoTe2}%
  \BibitemOpen
  \bibfield  {author} {\bibinfo {author} {\bibfnamefont {Z.}~\bibnamefont
  {Wang}}, \bibinfo {author} {\bibfnamefont {D.}~\bibnamefont {Gresch}},
  \bibinfo {author} {\bibfnamefont {A.~A.}\ \bibnamefont {Soluyanov}}, \bibinfo
  {author} {\bibfnamefont {W.}~\bibnamefont {Xie}}, \bibinfo {author}
  {\bibfnamefont {S.}~\bibnamefont {Kushwaha}}, \bibinfo {author}
  {\bibfnamefont {X.}~\bibnamefont {Dai}}, \bibinfo {author} {\bibfnamefont
  {M.}~\bibnamefont {Troyer}}, \bibinfo {author} {\bibfnamefont {R.~J.}\
  \bibnamefont {Cava}}, \ and\ \bibinfo {author} {\bibfnamefont {B.~A.}\
  \bibnamefont {Bernevig}},\ }\href {\doibase 10.1103/PhysRevLett.117.056805}
  {\bibfield  {journal} {\bibinfo  {journal} {Phys. Rev. Lett.}\ }\textbf
  {\bibinfo {volume} {117}},\ \bibinfo {pages} {056805} (\bibinfo {year}
  {2016}{\natexlab{b}})}\BibitemShut {NoStop}%
\bibitem [{\citenamefont {Tamai}\ \emph {et~al.}(2016)\citenamefont {Tamai},
  \citenamefont {Wu}, \citenamefont {Cucchi}, \citenamefont {Bruno},
  \citenamefont {Ricc\`o}, \citenamefont {Kim}, \citenamefont {Hoesch},
  \citenamefont {Barreteau}, \citenamefont {Giannini}, \citenamefont {Besnard},
  \citenamefont {Soluyanov},\ and\ \citenamefont {Baumberger}}]{Tamai2016}%
  \BibitemOpen
  \bibfield  {author} {\bibinfo {author} {\bibfnamefont {A.}~\bibnamefont
  {Tamai}}, \bibinfo {author} {\bibfnamefont {Q.~S.}\ \bibnamefont {Wu}},
  \bibinfo {author} {\bibfnamefont {I.}~\bibnamefont {Cucchi}}, \bibinfo
  {author} {\bibfnamefont {F.~Y.}\ \bibnamefont {Bruno}}, \bibinfo {author}
  {\bibfnamefont {S.}~\bibnamefont {Ricc\`o}}, \bibinfo {author} {\bibfnamefont
  {T.~K.}\ \bibnamefont {Kim}}, \bibinfo {author} {\bibfnamefont
  {M.}~\bibnamefont {Hoesch}}, \bibinfo {author} {\bibfnamefont
  {C.}~\bibnamefont {Barreteau}}, \bibinfo {author} {\bibfnamefont
  {E.}~\bibnamefont {Giannini}}, \bibinfo {author} {\bibfnamefont
  {C.}~\bibnamefont {Besnard}}, \bibinfo {author} {\bibfnamefont {A.~A.}\
  \bibnamefont {Soluyanov}}, \ and\ \bibinfo {author} {\bibfnamefont
  {F.}~\bibnamefont {Baumberger}},\ }\href {\doibase 10.1103/PhysRevX.6.031021}
  {\bibfield  {journal} {\bibinfo  {journal} {Phys. Rev. X}\ }\textbf {\bibinfo
  {volume} {6}},\ \bibinfo {pages} {031021} (\bibinfo {year}
  {2016})}\BibitemShut {NoStop}%
\bibitem [{\citenamefont {Deng}\ \emph {et~al.}(2016)\citenamefont {Deng},
  \citenamefont {Wan}, \citenamefont {Deng}, \citenamefont {Zhang},
  \citenamefont {Ding}, \citenamefont {Wang}, \citenamefont {Yan},
  \citenamefont {Huang}, \citenamefont {Zhang}, \citenamefont {Xu} \emph
  {et~al.}}]{deng2016experimental}%
  \BibitemOpen
  \bibfield  {author} {\bibinfo {author} {\bibfnamefont {K.}~\bibnamefont
  {Deng}}, \bibinfo {author} {\bibfnamefont {G.}~\bibnamefont {Wan}}, \bibinfo
  {author} {\bibfnamefont {P.}~\bibnamefont {Deng}}, \bibinfo {author}
  {\bibfnamefont {K.}~\bibnamefont {Zhang}}, \bibinfo {author} {\bibfnamefont
  {S.}~\bibnamefont {Ding}}, \bibinfo {author} {\bibfnamefont {E.}~\bibnamefont
  {Wang}}, \bibinfo {author} {\bibfnamefont {M.}~\bibnamefont {Yan}}, \bibinfo
  {author} {\bibfnamefont {H.}~\bibnamefont {Huang}}, \bibinfo {author}
  {\bibfnamefont {H.}~\bibnamefont {Zhang}}, \bibinfo {author} {\bibfnamefont
  {Z.}~\bibnamefont {Xu}},  \emph {et~al.},\ }\href {\doibase
  10.1038/nphys3871} {\bibfield  {journal} {\bibinfo  {journal} {Nat. Phys.}\
  }\textbf {\bibinfo {volume} {12}},\ \bibinfo {pages} {1105} (\bibinfo {year}
  {2016})}\BibitemShut {NoStop}%
\bibitem [{\citenamefont {Jiang}\ \emph {et~al.}(2017)\citenamefont {Jiang},
  \citenamefont {Liu}, \citenamefont {Sun}, \citenamefont {Yang}, \citenamefont
  {Rajamathi}, \citenamefont {Qi}, \citenamefont {Yang}, \citenamefont {Chen},
  \citenamefont {Peng}, \citenamefont {Hwang} \emph
  {et~al.}}]{jiang2017signature}%
  \BibitemOpen
  \bibfield  {author} {\bibinfo {author} {\bibfnamefont {J.}~\bibnamefont
  {Jiang}}, \bibinfo {author} {\bibfnamefont {Z.}~\bibnamefont {Liu}}, \bibinfo
  {author} {\bibfnamefont {Y.}~\bibnamefont {Sun}}, \bibinfo {author}
  {\bibfnamefont {H.}~\bibnamefont {Yang}}, \bibinfo {author} {\bibfnamefont
  {C.}~\bibnamefont {Rajamathi}}, \bibinfo {author} {\bibfnamefont
  {Y.}~\bibnamefont {Qi}}, \bibinfo {author} {\bibfnamefont {L.}~\bibnamefont
  {Yang}}, \bibinfo {author} {\bibfnamefont {C.}~\bibnamefont {Chen}}, \bibinfo
  {author} {\bibfnamefont {H.}~\bibnamefont {Peng}}, \bibinfo {author}
  {\bibfnamefont {C.}~\bibnamefont {Hwang}},  \emph {et~al.},\ }\href {\doibase
  10.1038/ncomms13973} {\bibfield  {journal} {\bibinfo  {journal} {Nat.
  commun.}\ }\textbf {\bibinfo {volume} {8}},\ \bibinfo {pages} {13973}
  (\bibinfo {year} {2017})}\BibitemShut {NoStop}%
\bibitem [{\citenamefont {Wang}\ \emph {et~al.}(2019)\citenamefont {Wang},
  \citenamefont {Wieder}, \citenamefont {Li}, \citenamefont {Yan},\ and\
  \citenamefont {Bernevig}}]{Wang2019Higher}%
  \BibitemOpen
  \bibfield  {author} {\bibinfo {author} {\bibfnamefont {Z.}~\bibnamefont
  {Wang}}, \bibinfo {author} {\bibfnamefont {B.~J.}\ \bibnamefont {Wieder}},
  \bibinfo {author} {\bibfnamefont {J.}~\bibnamefont {Li}}, \bibinfo {author}
  {\bibfnamefont {B.}~\bibnamefont {Yan}}, \ and\ \bibinfo {author}
  {\bibfnamefont {B.~A.}\ \bibnamefont {Bernevig}},\ }\href {\doibase
  10.1103/PhysRevLett.123.186401} {\bibfield  {journal} {\bibinfo  {journal}
  {Phys. Rev. Lett.}\ }\textbf {\bibinfo {volume} {123}},\ \bibinfo {pages}
  {186401} (\bibinfo {year} {2019})}\BibitemShut {NoStop}%
\bibitem [{\citenamefont {Qian}\ \emph {et~al.}(2014)\citenamefont {Qian},
  \citenamefont {Liu}, \citenamefont {Fu},\ and\ \citenamefont
  {Li}}]{qian2014quantum}%
  \BibitemOpen
  \bibfield  {author} {\bibinfo {author} {\bibfnamefont {X.}~\bibnamefont
  {Qian}}, \bibinfo {author} {\bibfnamefont {J.}~\bibnamefont {Liu}}, \bibinfo
  {author} {\bibfnamefont {L.}~\bibnamefont {Fu}}, \ and\ \bibinfo {author}
  {\bibfnamefont {J.}~\bibnamefont {Li}},\ }\href {\doibase
  10.1126/science.1256815} {\bibfield  {journal} {\bibinfo  {journal}
  {Science}\ }\textbf {\bibinfo {volume} {346}},\ \bibinfo {pages} {1344}
  (\bibinfo {year} {2014})}\BibitemShut {NoStop}%
\bibitem [{\citenamefont {Andrei}\ \emph {et~al.}(2021)\citenamefont {Andrei},
  \citenamefont {Efetov}, \citenamefont {Jarillo-Herrero}, \citenamefont
  {MacDonald}, \citenamefont {Mak}, \citenamefont {Senthil}, \citenamefont
  {Tutuc}, \citenamefont {Yazdani},\ and\ \citenamefont
  {Young}}]{andrei2021marvels}%
  \BibitemOpen
  \bibfield  {author} {\bibinfo {author} {\bibfnamefont {E.~Y.}\ \bibnamefont
  {Andrei}}, \bibinfo {author} {\bibfnamefont {D.~K.}\ \bibnamefont {Efetov}},
  \bibinfo {author} {\bibfnamefont {P.}~\bibnamefont {Jarillo-Herrero}},
  \bibinfo {author} {\bibfnamefont {A.~H.}\ \bibnamefont {MacDonald}}, \bibinfo
  {author} {\bibfnamefont {K.~F.}\ \bibnamefont {Mak}}, \bibinfo {author}
  {\bibfnamefont {T.}~\bibnamefont {Senthil}}, \bibinfo {author} {\bibfnamefont
  {E.}~\bibnamefont {Tutuc}}, \bibinfo {author} {\bibfnamefont
  {A.}~\bibnamefont {Yazdani}}, \ and\ \bibinfo {author} {\bibfnamefont
  {A.~F.}\ \bibnamefont {Young}},\ }\href {\doibase 10.1038/s41578-021-00284-1}
  {\bibfield  {journal} {\bibinfo  {journal} {Nat. Rev. Mat.}\ }\textbf
  {\bibinfo {volume} {6}},\ \bibinfo {pages} {201} (\bibinfo {year}
  {2021})}\BibitemShut {NoStop}%
\bibitem [{\citenamefont {Naik}\ and\ \citenamefont {Jain}(2018)}]{Naik2018}%
  \BibitemOpen
  \bibfield  {author} {\bibinfo {author} {\bibfnamefont {M.~H.}\ \bibnamefont
  {Naik}}\ and\ \bibinfo {author} {\bibfnamefont {M.}~\bibnamefont {Jain}},\
  }\href {\doibase 10.1103/PhysRevLett.121.266401} {\bibfield  {journal}
  {\bibinfo  {journal} {Phys. Rev. Lett.}\ }\textbf {\bibinfo {volume} {121}},\
  \bibinfo {pages} {266401} (\bibinfo {year} {2018})}\BibitemShut {NoStop}%
\bibitem [{\citenamefont {Wu}\ \emph {et~al.}(2018{\natexlab{a}})\citenamefont
  {Wu}, \citenamefont {Lovorn}, \citenamefont {Tutuc},\ and\ \citenamefont
  {MacDonald}}]{Wu2018Hubbard}%
  \BibitemOpen
  \bibfield  {author} {\bibinfo {author} {\bibfnamefont {F.}~\bibnamefont
  {Wu}}, \bibinfo {author} {\bibfnamefont {T.}~\bibnamefont {Lovorn}}, \bibinfo
  {author} {\bibfnamefont {E.}~\bibnamefont {Tutuc}}, \ and\ \bibinfo {author}
  {\bibfnamefont {A.~H.}\ \bibnamefont {MacDonald}},\ }\href {\doibase
  10.1103/PhysRevLett.121.026402} {\bibfield  {journal} {\bibinfo  {journal}
  {Phys. Rev. Lett.}\ }\textbf {\bibinfo {volume} {121}},\ \bibinfo {pages}
  {026402} (\bibinfo {year} {2018}{\natexlab{a}})}\BibitemShut {NoStop}%
\bibitem [{\citenamefont {Zang}\ \emph {et~al.}(2022)\citenamefont {Zang},
  \citenamefont {Wang}, \citenamefont {Cano}, \citenamefont {Georges},\ and\
  \citenamefont {Millis}}]{Zang2022}%
  \BibitemOpen
  \bibfield  {author} {\bibinfo {author} {\bibfnamefont {J.}~\bibnamefont
  {Zang}}, \bibinfo {author} {\bibfnamefont {J.}~\bibnamefont {Wang}}, \bibinfo
  {author} {\bibfnamefont {J.}~\bibnamefont {Cano}}, \bibinfo {author}
  {\bibfnamefont {A.}~\bibnamefont {Georges}}, \ and\ \bibinfo {author}
  {\bibfnamefont {A.~J.}\ \bibnamefont {Millis}},\ }\href {\doibase
  10.1103/PhysRevX.12.021064} {\bibfield  {journal} {\bibinfo  {journal} {Phys.
  Rev. X}\ }\textbf {\bibinfo {volume} {12}},\ \bibinfo {pages} {021064}
  (\bibinfo {year} {2022})}\BibitemShut {NoStop}%
\bibitem [{\citenamefont {Zhang}\ \emph {et~al.}(2017)\citenamefont {Zhang},
  \citenamefont {Chuu}, \citenamefont {Ren}, \citenamefont {Li}, \citenamefont
  {Li}, \citenamefont {Jin}, \citenamefont {Chou},\ and\ \citenamefont
  {Shih}}]{zhang2017interlayer}%
  \BibitemOpen
  \bibfield  {author} {\bibinfo {author} {\bibfnamefont {C.}~\bibnamefont
  {Zhang}}, \bibinfo {author} {\bibfnamefont {C.-P.}\ \bibnamefont {Chuu}},
  \bibinfo {author} {\bibfnamefont {X.}~\bibnamefont {Ren}}, \bibinfo {author}
  {\bibfnamefont {M.-Y.}\ \bibnamefont {Li}}, \bibinfo {author} {\bibfnamefont
  {L.-J.}\ \bibnamefont {Li}}, \bibinfo {author} {\bibfnamefont
  {C.}~\bibnamefont {Jin}}, \bibinfo {author} {\bibfnamefont {M.-Y.}\
  \bibnamefont {Chou}}, \ and\ \bibinfo {author} {\bibfnamefont {C.-K.}\
  \bibnamefont {Shih}},\ }\href {\doibase 10.1126/sciadv.1601459} {\bibfield
  {journal} {\bibinfo  {journal} {Sci. Adv.}\ }\textbf {\bibinfo {volume}
  {3}},\ \bibinfo {pages} {e1601459} (\bibinfo {year} {2017})}\BibitemShut
  {NoStop}%
\bibitem [{\citenamefont {Wang}\ \emph {et~al.}(2020)\citenamefont {Wang},
  \citenamefont {Shih}, \citenamefont {Ghiotto}, \citenamefont {Xian},
  \citenamefont {Rhodes}, \citenamefont {Tan}, \citenamefont {Claassen},
  \citenamefont {Kennes}, \citenamefont {Bai}, \citenamefont {Kim} \emph
  {et~al.}}]{wang2020correlated}%
  \BibitemOpen
  \bibfield  {author} {\bibinfo {author} {\bibfnamefont {L.}~\bibnamefont
  {Wang}}, \bibinfo {author} {\bibfnamefont {E.-M.}\ \bibnamefont {Shih}},
  \bibinfo {author} {\bibfnamefont {A.}~\bibnamefont {Ghiotto}}, \bibinfo
  {author} {\bibfnamefont {L.}~\bibnamefont {Xian}}, \bibinfo {author}
  {\bibfnamefont {D.~A.}\ \bibnamefont {Rhodes}}, \bibinfo {author}
  {\bibfnamefont {C.}~\bibnamefont {Tan}}, \bibinfo {author} {\bibfnamefont
  {M.}~\bibnamefont {Claassen}}, \bibinfo {author} {\bibfnamefont {D.~M.}\
  \bibnamefont {Kennes}}, \bibinfo {author} {\bibfnamefont {Y.}~\bibnamefont
  {Bai}}, \bibinfo {author} {\bibfnamefont {B.}~\bibnamefont {Kim}},  \emph
  {et~al.},\ }\href {\doibase 10.1038/s41563-020-0708-6} {\bibfield  {journal}
  {\bibinfo  {journal} {Nat. mater.}\ }\textbf {\bibinfo {volume} {19}},\
  \bibinfo {pages} {861} (\bibinfo {year} {2020})}\BibitemShut {NoStop}%
\bibitem [{\citenamefont {Wang}\ \emph {et~al.}(2015)\citenamefont {Wang},
  \citenamefont {Chen},\ and\ \citenamefont {Wang}}]{wang2015electronic}%
  \BibitemOpen
  \bibfield  {author} {\bibinfo {author} {\bibfnamefont {Z.}~\bibnamefont
  {Wang}}, \bibinfo {author} {\bibfnamefont {Q.}~\bibnamefont {Chen}}, \ and\
  \bibinfo {author} {\bibfnamefont {J.}~\bibnamefont {Wang}},\ }\href {\doibase
  10.1021/jp507751p} {\bibfield  {journal} {\bibinfo  {journal} {J. Phys. Chem.
  C}\ }\textbf {\bibinfo {volume} {119}},\ \bibinfo {pages} {4752} (\bibinfo
  {year} {2015})}\BibitemShut {NoStop}%
\bibitem [{\citenamefont {Castro~Neto}(2001)}]{Castro2001}%
  \BibitemOpen
  \bibfield  {author} {\bibinfo {author} {\bibfnamefont {A.~H.}\ \bibnamefont
  {Castro~Neto}},\ }\href {\doibase 10.1103/PhysRevLett.86.4382} {\bibfield
  {journal} {\bibinfo  {journal} {Phys. Rev. Lett.}\ }\textbf {\bibinfo
  {volume} {86}},\ \bibinfo {pages} {4382} (\bibinfo {year}
  {2001})}\BibitemShut {NoStop}%
\bibitem [{\citenamefont {Yuan}\ \emph {et~al.}(2014)\citenamefont {Yuan},
  \citenamefont {Mak},\ and\ \citenamefont {Law}}]{Yuan2014TopSup}%
  \BibitemOpen
  \bibfield  {author} {\bibinfo {author} {\bibfnamefont {N.~F.~Q.}\
  \bibnamefont {Yuan}}, \bibinfo {author} {\bibfnamefont {K.~F.}\ \bibnamefont
  {Mak}}, \ and\ \bibinfo {author} {\bibfnamefont {K.~T.}\ \bibnamefont
  {Law}},\ }\href {\doibase 10.1103/PhysRevLett.113.097001} {\bibfield
  {journal} {\bibinfo  {journal} {Phys. Rev. Lett.}\ }\textbf {\bibinfo
  {volume} {113}},\ \bibinfo {pages} {097001} (\bibinfo {year}
  {2014})}\BibitemShut {NoStop}%
\bibitem [{\citenamefont {Chu}\ \emph {et~al.}(2014)\citenamefont {Chu},
  \citenamefont {Liu}, \citenamefont {Yao}, \citenamefont {Xu}, \citenamefont
  {Xiao},\ and\ \citenamefont {Zhang}}]{Chu2014}%
  \BibitemOpen
  \bibfield  {author} {\bibinfo {author} {\bibfnamefont {R.-L.}\ \bibnamefont
  {Chu}}, \bibinfo {author} {\bibfnamefont {G.-B.}\ \bibnamefont {Liu}},
  \bibinfo {author} {\bibfnamefont {W.}~\bibnamefont {Yao}}, \bibinfo {author}
  {\bibfnamefont {X.}~\bibnamefont {Xu}}, \bibinfo {author} {\bibfnamefont
  {D.}~\bibnamefont {Xiao}}, \ and\ \bibinfo {author} {\bibfnamefont
  {C.}~\bibnamefont {Zhang}},\ }\href {\doibase 10.1103/PhysRevB.89.155317}
  {\bibfield  {journal} {\bibinfo  {journal} {Phys. Rev. B}\ }\textbf {\bibinfo
  {volume} {89}},\ \bibinfo {pages} {155317} (\bibinfo {year}
  {2014})}\BibitemShut {NoStop}%
\bibitem [{\citenamefont {Zhou}\ \emph {et~al.}(2016)\citenamefont {Zhou},
  \citenamefont {Yuan}, \citenamefont {Jiang},\ and\ \citenamefont
  {Law}}]{Zhou2016}%
  \BibitemOpen
  \bibfield  {author} {\bibinfo {author} {\bibfnamefont {B.~T.}\ \bibnamefont
  {Zhou}}, \bibinfo {author} {\bibfnamefont {N.~F.~Q.}\ \bibnamefont {Yuan}},
  \bibinfo {author} {\bibfnamefont {H.-L.}\ \bibnamefont {Jiang}}, \ and\
  \bibinfo {author} {\bibfnamefont {K.~T.}\ \bibnamefont {Law}},\ }\href
  {\doibase 10.1103/PhysRevB.93.180501} {\bibfield  {journal} {\bibinfo
  {journal} {Phys. Rev. B}\ }\textbf {\bibinfo {volume} {93}},\ \bibinfo
  {pages} {180501} (\bibinfo {year} {2016})}\BibitemShut {NoStop}%
\bibitem [{\citenamefont {Hsu}\ \emph {et~al.}(2017)\citenamefont {Hsu},
  \citenamefont {Vaezi}, \citenamefont {Fischer},\ and\ \citenamefont
  {Kim}}]{hsu2017topological}%
  \BibitemOpen
  \bibfield  {author} {\bibinfo {author} {\bibfnamefont {Y.-T.}\ \bibnamefont
  {Hsu}}, \bibinfo {author} {\bibfnamefont {A.}~\bibnamefont {Vaezi}}, \bibinfo
  {author} {\bibfnamefont {M.~H.}\ \bibnamefont {Fischer}}, \ and\ \bibinfo
  {author} {\bibfnamefont {E.-A.}\ \bibnamefont {Kim}},\ }\href {\doibase
  10.1038/ncomms14985} {\bibfield  {journal} {\bibinfo  {journal} {Nature
  communications}\ }\textbf {\bibinfo {volume} {8}},\ \bibinfo {pages} {14985}
  (\bibinfo {year} {2017})}\BibitemShut {NoStop}%
\bibitem [{\citenamefont {Liu}(2017)}]{CXLiu2017}%
  \BibitemOpen
  \bibfield  {author} {\bibinfo {author} {\bibfnamefont {C.-X.}\ \bibnamefont
  {Liu}},\ }\href {\doibase 10.1103/PhysRevLett.118.087001} {\bibfield
  {journal} {\bibinfo  {journal} {Phys. Rev. Lett.}\ }\textbf {\bibinfo
  {volume} {118}},\ \bibinfo {pages} {087001} (\bibinfo {year}
  {2017})}\BibitemShut {NoStop}%
\bibitem [{\citenamefont {Li}\ \emph {et~al.}(2021)\citenamefont {Li},
  \citenamefont {Zheng}, \citenamefont {Fang}, \citenamefont {Zhang},
  \citenamefont {Chen}, \citenamefont {Chen}, \citenamefont {Liang},
  \citenamefont {Shi}, \citenamefont {Pei}, \citenamefont {Xu} \emph
  {et~al.}}]{li2021observation}%
  \BibitemOpen
  \bibfield  {author} {\bibinfo {author} {\bibfnamefont {Y.}~\bibnamefont
  {Li}}, \bibinfo {author} {\bibfnamefont {H.}~\bibnamefont {Zheng}}, \bibinfo
  {author} {\bibfnamefont {Y.}~\bibnamefont {Fang}}, \bibinfo {author}
  {\bibfnamefont {D.}~\bibnamefont {Zhang}}, \bibinfo {author} {\bibfnamefont
  {Y.}~\bibnamefont {Chen}}, \bibinfo {author} {\bibfnamefont {C.}~\bibnamefont
  {Chen}}, \bibinfo {author} {\bibfnamefont {A.}~\bibnamefont {Liang}},
  \bibinfo {author} {\bibfnamefont {W.}~\bibnamefont {Shi}}, \bibinfo {author}
  {\bibfnamefont {D.}~\bibnamefont {Pei}}, \bibinfo {author} {\bibfnamefont
  {L.}~\bibnamefont {Xu}},  \emph {et~al.},\ }\href {\doibase
  10.1038/s41467-021-23076-1} {\bibfield  {journal} {\bibinfo  {journal} {Nat.
  Commun.}\ }\textbf {\bibinfo {volume} {12}},\ \bibinfo {pages} {2874}
  (\bibinfo {year} {2021})}\BibitemShut {NoStop}%
\bibitem [{\citenamefont {Wang}\ \emph {et~al.}(2012)\citenamefont {Wang},
  \citenamefont {Kalantar-Zadeh}, \citenamefont {Kis}, \citenamefont
  {Coleman},\ and\ \citenamefont {Strano}}]{wang2012electronics}%
  \BibitemOpen
  \bibfield  {author} {\bibinfo {author} {\bibfnamefont {Q.~H.}\ \bibnamefont
  {Wang}}, \bibinfo {author} {\bibfnamefont {K.}~\bibnamefont
  {Kalantar-Zadeh}}, \bibinfo {author} {\bibfnamefont {A.}~\bibnamefont {Kis}},
  \bibinfo {author} {\bibfnamefont {J.~N.}\ \bibnamefont {Coleman}}, \ and\
  \bibinfo {author} {\bibfnamefont {M.~S.}\ \bibnamefont {Strano}},\ }\href
  {\doibase 10.1038/nnano.2012.193} {\bibfield  {journal} {\bibinfo  {journal}
  {Nat. Nantechnol.}\ }\textbf {\bibinfo {volume} {7}},\ \bibinfo {pages} {699}
  (\bibinfo {year} {2012})}\BibitemShut {NoStop}%
\bibitem [{\citenamefont {Manzeli}\ \emph {et~al.}(2017)\citenamefont
  {Manzeli}, \citenamefont {Ovchinnikov}, \citenamefont {Pasquier},
  \citenamefont {Yazyev},\ and\ \citenamefont {Kis}}]{manzeli20172d}%
  \BibitemOpen
  \bibfield  {author} {\bibinfo {author} {\bibfnamefont {S.}~\bibnamefont
  {Manzeli}}, \bibinfo {author} {\bibfnamefont {D.}~\bibnamefont
  {Ovchinnikov}}, \bibinfo {author} {\bibfnamefont {D.}~\bibnamefont
  {Pasquier}}, \bibinfo {author} {\bibfnamefont {O.~V.}\ \bibnamefont
  {Yazyev}}, \ and\ \bibinfo {author} {\bibfnamefont {A.}~\bibnamefont {Kis}},\
  }\href {\doibase 10.1038/natrevmats.2017.33} {\bibfield  {journal} {\bibinfo
  {journal} {Nat. Rev. Mater.}\ }\textbf {\bibinfo {volume} {2}},\ \bibinfo
  {pages} {1} (\bibinfo {year} {2017})}\BibitemShut {NoStop}%
\bibitem [{\citenamefont {Tsirkin}(2021)}]{tsirkin2021high}%
  \BibitemOpen
  \bibfield  {author} {\bibinfo {author} {\bibfnamefont {S.~S.}\ \bibnamefont
  {Tsirkin}},\ }\href {\doibase 10.1038/s41524-021-00498-5} {\bibfield
  {journal} {\bibinfo  {journal} {npj Computational Materials}\ }\textbf
  {\bibinfo {volume} {7}},\ \bibinfo {pages} {1} (\bibinfo {year}
  {2021})}\BibitemShut {NoStop}%
\bibitem [{\citenamefont {Wu}\ \emph {et~al.}(2018{\natexlab{b}})\citenamefont
  {Wu}, \citenamefont {Zhang}, \citenamefont {Song}, \citenamefont {Troyer},\
  and\ \citenamefont {Soluyanov}}]{WU2017}%
  \BibitemOpen
  \bibfield  {author} {\bibinfo {author} {\bibfnamefont {Q.}~\bibnamefont
  {Wu}}, \bibinfo {author} {\bibfnamefont {S.}~\bibnamefont {Zhang}}, \bibinfo
  {author} {\bibfnamefont {H.-F.}\ \bibnamefont {Song}}, \bibinfo {author}
  {\bibfnamefont {M.}~\bibnamefont {Troyer}}, \ and\ \bibinfo {author}
  {\bibfnamefont {A.~A.}\ \bibnamefont {Soluyanov}},\ }\href {\doibase
  https://doi.org/10.1016/j.cpc.2017.09.033} {\bibfield  {journal} {\bibinfo
  {journal} {Computer Physics Communications}\ }\textbf {\bibinfo {volume}
  {224}},\ \bibinfo {pages} {405 } (\bibinfo {year}
  {2018}{\natexlab{b}})}\BibitemShut {NoStop}%
\bibitem [{\citenamefont {Sancho}\ \emph {et~al.}(1985)\citenamefont {Sancho},
  \citenamefont {Sancho}, \citenamefont {Sancho},\ and\ \citenamefont
  {Rubio}}]{sancho1985highly}%
  \BibitemOpen
  \bibfield  {author} {\bibinfo {author} {\bibfnamefont {M.~L.}\ \bibnamefont
  {Sancho}}, \bibinfo {author} {\bibfnamefont {J.~L.}\ \bibnamefont {Sancho}},
  \bibinfo {author} {\bibfnamefont {J.~L.}\ \bibnamefont {Sancho}}, \ and\
  \bibinfo {author} {\bibfnamefont {J.}~\bibnamefont {Rubio}},\ }\href
  {\doibase 10.1088/0305-4608/15/4/009} {\bibfield  {journal} {\bibinfo
  {journal} {J. Phys. F: Met. Phys.}\ }\textbf {\bibinfo {volume} {15}},\
  \bibinfo {pages} {851} (\bibinfo {year} {1985})}\BibitemShut {NoStop}%
\bibitem [{\citenamefont {Destraz}\ \emph {et~al.}(2020)\citenamefont
  {Destraz}, \citenamefont {Das}, \citenamefont {Tsirkin}, \citenamefont {Xu},
  \citenamefont {Neupert}, \citenamefont {Chang}, \citenamefont {Schilling},
  \citenamefont {Grushin}, \citenamefont {Kohlbrecher}, \citenamefont {Keller}
  \emph {et~al.}}]{destraz2020magnetism}%
  \BibitemOpen
  \bibfield  {author} {\bibinfo {author} {\bibfnamefont {D.}~\bibnamefont
  {Destraz}}, \bibinfo {author} {\bibfnamefont {L.}~\bibnamefont {Das}},
  \bibinfo {author} {\bibfnamefont {S.~S.}\ \bibnamefont {Tsirkin}}, \bibinfo
  {author} {\bibfnamefont {Y.}~\bibnamefont {Xu}}, \bibinfo {author}
  {\bibfnamefont {T.}~\bibnamefont {Neupert}}, \bibinfo {author} {\bibfnamefont
  {J.}~\bibnamefont {Chang}}, \bibinfo {author} {\bibfnamefont
  {A.}~\bibnamefont {Schilling}}, \bibinfo {author} {\bibfnamefont {A.~G.}\
  \bibnamefont {Grushin}}, \bibinfo {author} {\bibfnamefont {J.}~\bibnamefont
  {Kohlbrecher}}, \bibinfo {author} {\bibfnamefont {L.}~\bibnamefont {Keller}},
   \emph {et~al.},\ }\href {\doibase 10.1038/s41535-019-0207-7} {\bibfield
  {journal} {\bibinfo  {journal} {npj Quantum Mater.}\ }\textbf {\bibinfo
  {volume} {5}},\ \bibinfo {pages} {5} (\bibinfo {year} {2020})}\BibitemShut
  {NoStop}%
\bibitem [{\citenamefont {Po}\ \emph {et~al.}(2018)\citenamefont {Po},
  \citenamefont {Watanabe},\ and\ \citenamefont {Vishwanath}}]{FragilePo}%
  \BibitemOpen
  \bibfield  {author} {\bibinfo {author} {\bibfnamefont {H.~C.}\ \bibnamefont
  {Po}}, \bibinfo {author} {\bibfnamefont {H.}~\bibnamefont {Watanabe}}, \ and\
  \bibinfo {author} {\bibfnamefont {A.}~\bibnamefont {Vishwanath}},\ }\href
  {\doibase 10.1103/PhysRevLett.121.126402} {\bibfield  {journal} {\bibinfo
  {journal} {Phys. Rev. Lett.}\ }\textbf {\bibinfo {volume} {121}},\ \bibinfo
  {pages} {126402} (\bibinfo {year} {2018})}\BibitemShut {NoStop}%
\bibitem [{\citenamefont {Song}\ \emph {et~al.}(2020)\citenamefont {Song},
  \citenamefont {Elcoro},\ and\ \citenamefont {Bernevig}}]{song2020twisted}%
  \BibitemOpen
  \bibfield  {author} {\bibinfo {author} {\bibfnamefont {Z.-D.}\ \bibnamefont
  {Song}}, \bibinfo {author} {\bibfnamefont {L.}~\bibnamefont {Elcoro}}, \ and\
  \bibinfo {author} {\bibfnamefont {B.~A.}\ \bibnamefont {Bernevig}},\ }\href
  {\doibase 10.1126/science.aaz7650} {\bibfield  {journal} {\bibinfo  {journal}
  {Science}\ }\textbf {\bibinfo {volume} {367}},\ \bibinfo {pages} {794}
  (\bibinfo {year} {2020})}\BibitemShut {NoStop}%
\bibitem [{\citenamefont {Wieder}\ \emph {et~al.}(2020)\citenamefont {Wieder},
  \citenamefont {Wang}, \citenamefont {Cano}, \citenamefont {Dai},
  \citenamefont {Schoop}, \citenamefont {Bradlyn},\ and\ \citenamefont
  {Bernevig}}]{wieder2020strong}%
  \BibitemOpen
  \bibfield  {author} {\bibinfo {author} {\bibfnamefont {B.~J.}\ \bibnamefont
  {Wieder}}, \bibinfo {author} {\bibfnamefont {Z.}~\bibnamefont {Wang}},
  \bibinfo {author} {\bibfnamefont {J.}~\bibnamefont {Cano}}, \bibinfo {author}
  {\bibfnamefont {X.}~\bibnamefont {Dai}}, \bibinfo {author} {\bibfnamefont
  {L.~M.}\ \bibnamefont {Schoop}}, \bibinfo {author} {\bibfnamefont
  {B.}~\bibnamefont {Bradlyn}}, \ and\ \bibinfo {author} {\bibfnamefont
  {B.~A.}\ \bibnamefont {Bernevig}},\ }\href {\doibase
  10.1038/s41467-020-14443-5} {\bibfield  {journal} {\bibinfo  {journal} {Nat.
  Commun.}\ }\textbf {\bibinfo {volume} {11}},\ \bibinfo {pages} {627}
  (\bibinfo {year} {2020})}\BibitemShut {NoStop}%
\end{thebibliography}%
\par 
\textbf{Acknowledgements  }
\par 
This work was supported by the National Science Foundation MRSEC program (DMR-1720319) at the Materials Research Center of Northwestern University, and the start up funds of P. G. provided by the Northwestern University. This research was also supported in part by the National Science Foundation under Grant No. NSF PHY-1748958.

\appendix

\section*{Appendix A: Determination of invariant and spin-pumping}\label{app:A}
\begin{figure*}
\centering
\subfigure[]{
\includegraphics[scale=0.45]{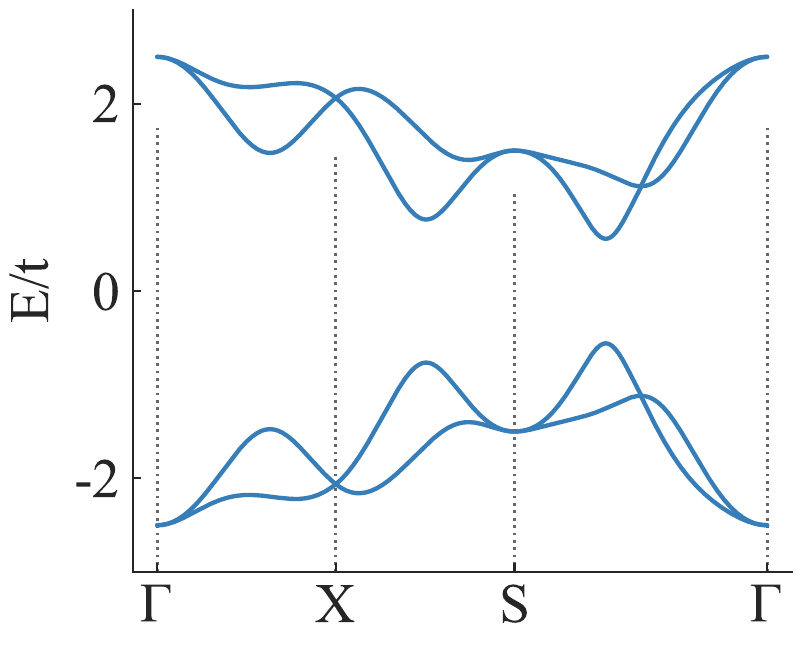}
\label{fig:FuBands}}
\subfigure[]{
\includegraphics[scale=0.36]{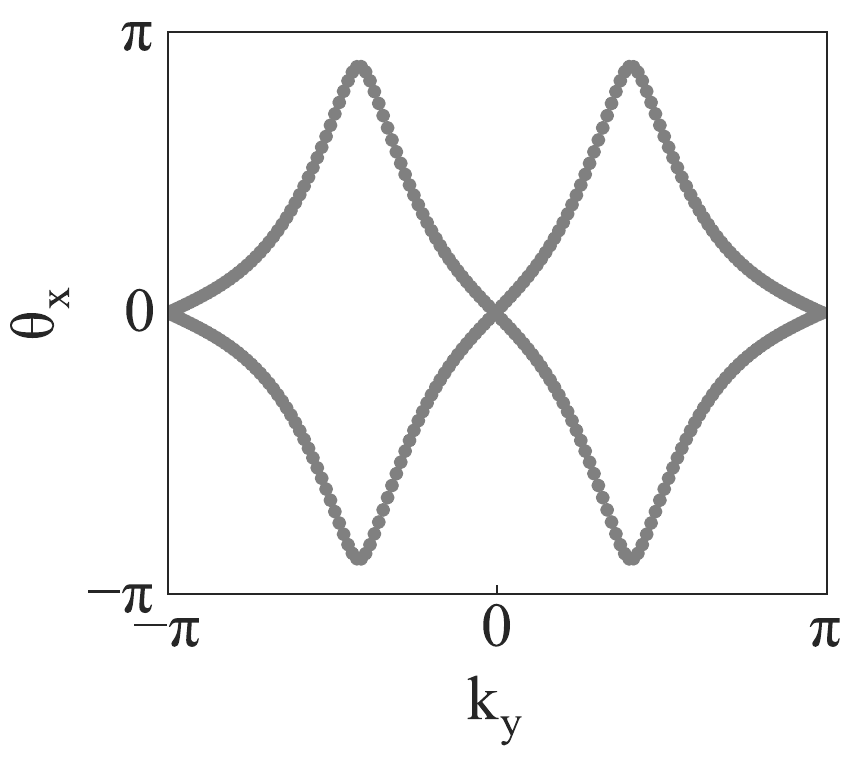}
\label{fig:FuWCC}}
\subfigure[]{
\includegraphics[scale=0.4]{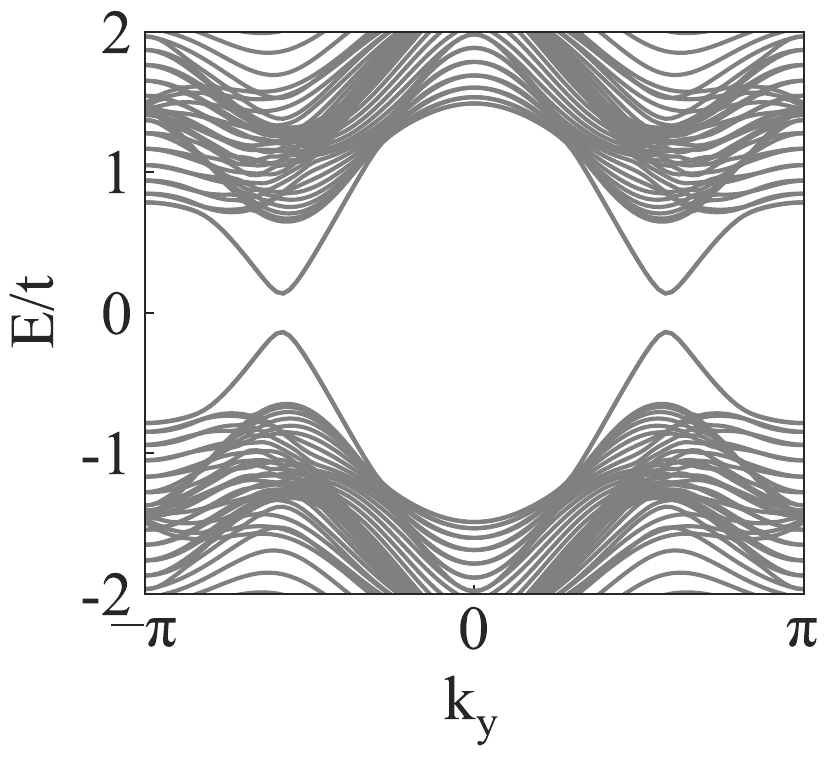}
\label{fig:FuSlab}}
\subfigure[]{
\includegraphics[scale=0.33]{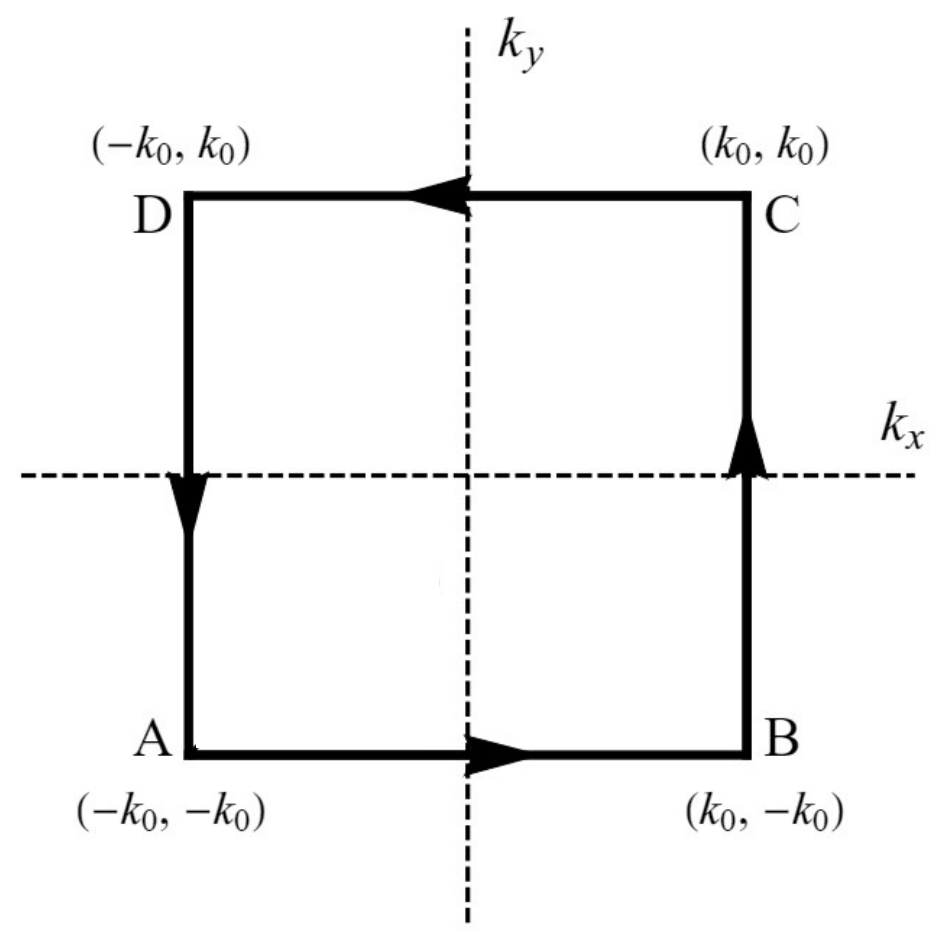}
\label{fig:FuPWLSchem}}
\subfigure[]{
\includegraphics[scale=0.28]{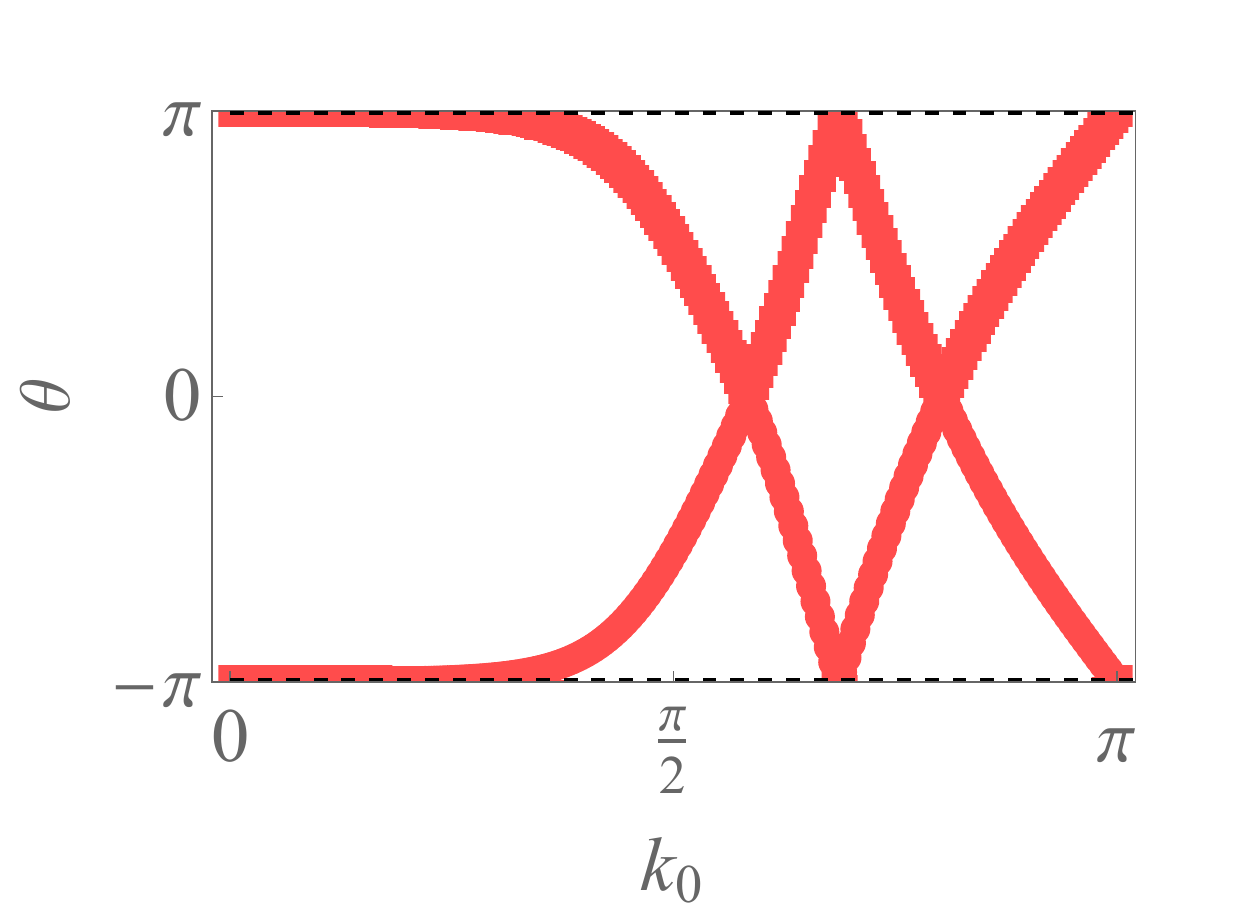}
\label{fig:FuPWLWCC}}
\textbf{\subfigure[]{
\includegraphics[scale=0.28]{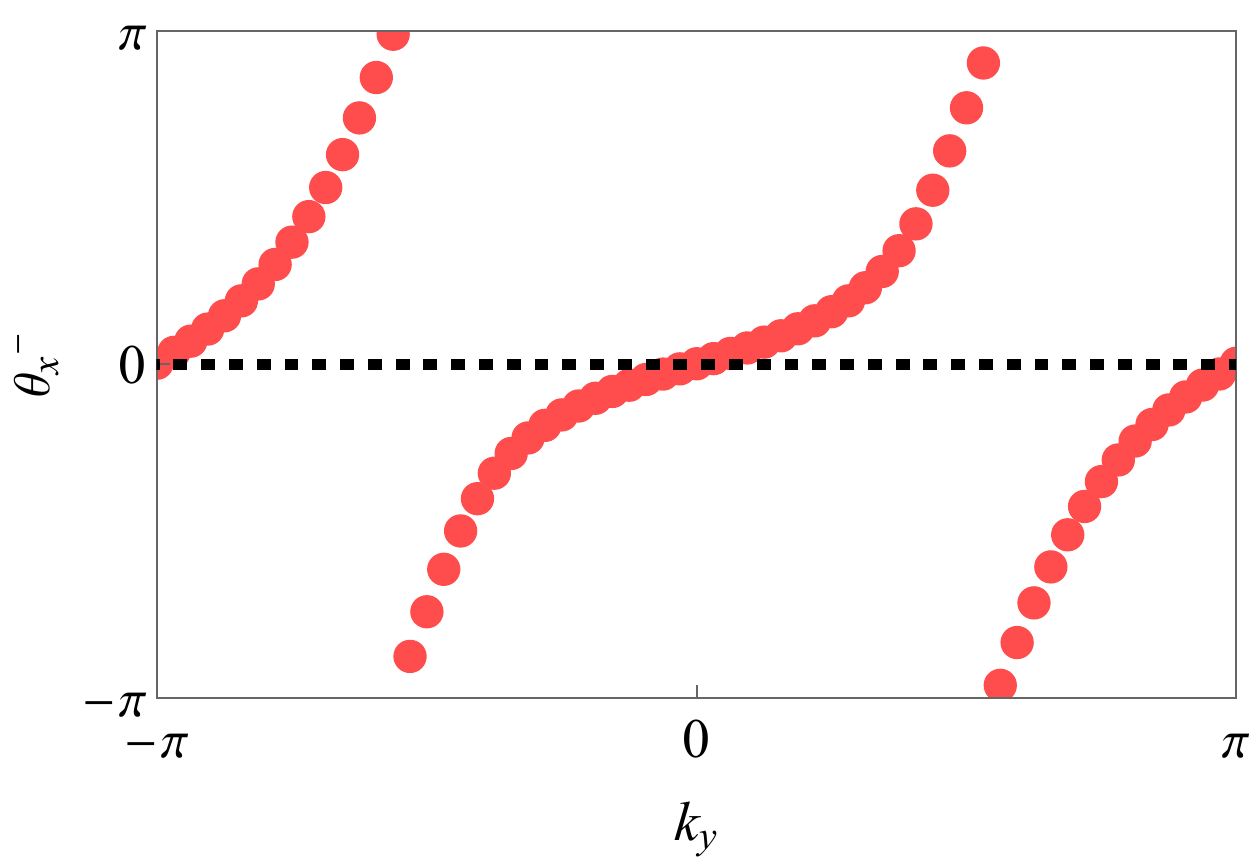}
\label{fig:FuPSP}}}
\subfigure[]{
\includegraphics[scale=0.28]{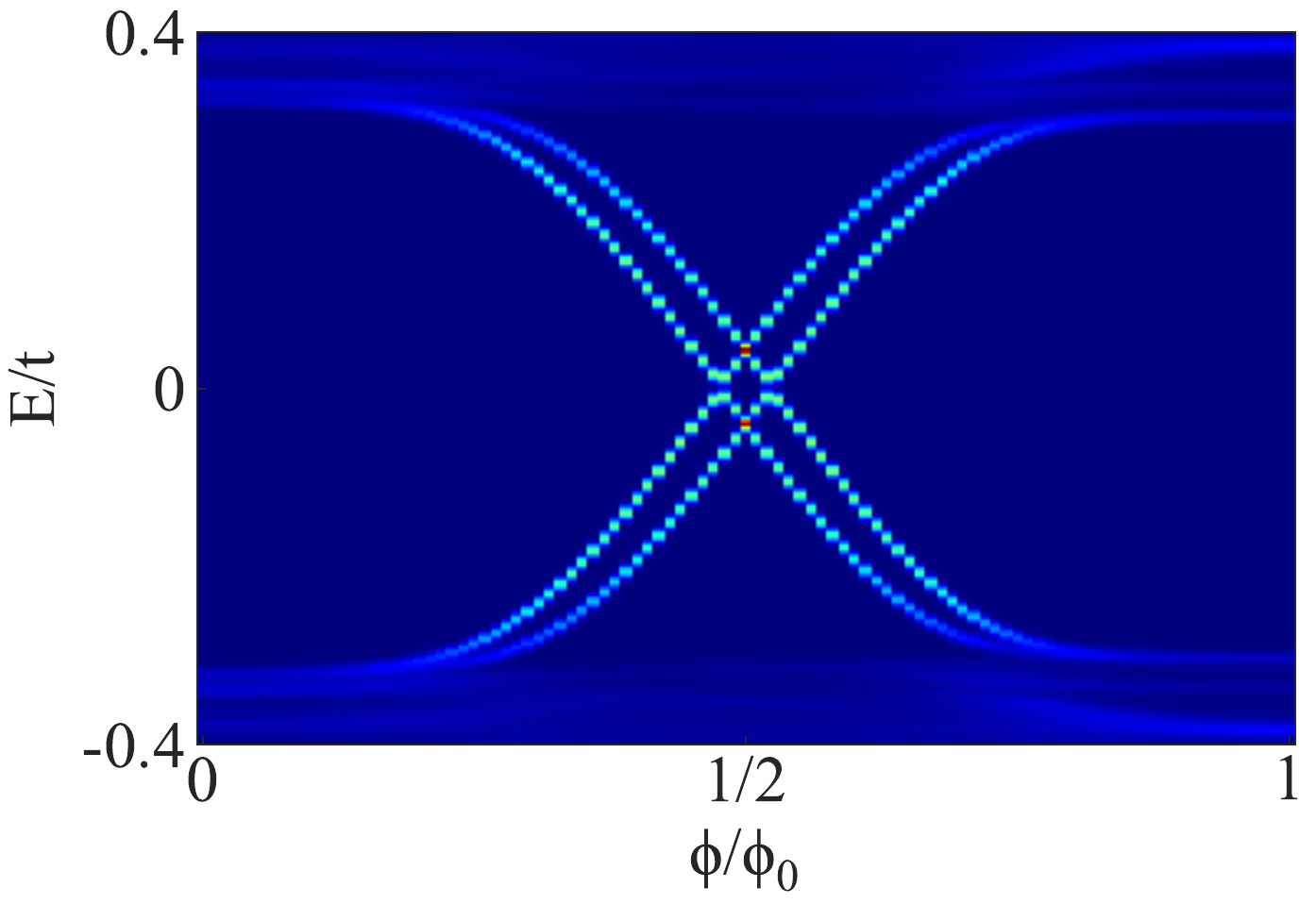}
\label{fig:FuStateEnergy}}
\subfigure[]{
\includegraphics[scale=0.28]{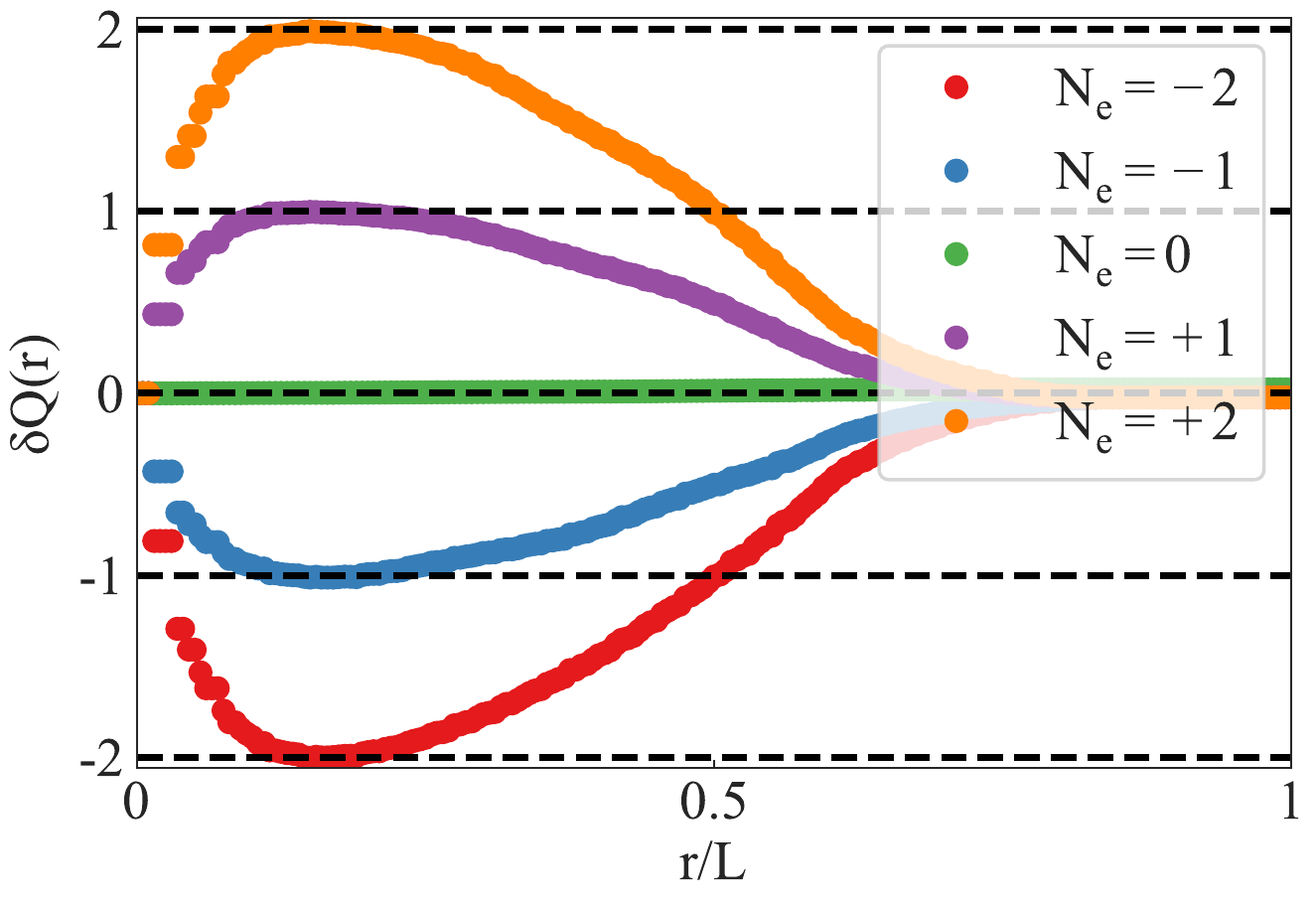}
\label{fig:FuInduced}}
\caption{(a) Band structure along high-symmetry path for tight-binding model given by eq. \eqref{eq:FuModel}. (b) Wannier center charge spectra for occupied bands. The spectra is gapped, disallowing assignment of non-trivial topology. (c) Spectra considering a slab of 40 unit cells along the $\hat{x}$ direction and periodic boundary conditions along $\hat{y}$. No determination about bulk topology can be made due to the absence of gapless states. (d) Schematic of path for calculation of in-plane Wilson loop for determining magnitude of ground-state spin-Chern number, $|\mathcal{C}_{s,G}|$ as a function of $k_{0}: 0\rightarrow \pi$. (e) In-plane loop Wannier center charge (WCC) spectra for eq. \eqref{eq:FuModel}. The spectra clearly demonstrates a double winding, corresponding to a ground state bulk invariant, $|\mathcal{C}_{s,G}|=2$.(f) Wannier center charge spectra for spin-resolved Wilson loop along $\hat{x}$ axis as a function of transverse momenta, $k_{y}$, displaying a double winding, indicating a spin Chern number $|\mathcal{C}_{s,G}|=2$. (g) Local density of states on the flux tube as a function of the flux strength, demonstrating the presence of vortex bound modes, but the absence of spectral-flow.  (h) Induced charge as a function of filling the VBMs seen in (g). The results demonstrate the robust nature of spin-charge separation upon breaking the degeneracy of the mid-gap VBMs at $\pi$-flux. } 
\label{fig:FuModel}
\end{figure*}

\begin{figure*}
\centering
\subfigure[]{
\includegraphics[scale=0.4]{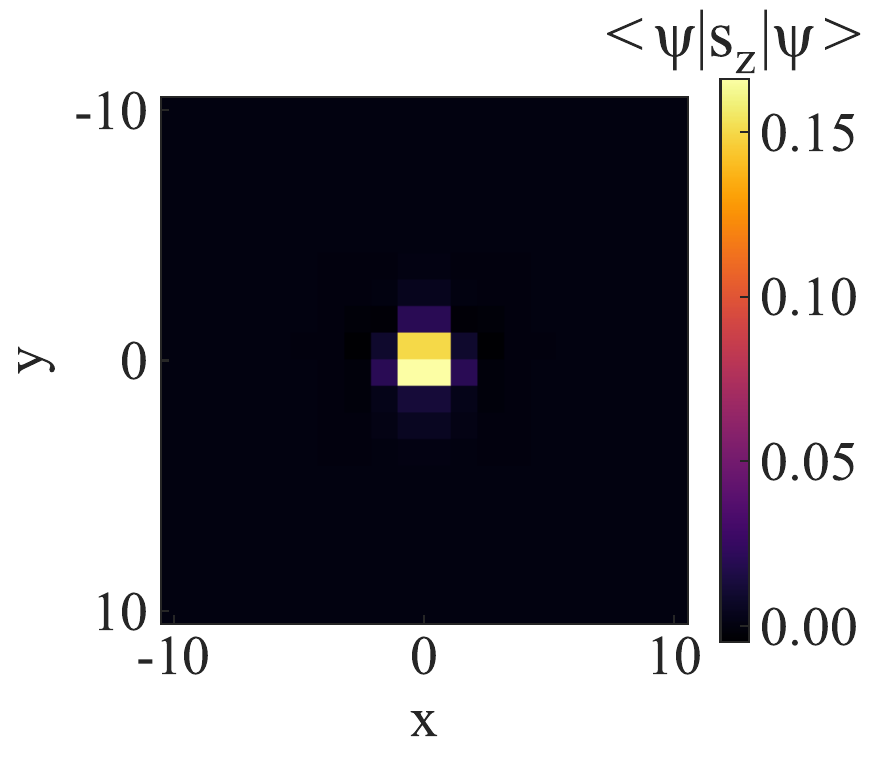}
\label{fig:state1sz}}
\textbf{\subfigure[]{
\includegraphics[scale=0.4]{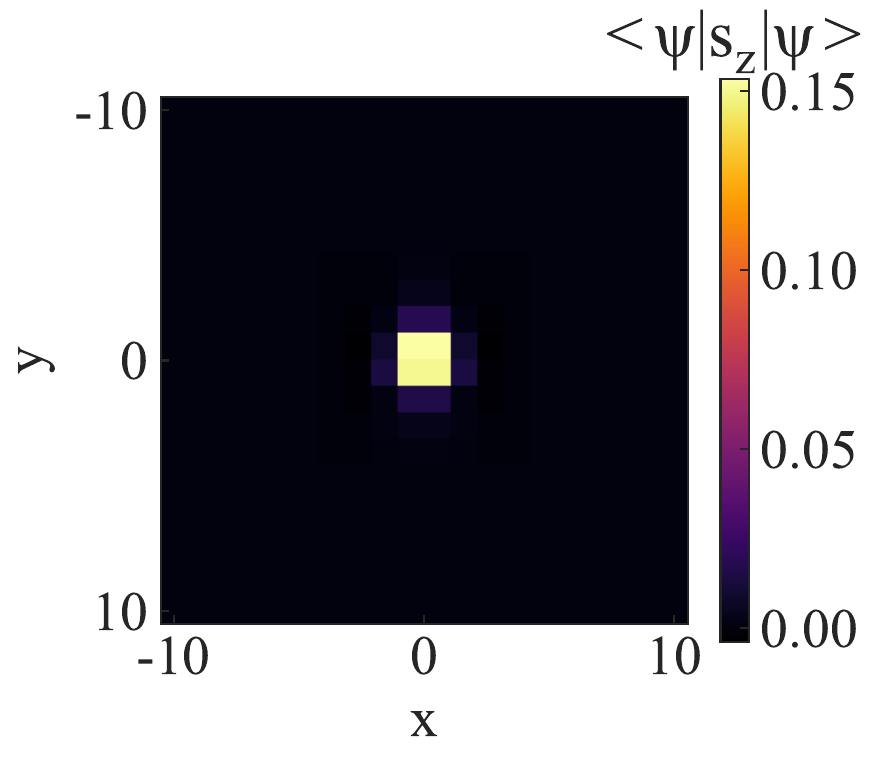}
\label{fig:state2sz}}}
\subfigure[]{
\includegraphics[scale=0.4]{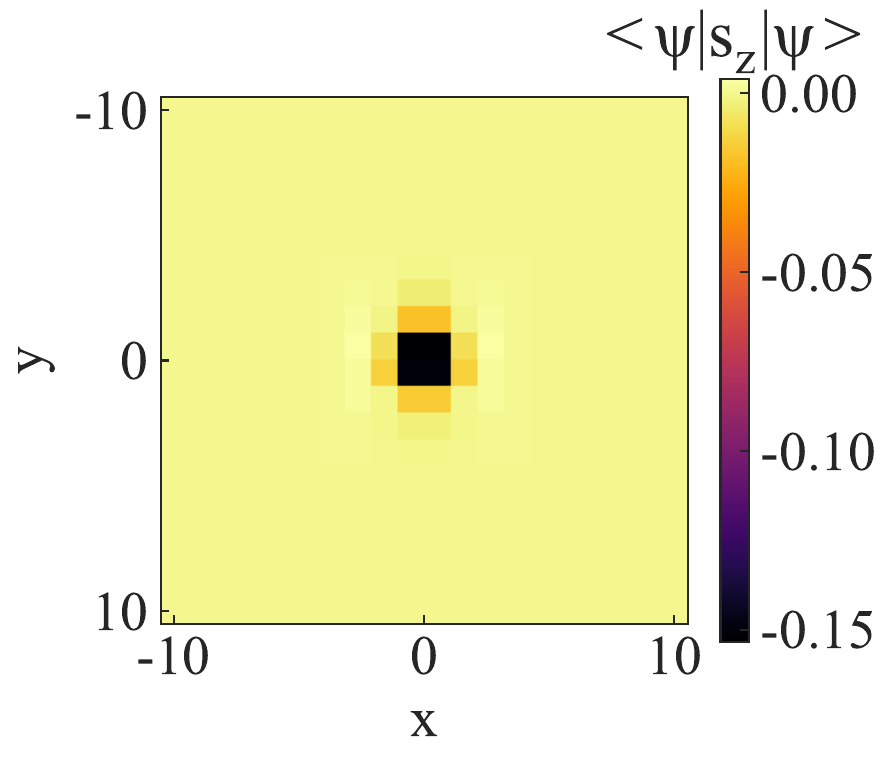}
\label{fig:state3sz}}
\subfigure[]{
\includegraphics[scale=0.4]{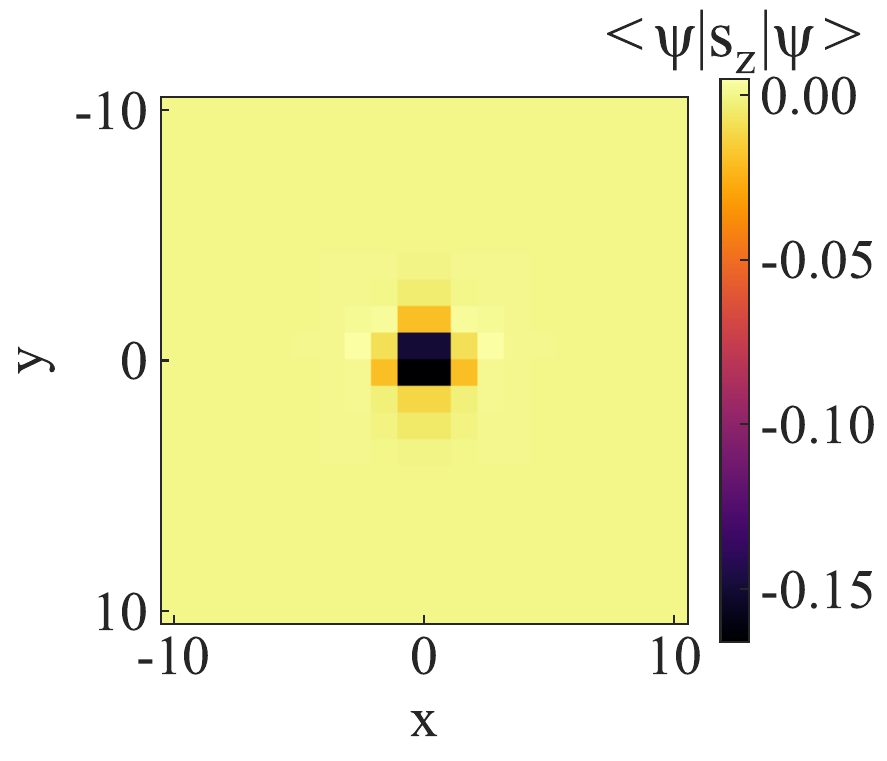}
\label{fig:stat4sz}}
\subfigure[]{
\includegraphics[scale=0.4]{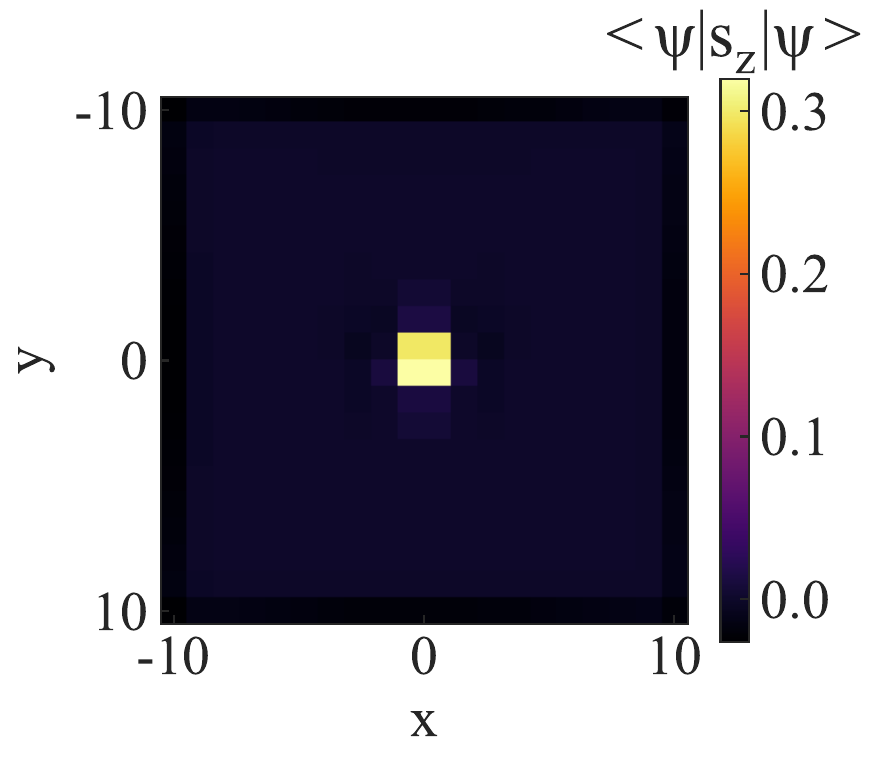}
\label{fig:halffillsz}}
\subfigure[]{
\includegraphics[scale=0.4]{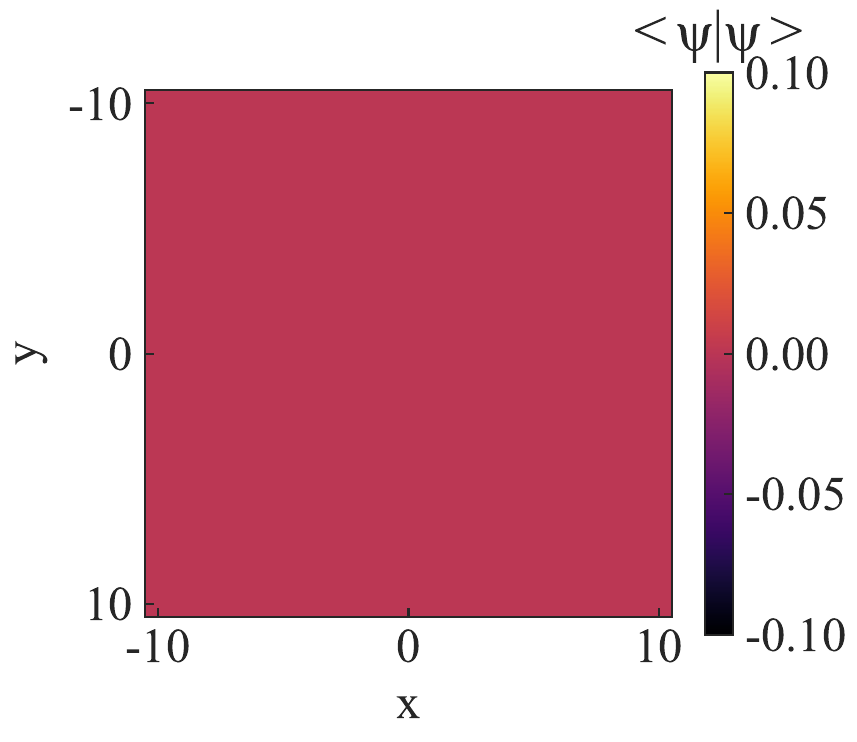}
\label{fig:halffillcharge}}
\subfigure[]{
\includegraphics[scale=0.4]{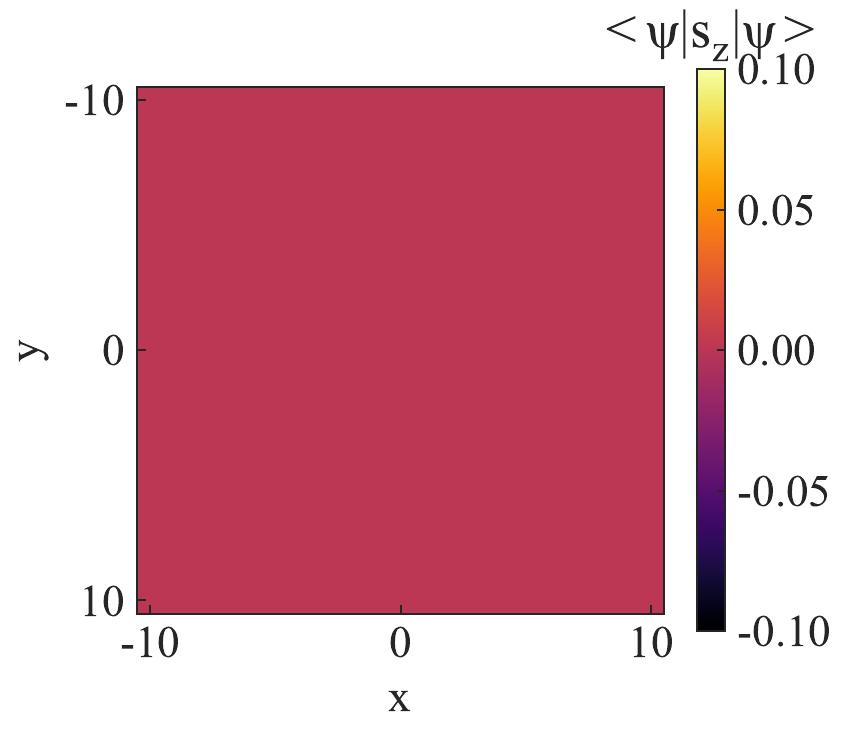}
\label{fig:2dopedsz}}
\subfigure[]{
\includegraphics[scale=0.4]{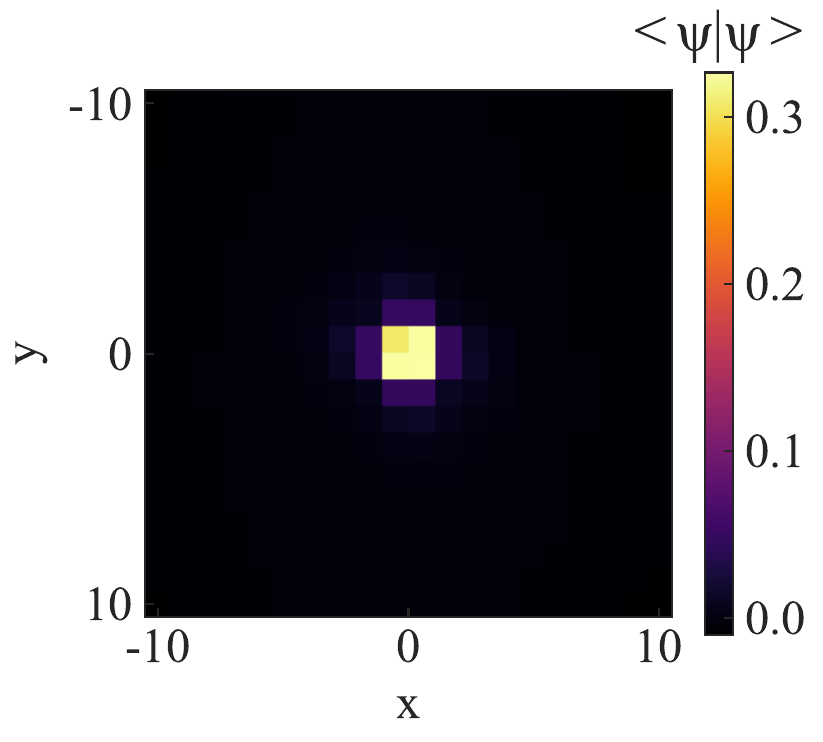}
\label{fig:2dopedcharge}}
\caption{(a-d) Real-space expectation value of of $\hat{s}=s_{z}$ for the four VBMs at $\phi=\phi_{0}/2$ in Fig. \eqref{fig:FuStateEnergy}, from lowest energy (a) to highest energy (d). The two VBMs below the Fermi energy (a,b) and the two VBMs above the Fermi energy (c,d) display equal and opposite structures, indicating that the four VBMs compose two Kramers pairs with Kramers partners separated by the Fermi energy. Due to this separation of Kramers partners, there exists a finite induced spin on the vortex at half-filling, shown in (e), although it need not be quantized along a given spin-axis in the presence of spin-orbit coupling. Nevertheless, spin-charge-separation is achieved as the induced charge at half-filling vanishes at half-filling, shown in (f). If we dope by $N_{e}=|\mathcal{C}_{s,G}|=2$ states, such that the vortex acquires a quantized induced charge, as calculated in Fig. \eqref{fig:FuInduced} with density shown in (g), we observe in (h) that induced spin on the vortex vanishes. In this way, as a function of filling the vortex can acquire no induced charge but finite induced spin, or no induced spin and quantized induced charge. } 
\label{fig:InducedSpin}
\end{figure*}

\begin{figure*}
\centering
\subfigure[]{
\includegraphics[scale=0.25]{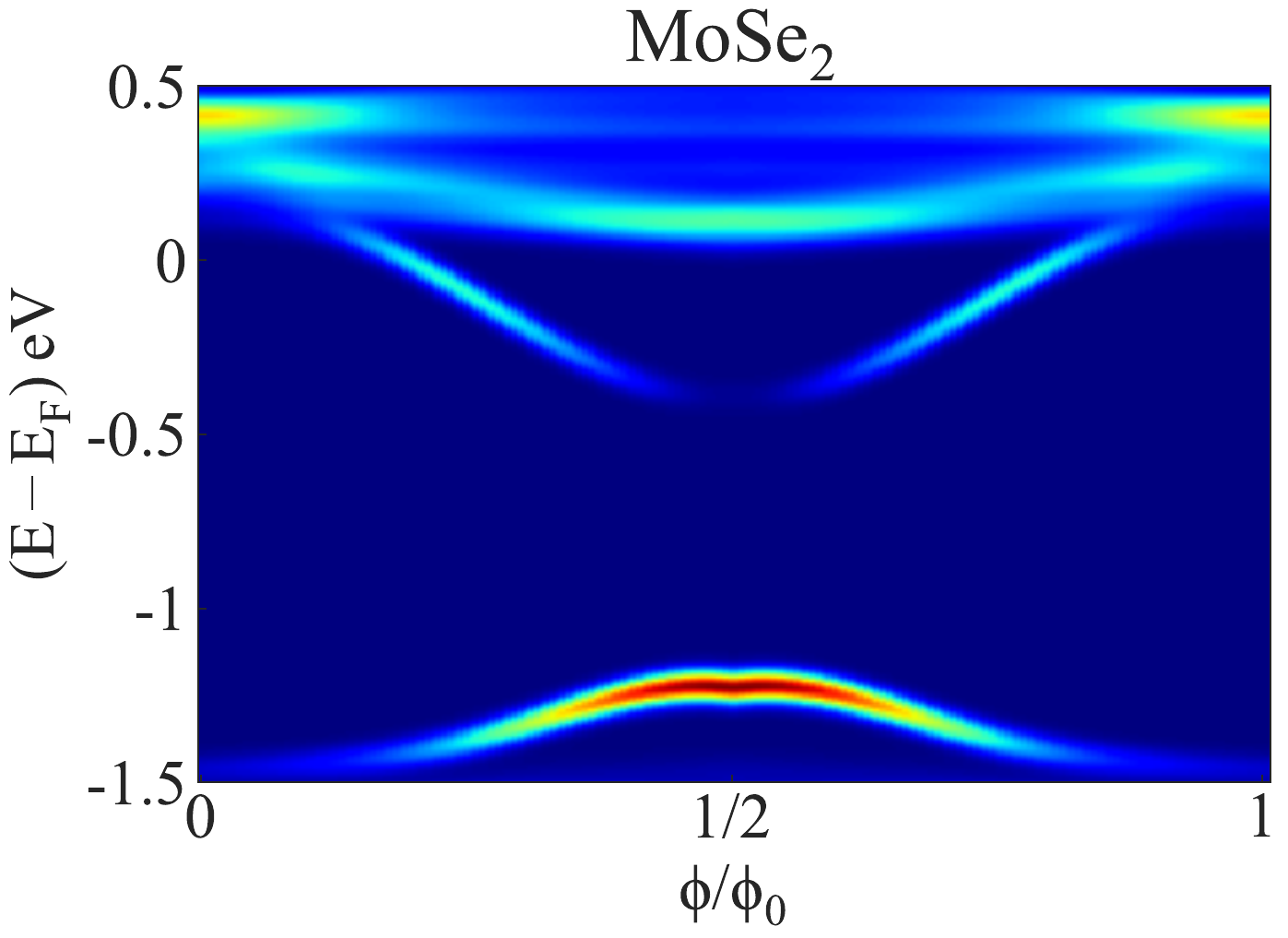}
\label{fig:}}
\subfigure[]{
\includegraphics[scale=0.25]{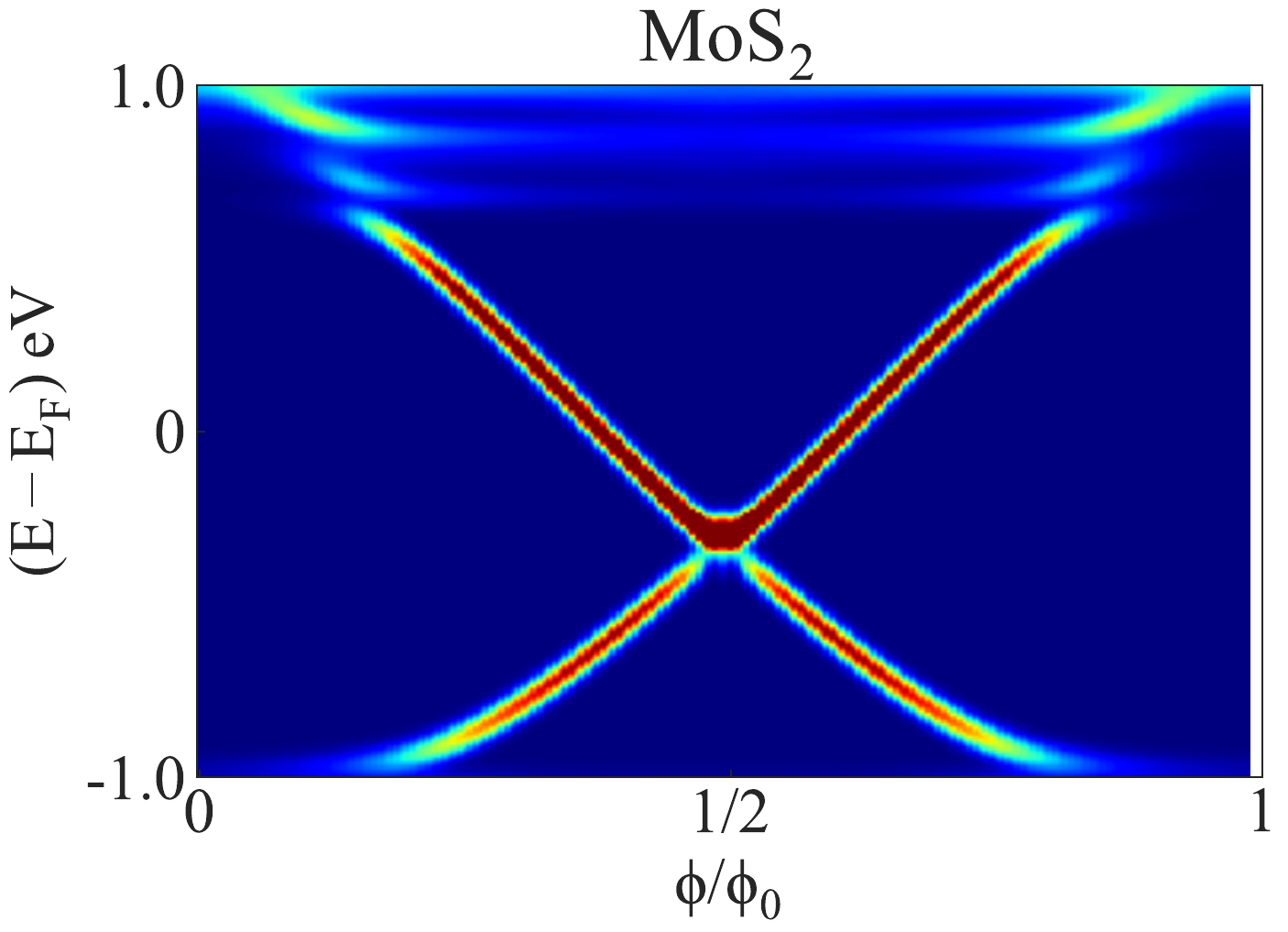}
\label{fig:}}
\subfigure[]{
\includegraphics[scale=0.25]{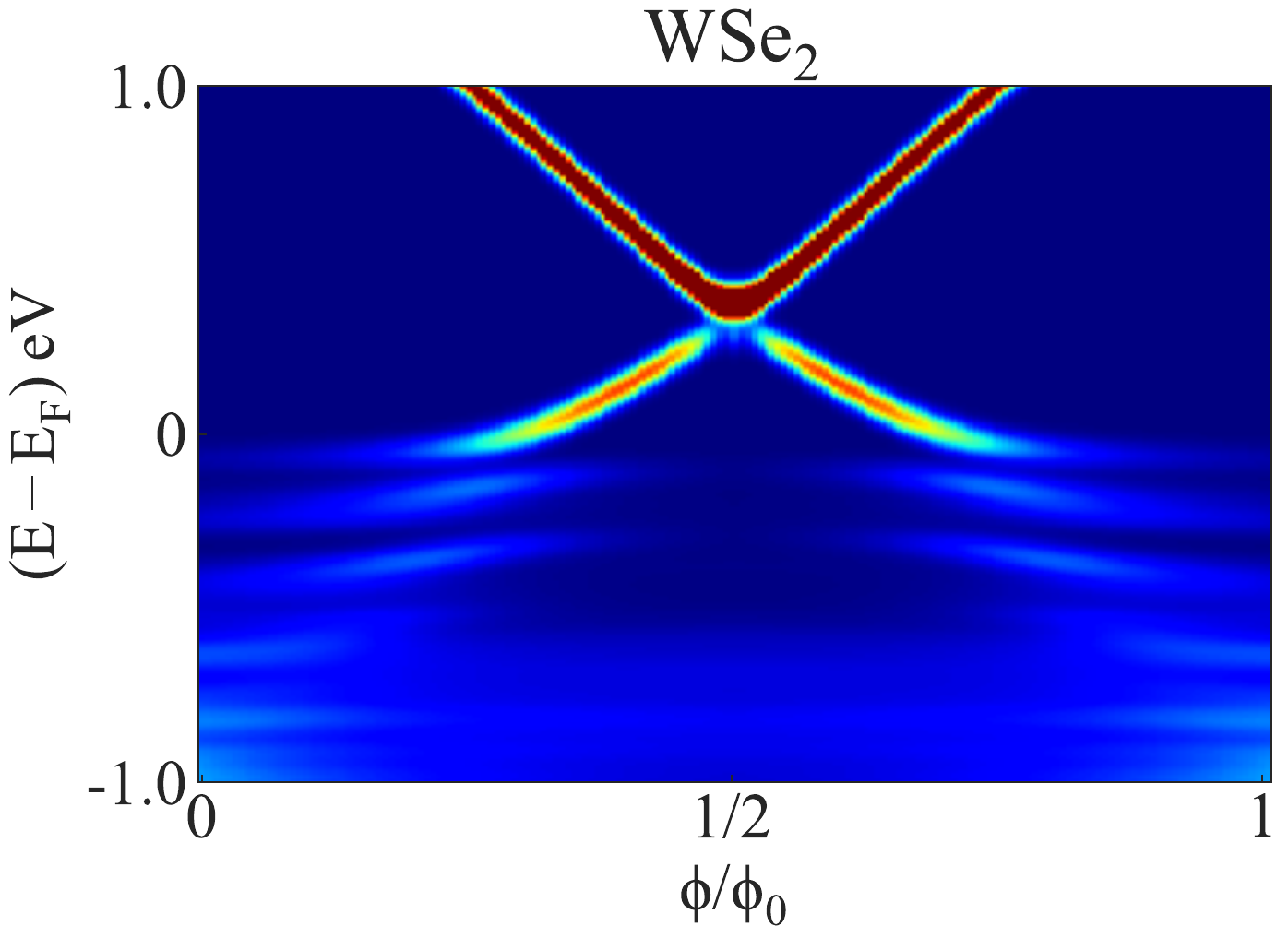}
\label{fig:}}
\subfigure[]{
\includegraphics[scale=0.25]{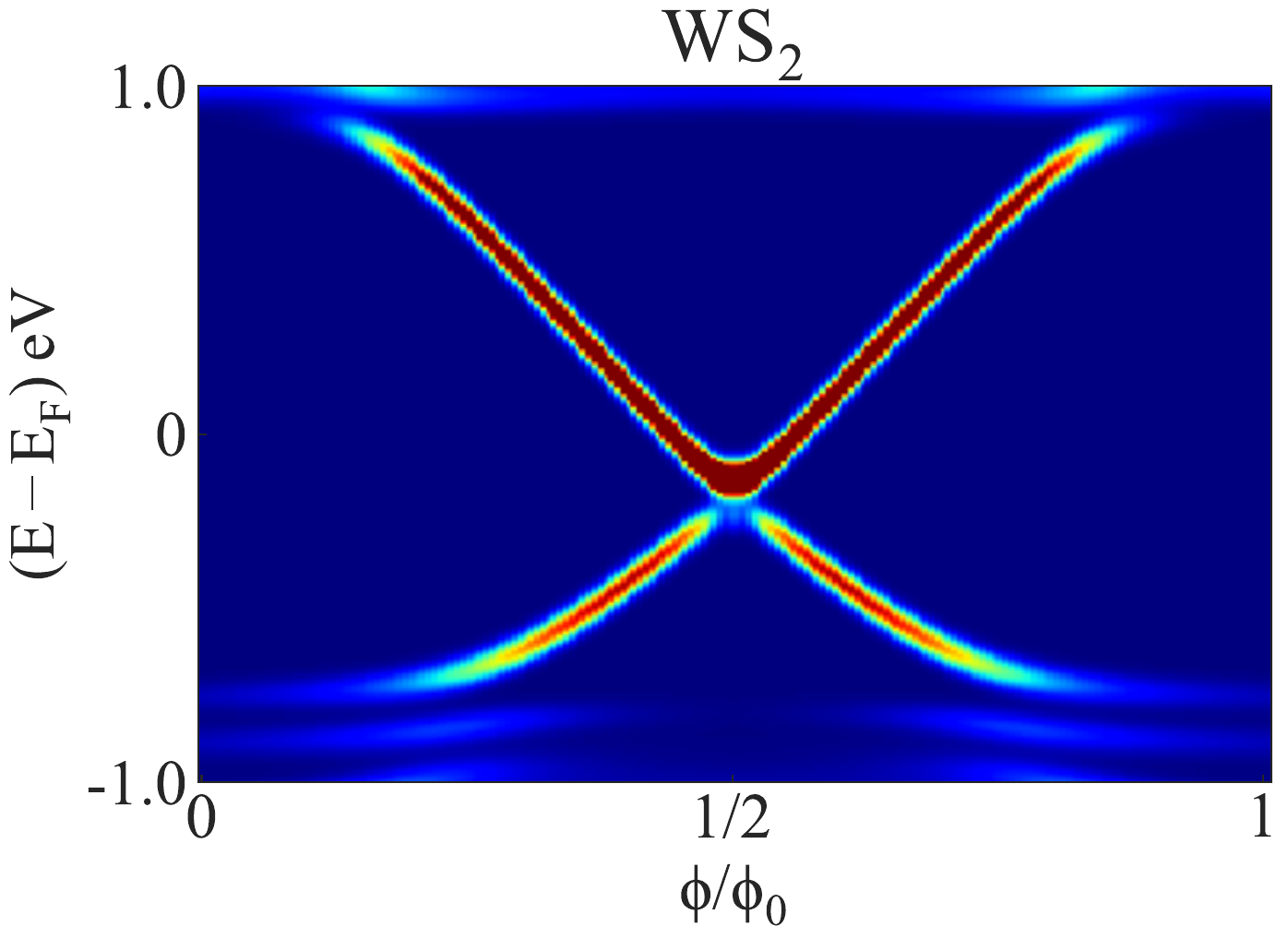}
\label{fig:}}
\subfigure[]{
\includegraphics[scale=0.25]{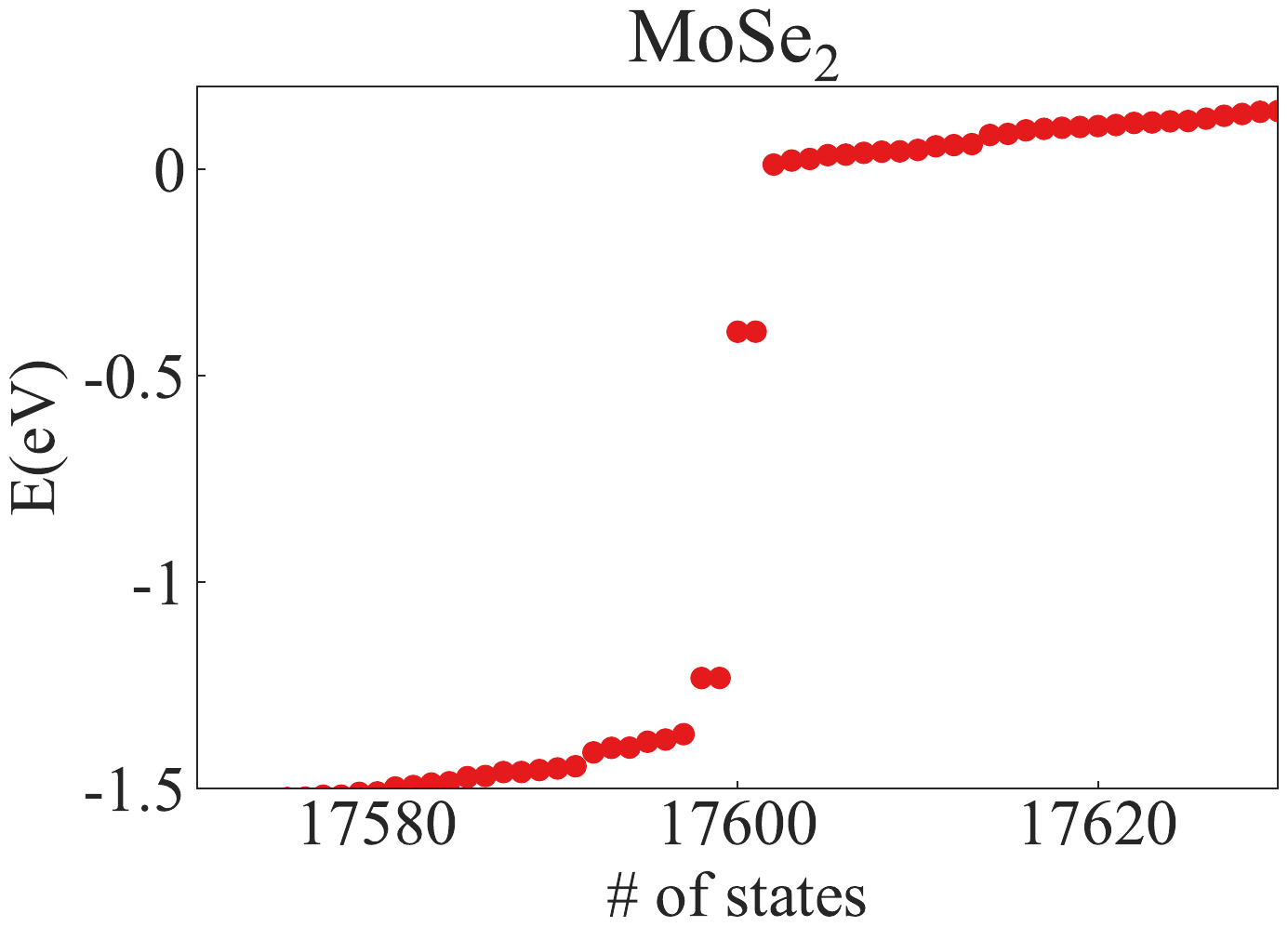}
\label{fig:}}
\subfigure[]{
\includegraphics[scale=0.25]{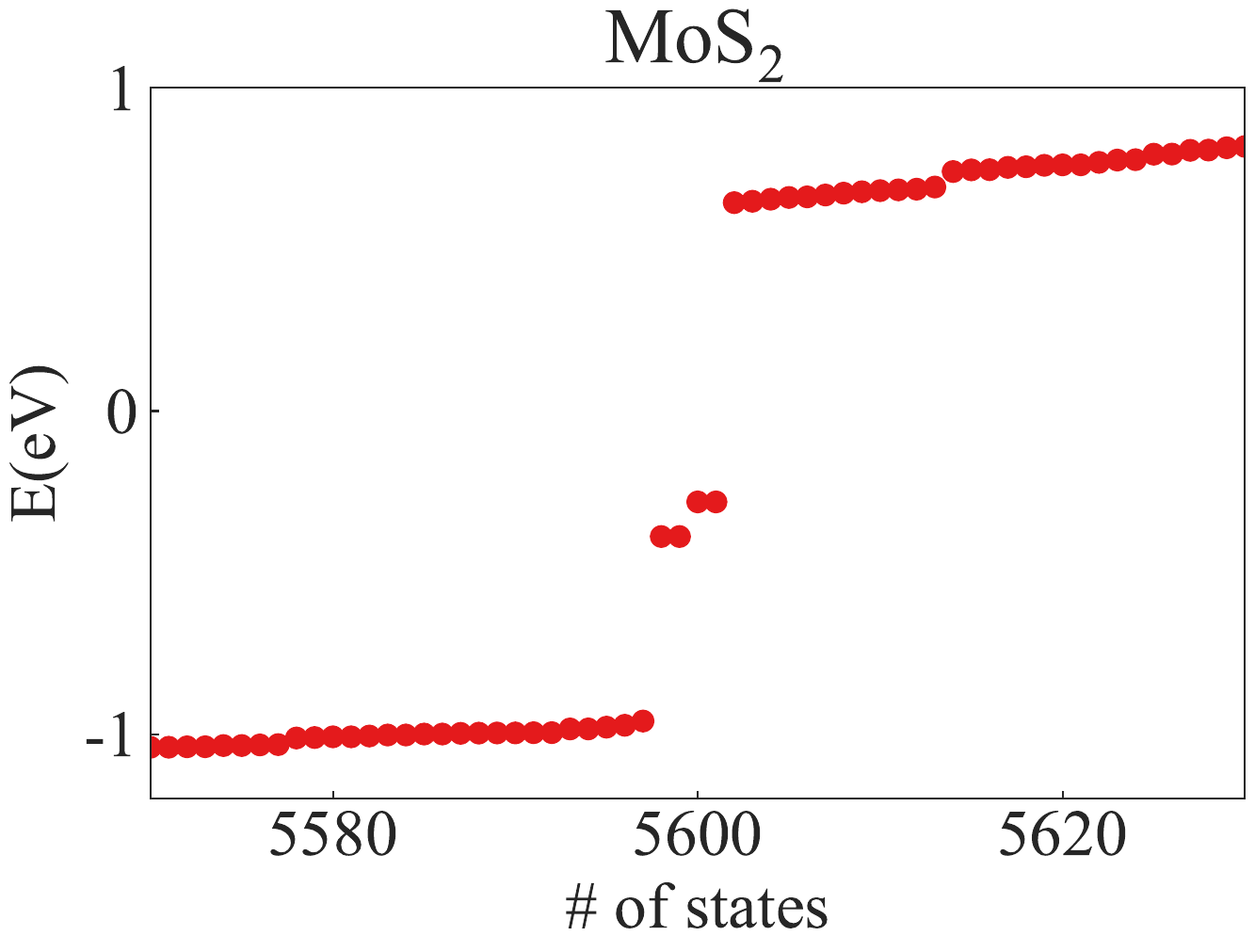}
\label{fig:}}
\subfigure[]{
\includegraphics[scale=0.25]{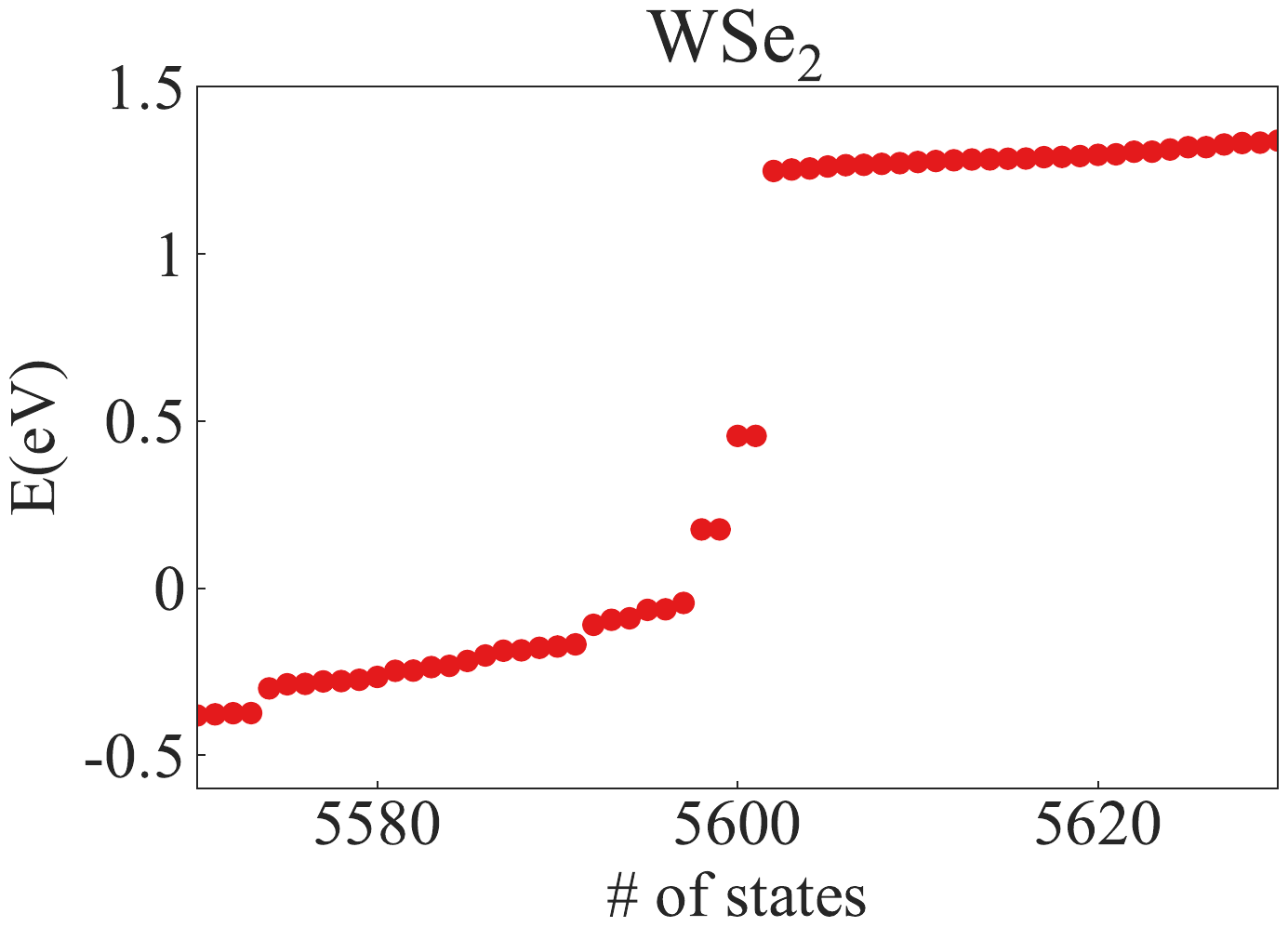}
\label{fig:}}
\subfigure[]{
\includegraphics[scale=0.25]{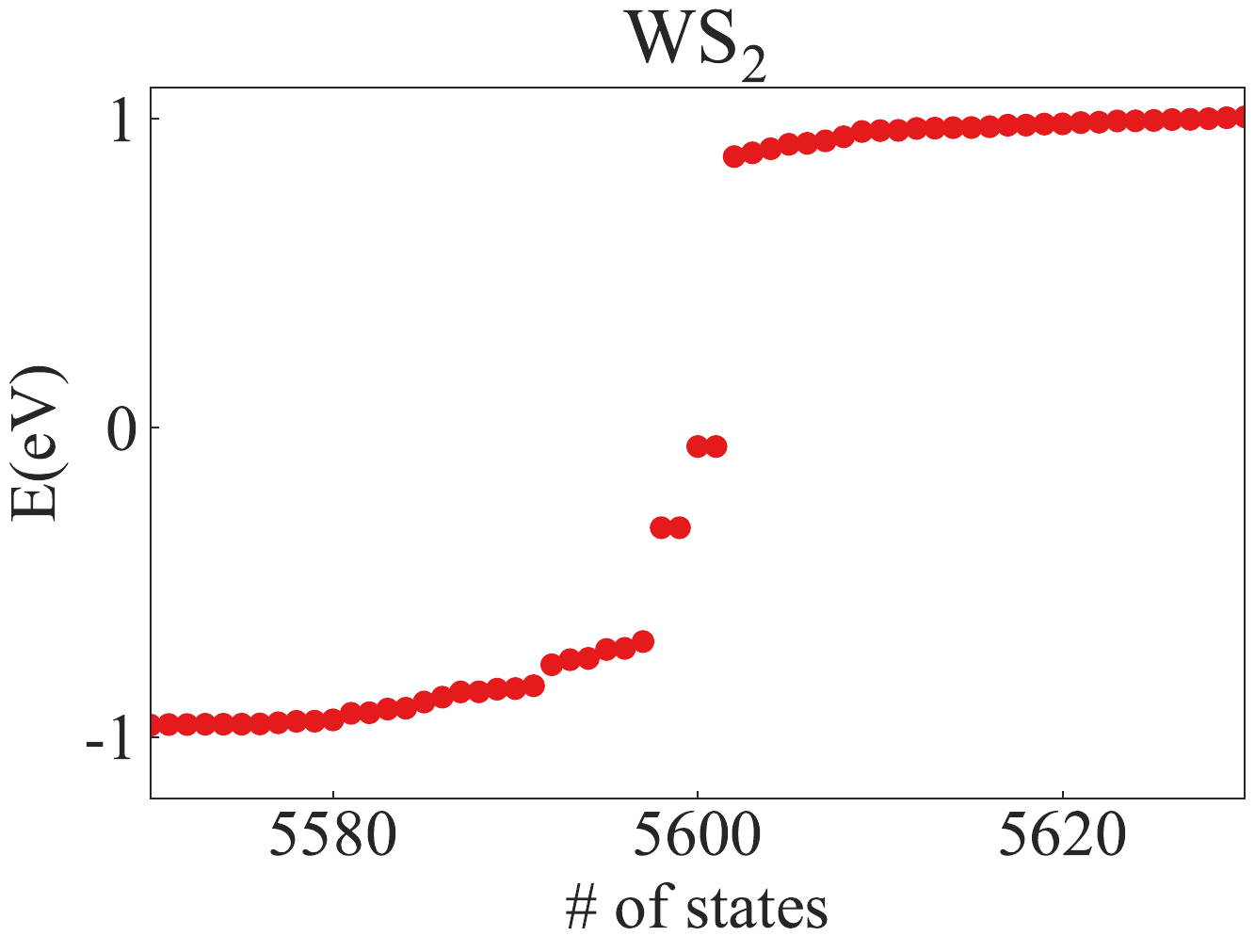}
\label{fig:}}
\subfigure[]{
\includegraphics[scale=0.25]{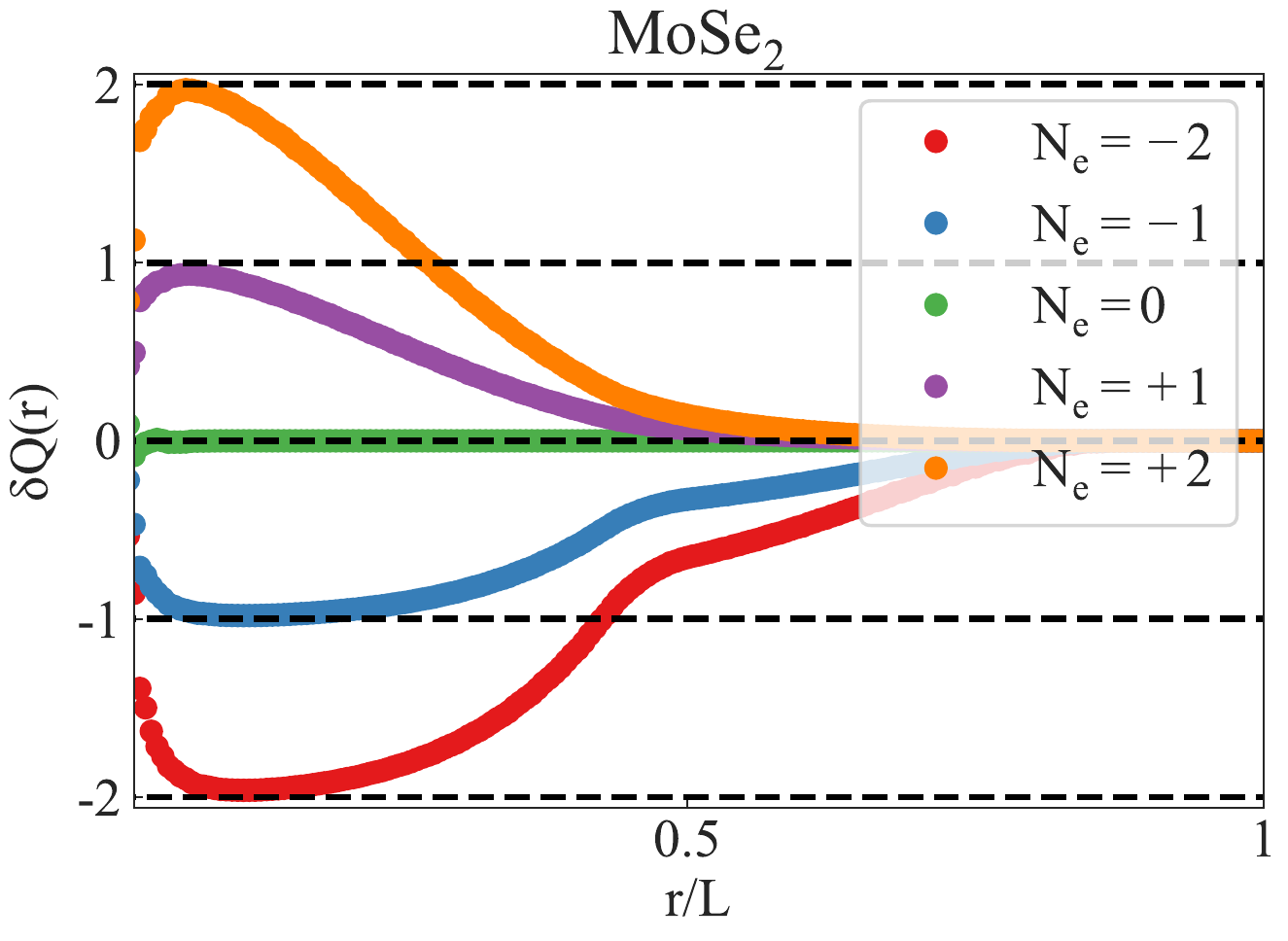}
\label{fig:}}
\subfigure[]{
\includegraphics[scale=0.25]{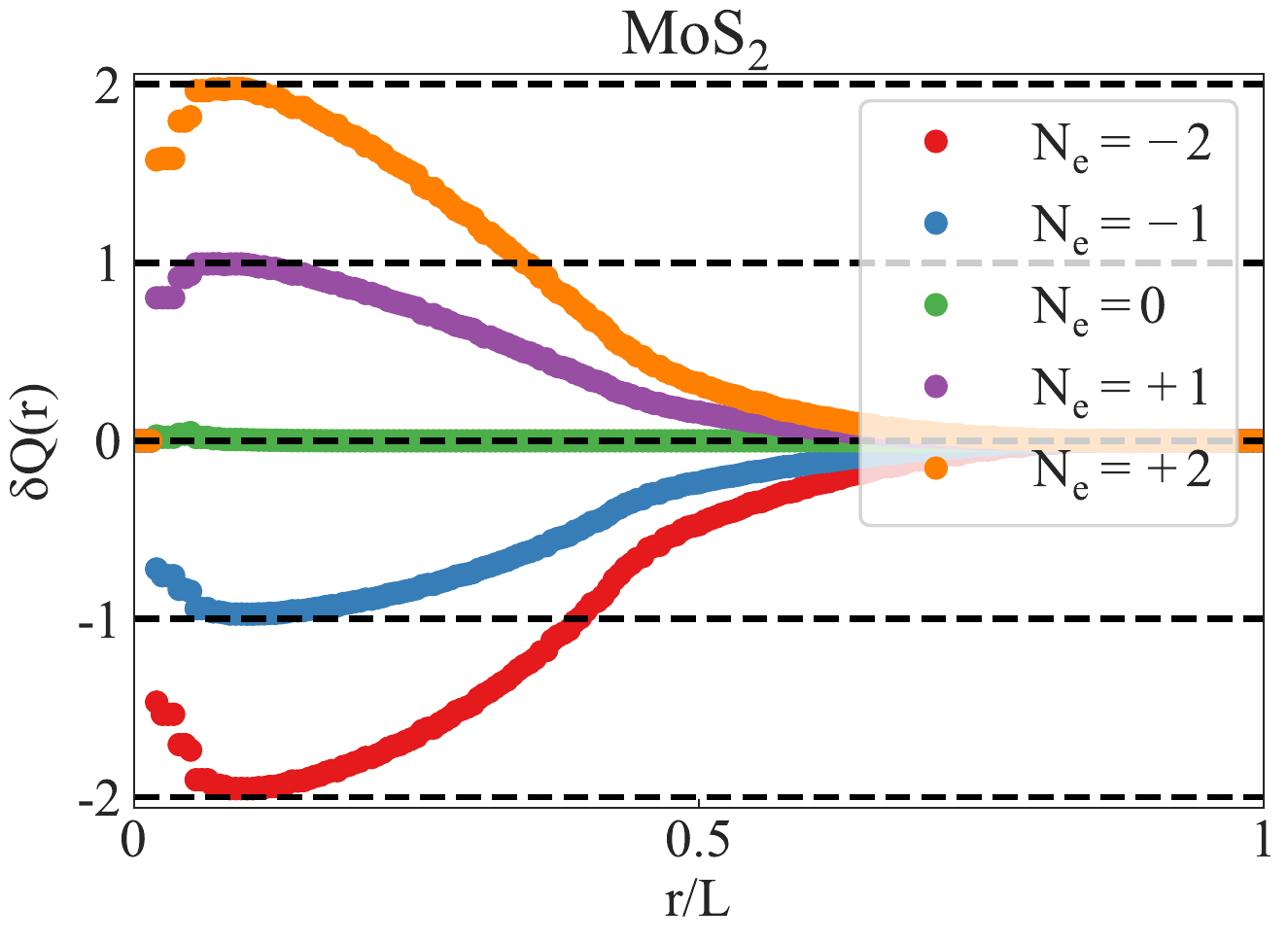}
\label{fig:}}
\subfigure[]{
\includegraphics[scale=0.25]{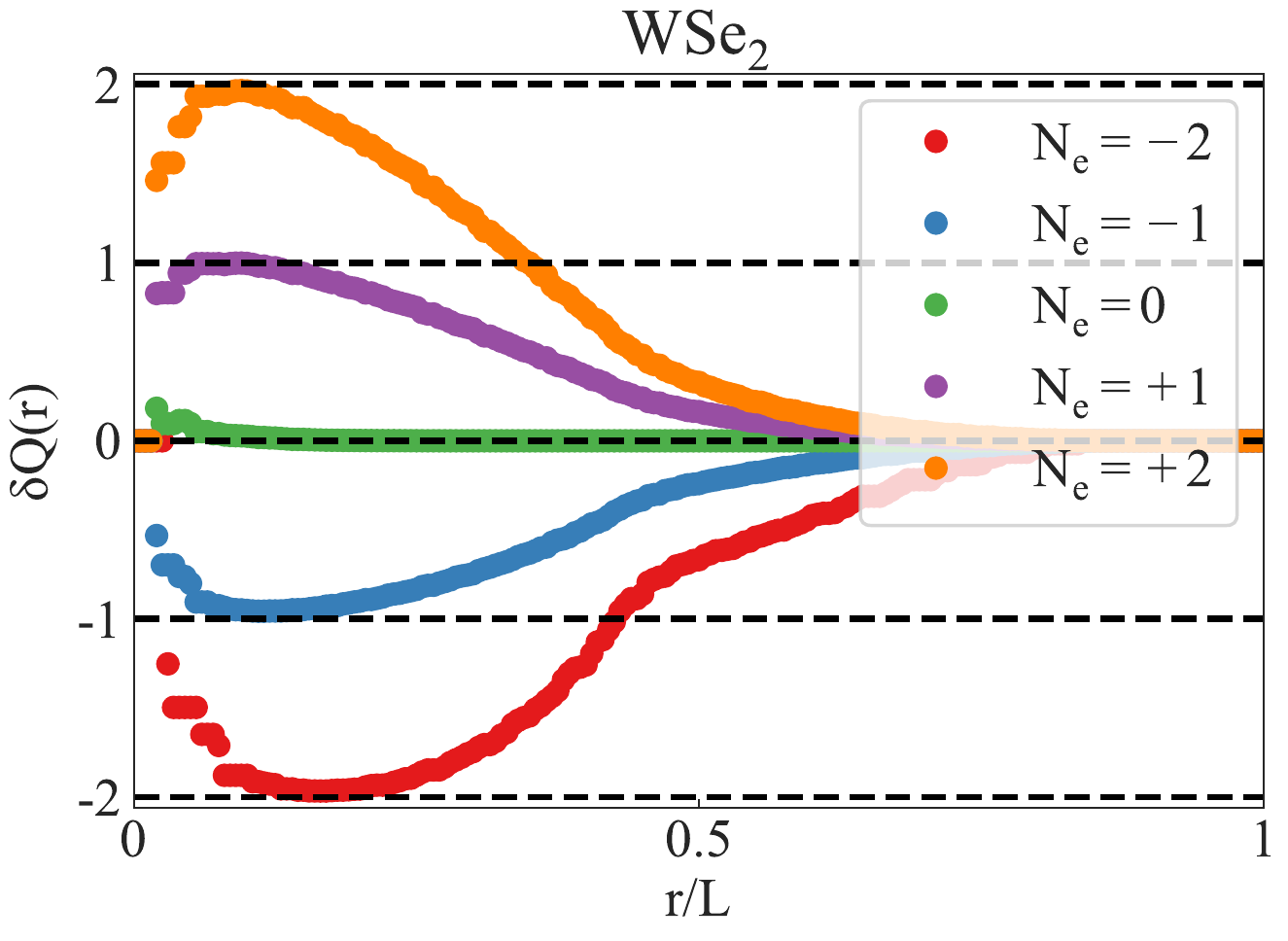}
\label{fig:}}
\subfigure[]{
\includegraphics[scale=0.25]{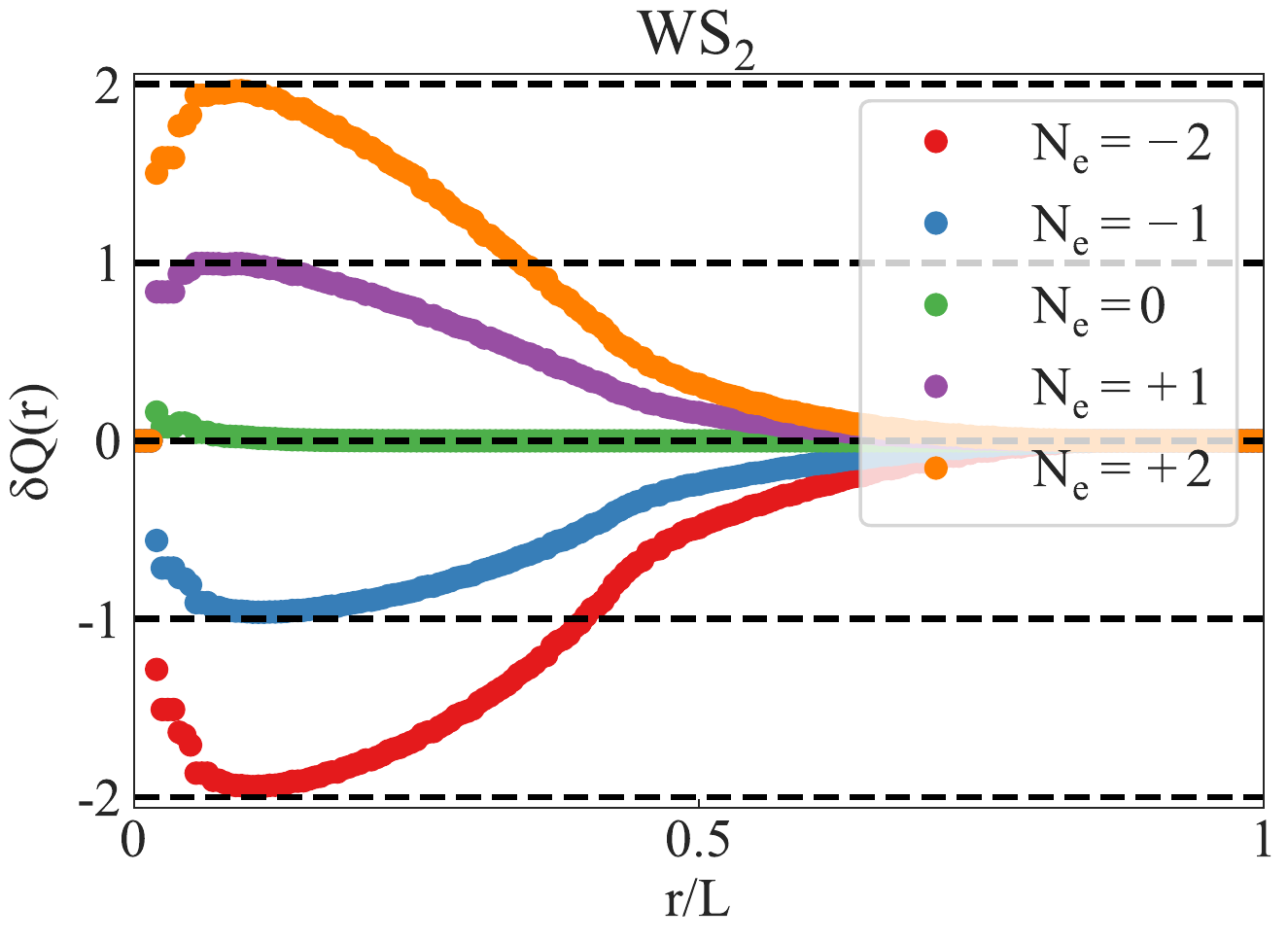}
\label{fig:}}
\caption{(a)-(d) Local density of states on vortex for 1H-MX$_{2}$ TMDs admitting ground state spin Chern number, $|\mathcal{C}_{s,G}|=2$, as a function of flux strength. (e)-(h) State vs. energy  upon insertion of $\pi$-flux vortex for a system of 20 $\times$ 20 unit cells. The results display clearly a gap between pairs of mid-gap vortex-bound modes. (i)-(l) Induced charge on the vortex as a function of doping away from half-filling of vortex bound modes (VBMs) by $N_{e}$ states. Results display robust nature of spin-charge separation and relation, $\delta Q= |\mathcal{C}_{s,G}|\times e$.}
\label{fig:VortexMX2}
\end{figure*}

\begin{figure*}
\centering
\subfigure[]{
\includegraphics[scale=0.3]{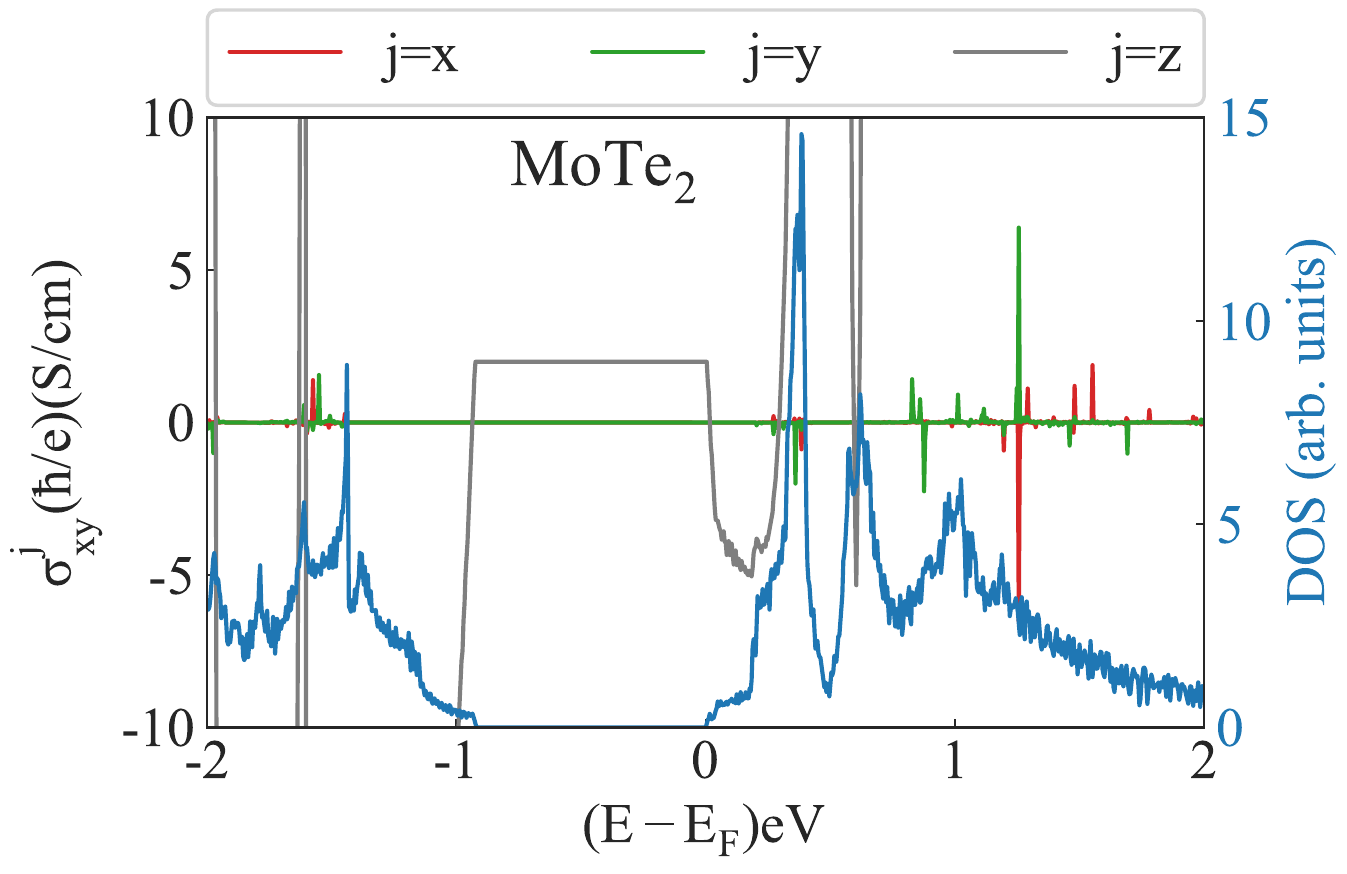}
\label{fig:}}
\subfigure[]{
\includegraphics[scale=0.3]{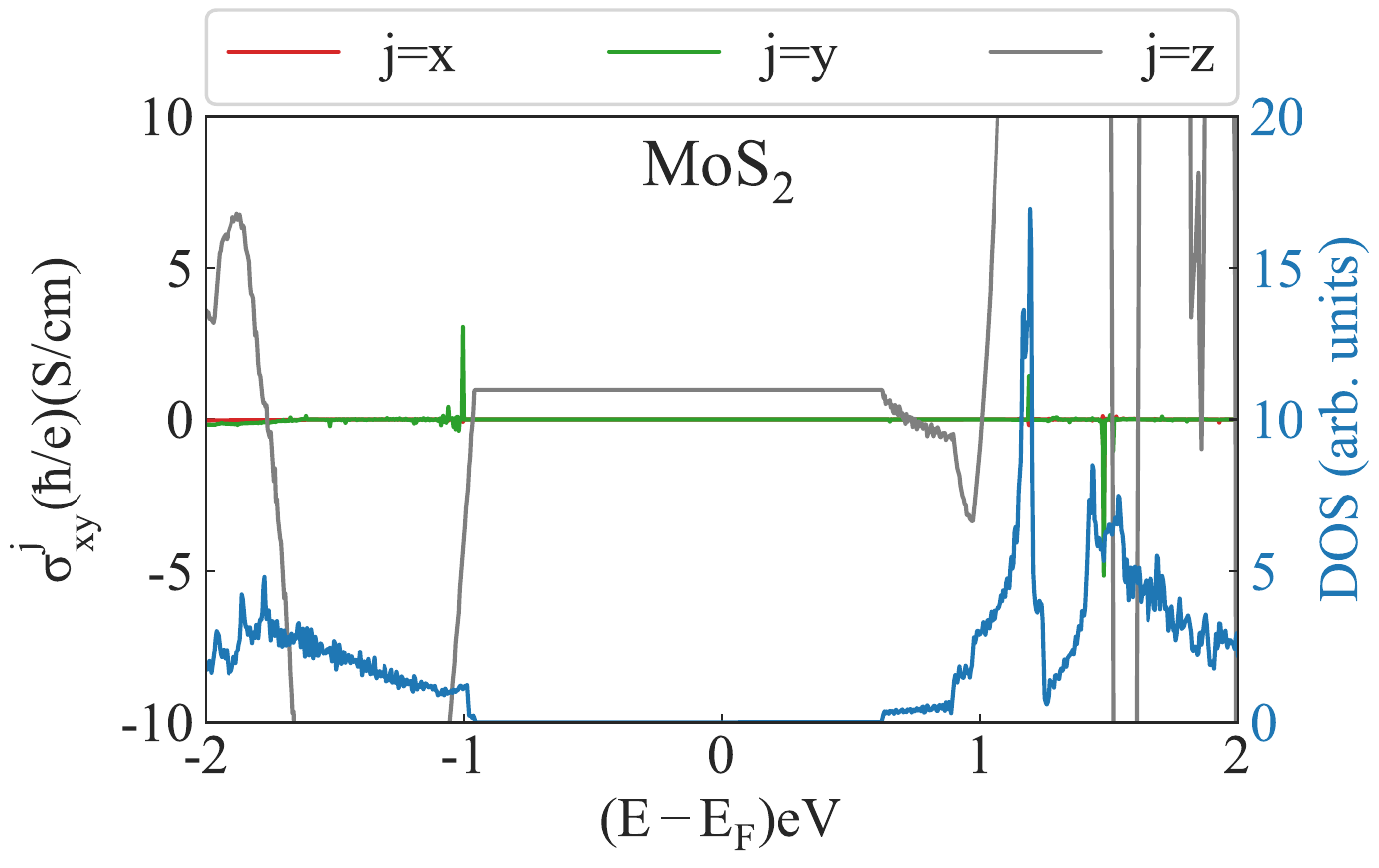}
\label{fig:}}
\subfigure[]{
\includegraphics[scale=0.3]{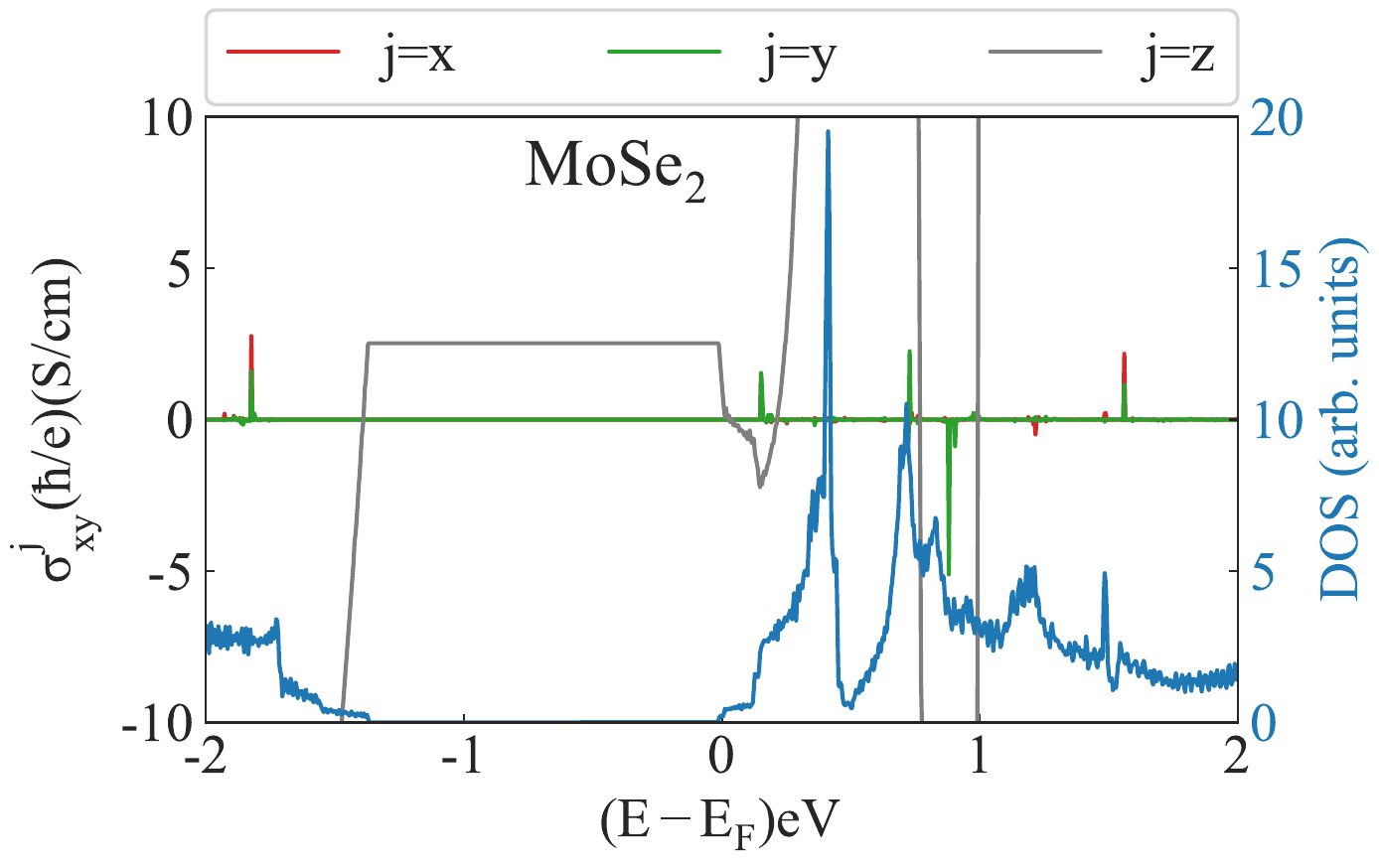}
\label{fig:}}
\subfigure[]{
\includegraphics[scale=0.3]{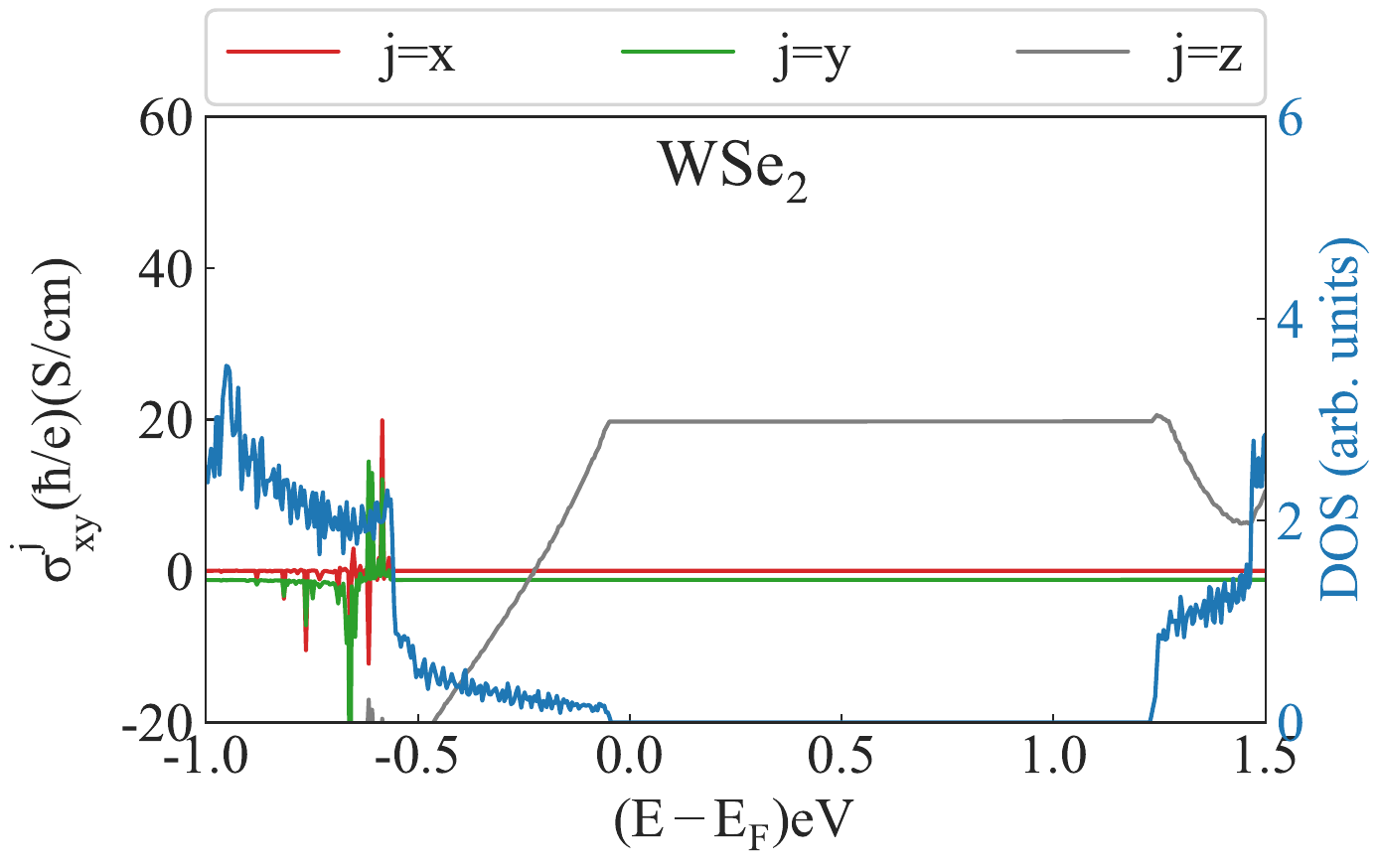}
\label{fig:}}
\subfigure[]{
\includegraphics[scale=0.3]{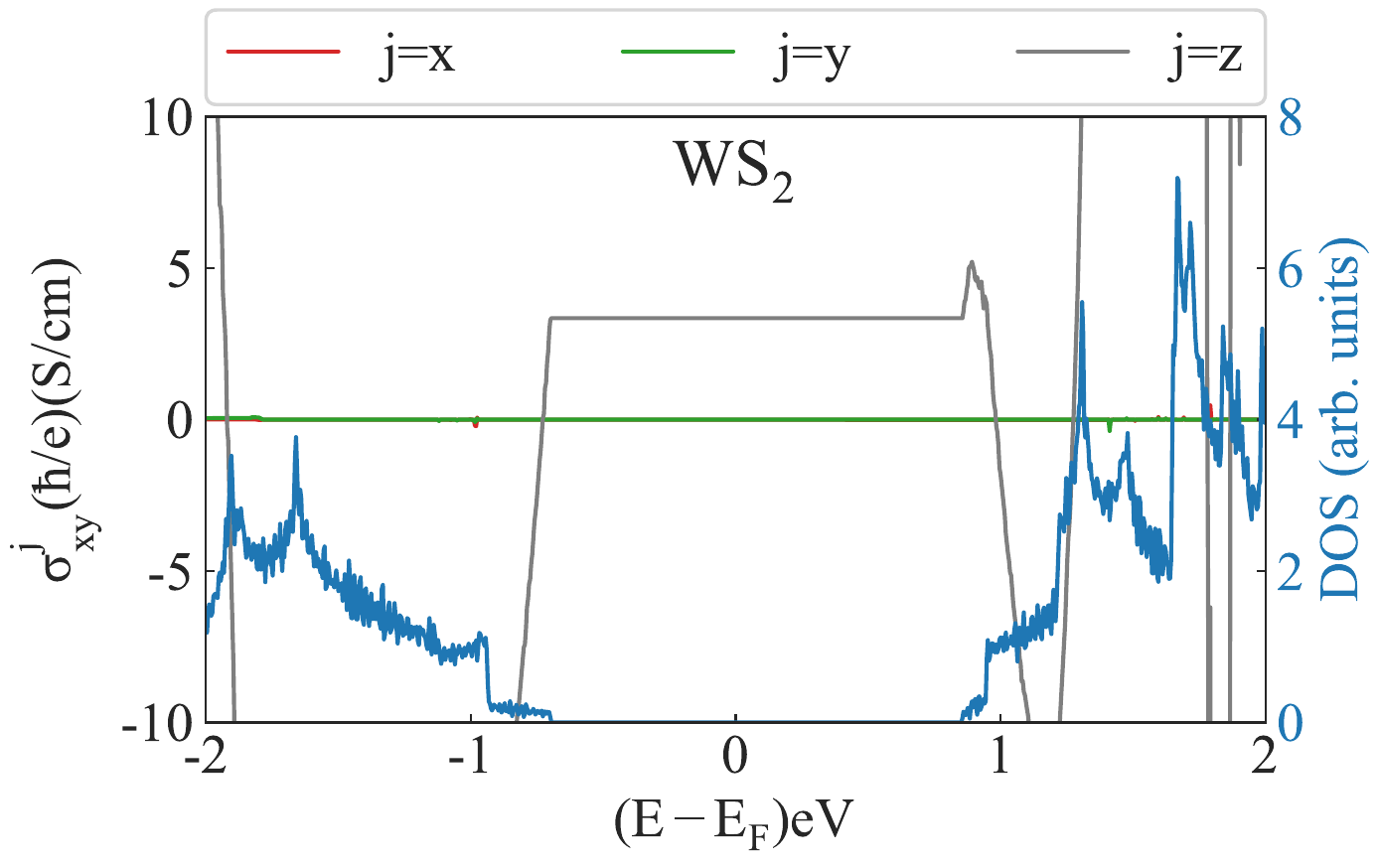}
\label{fig:}}
\subfigure[]{
\includegraphics[scale=0.3]{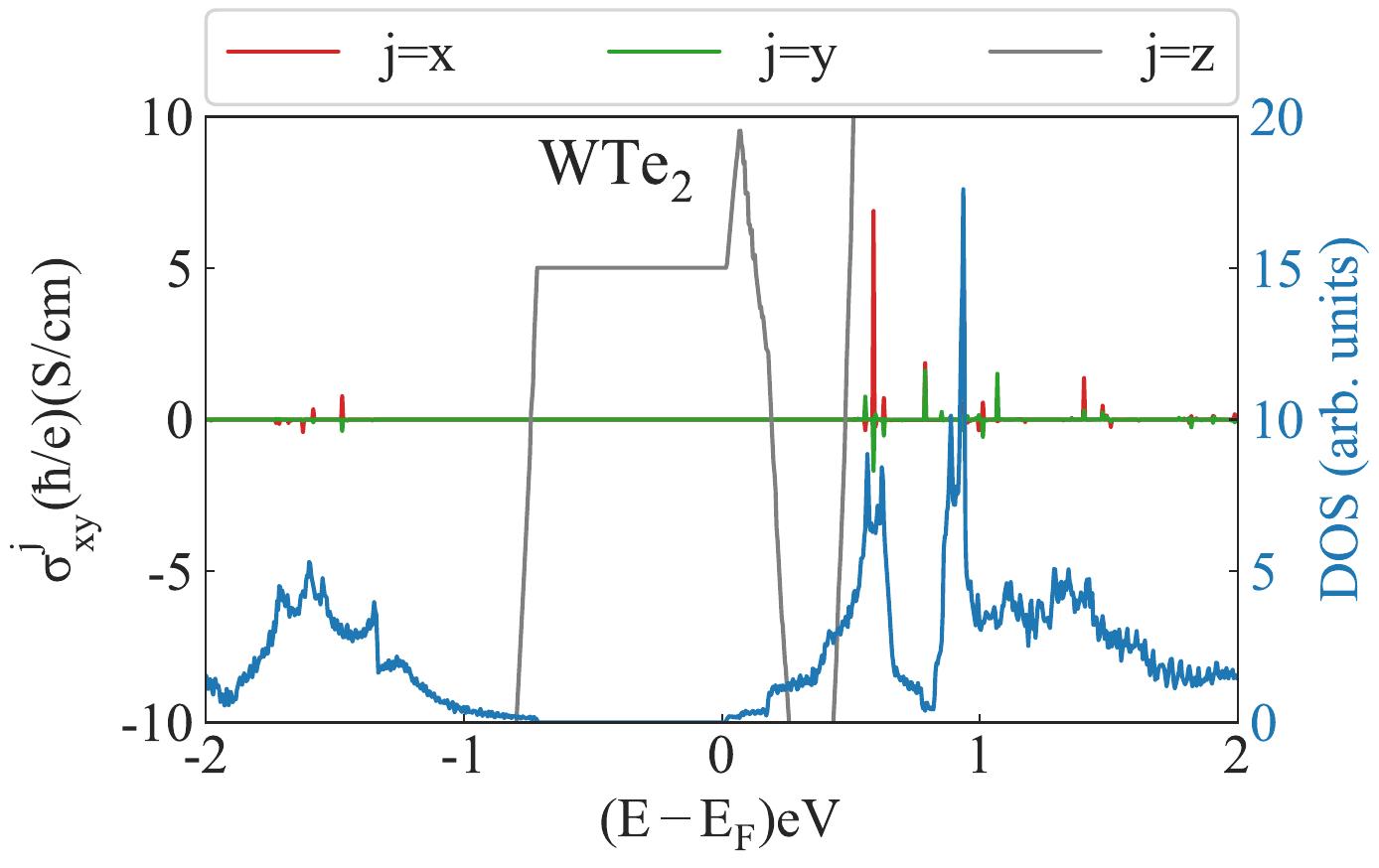}
\label{fig:}}
\subfigure[]{
\includegraphics[scale=0.3]{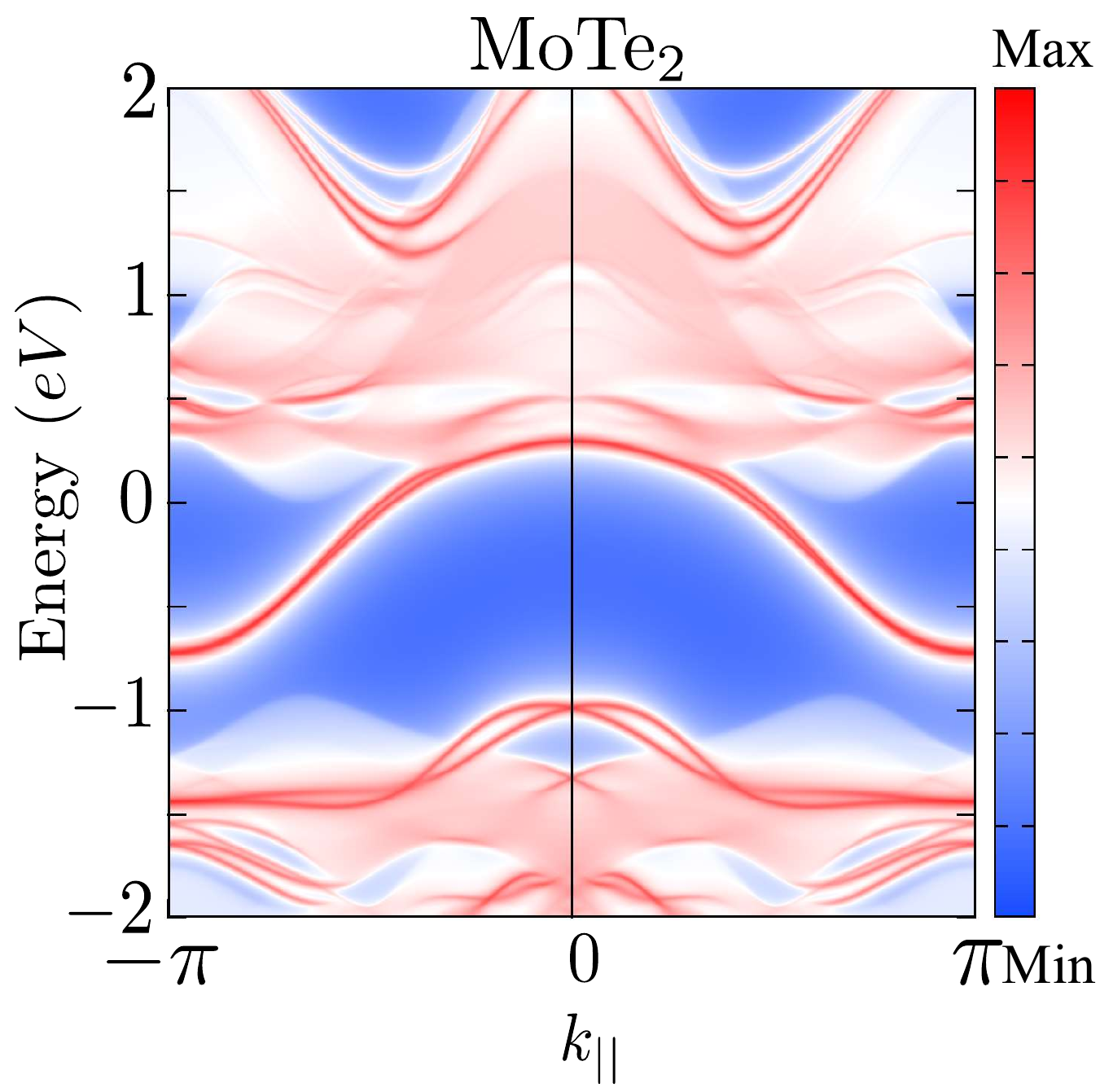}
\label{fig:}}
\subfigure[]{
\includegraphics[scale=0.3]{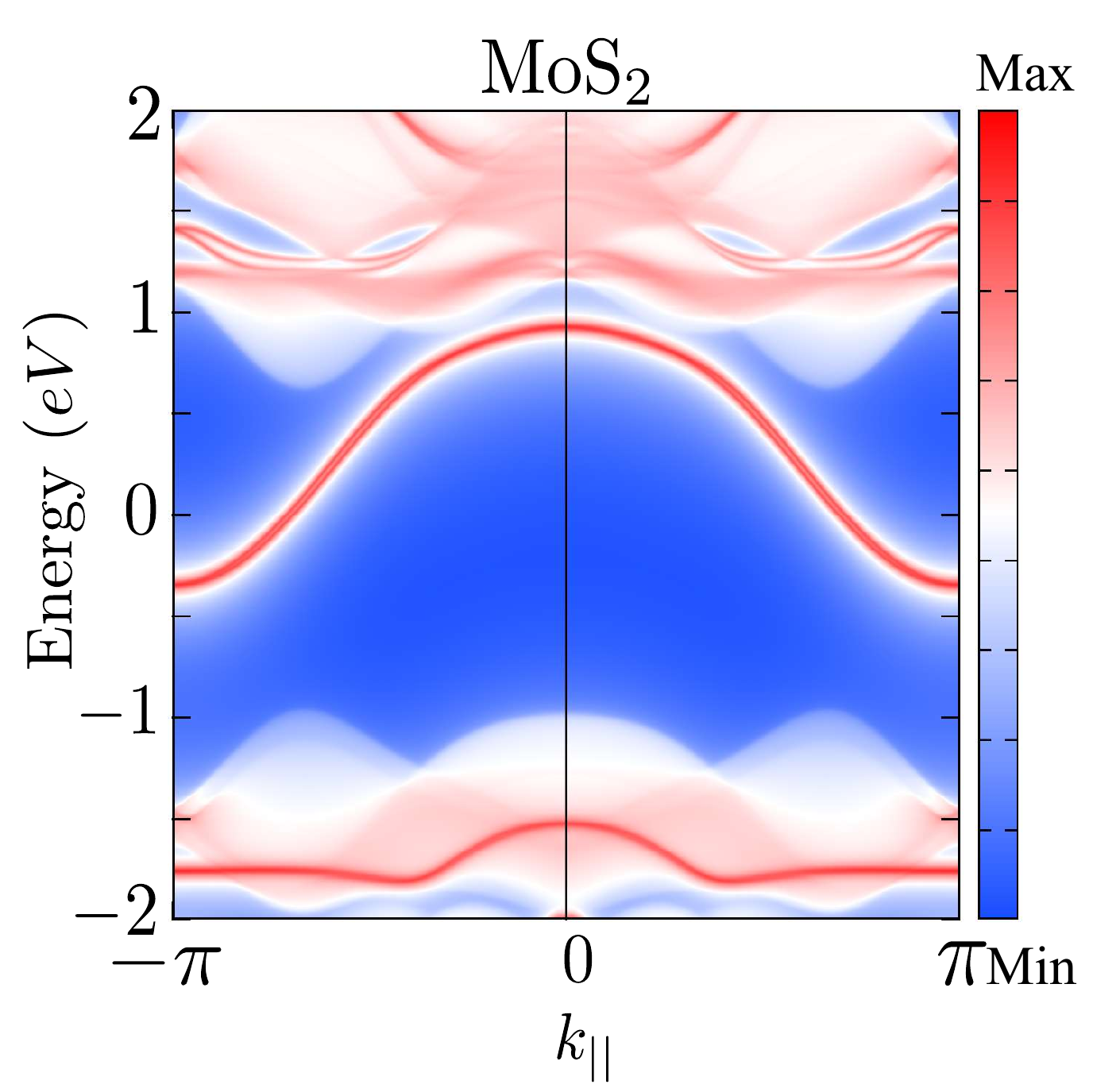}
\label{fig:}}
\subfigure[]{
\includegraphics[scale=0.3]{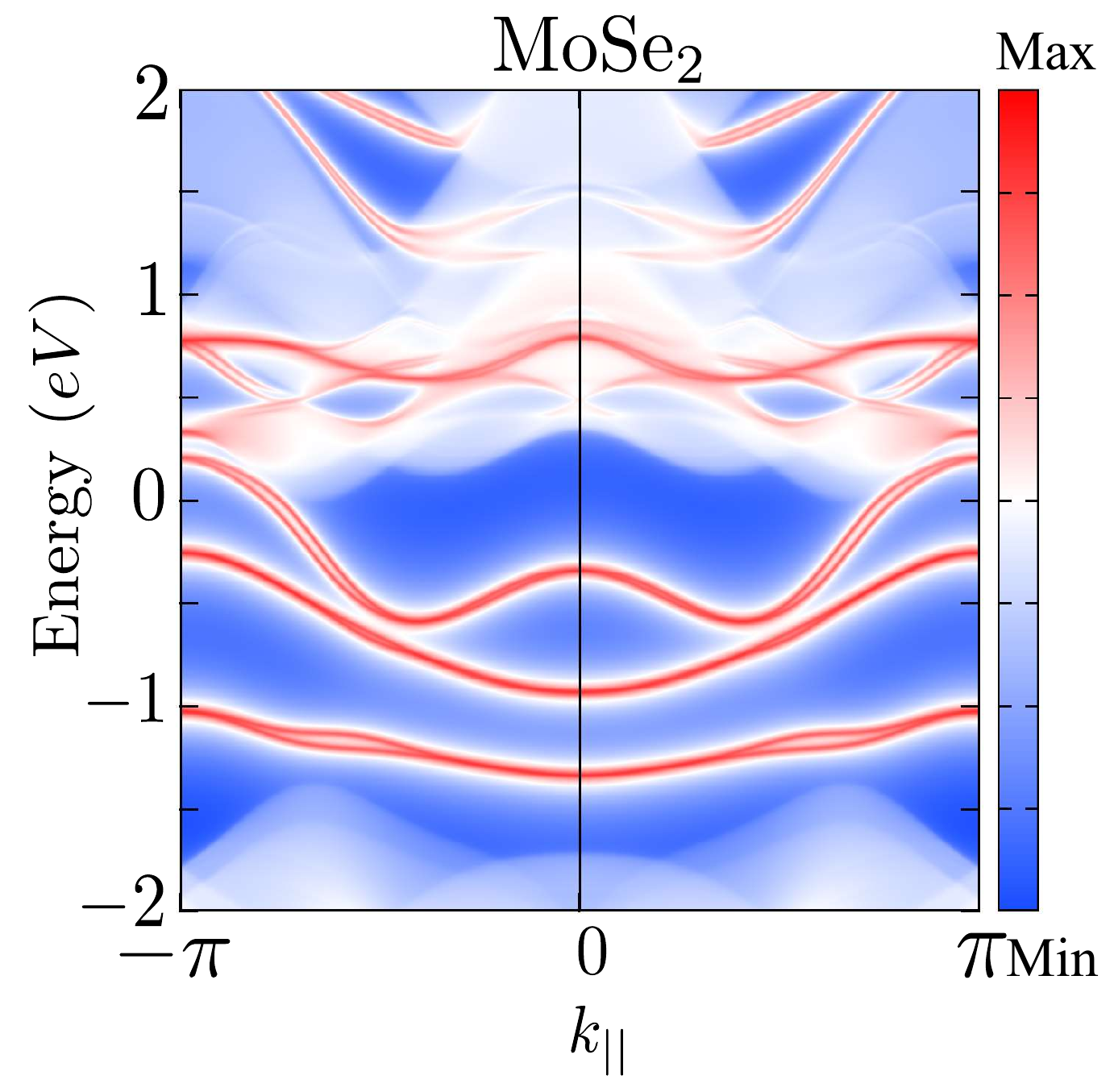}
\label{fig:}}
\subfigure[]{
\includegraphics[scale=0.3]{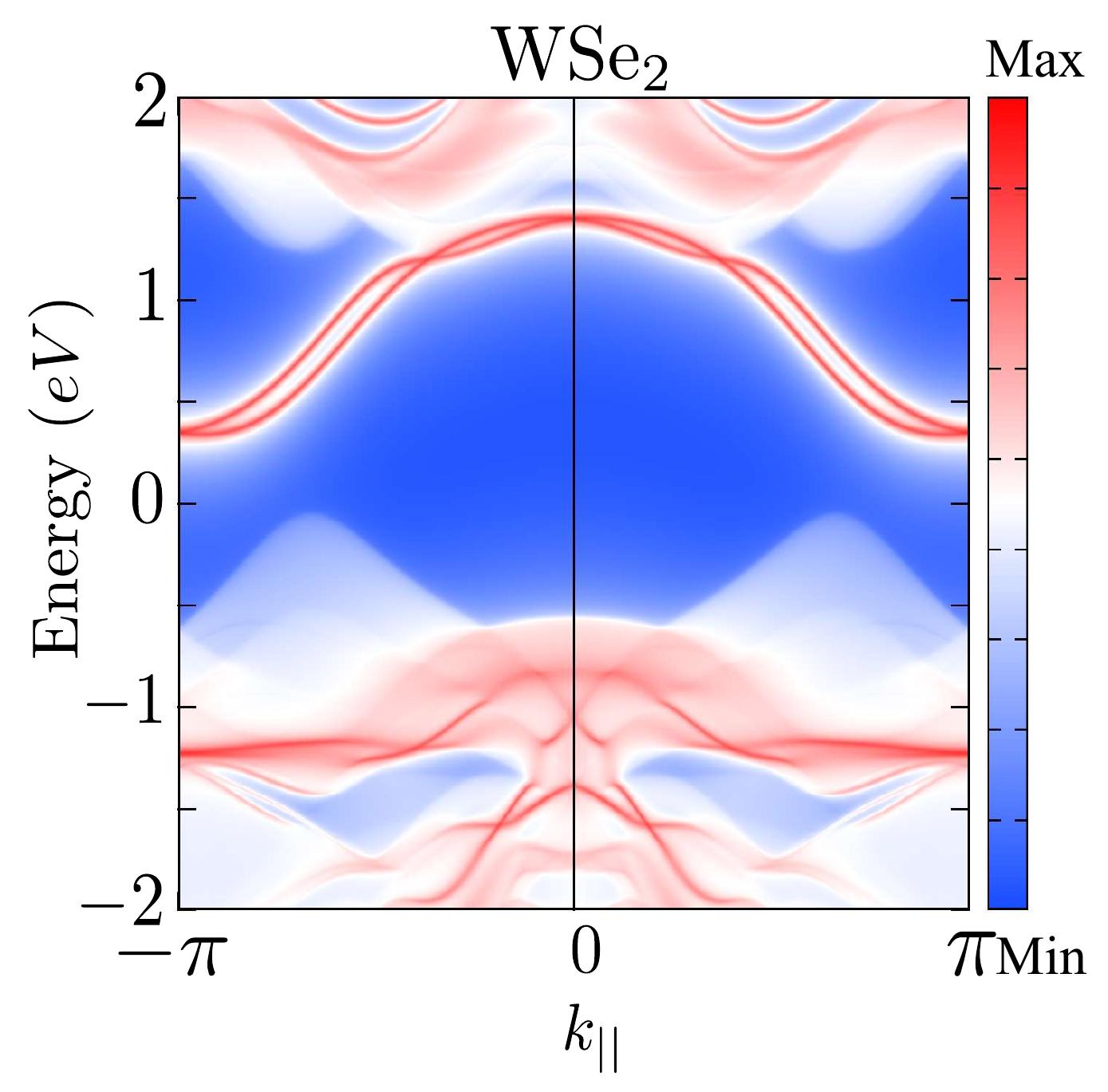}
\label{fig:}}
\subfigure[]{
\includegraphics[scale=0.3]{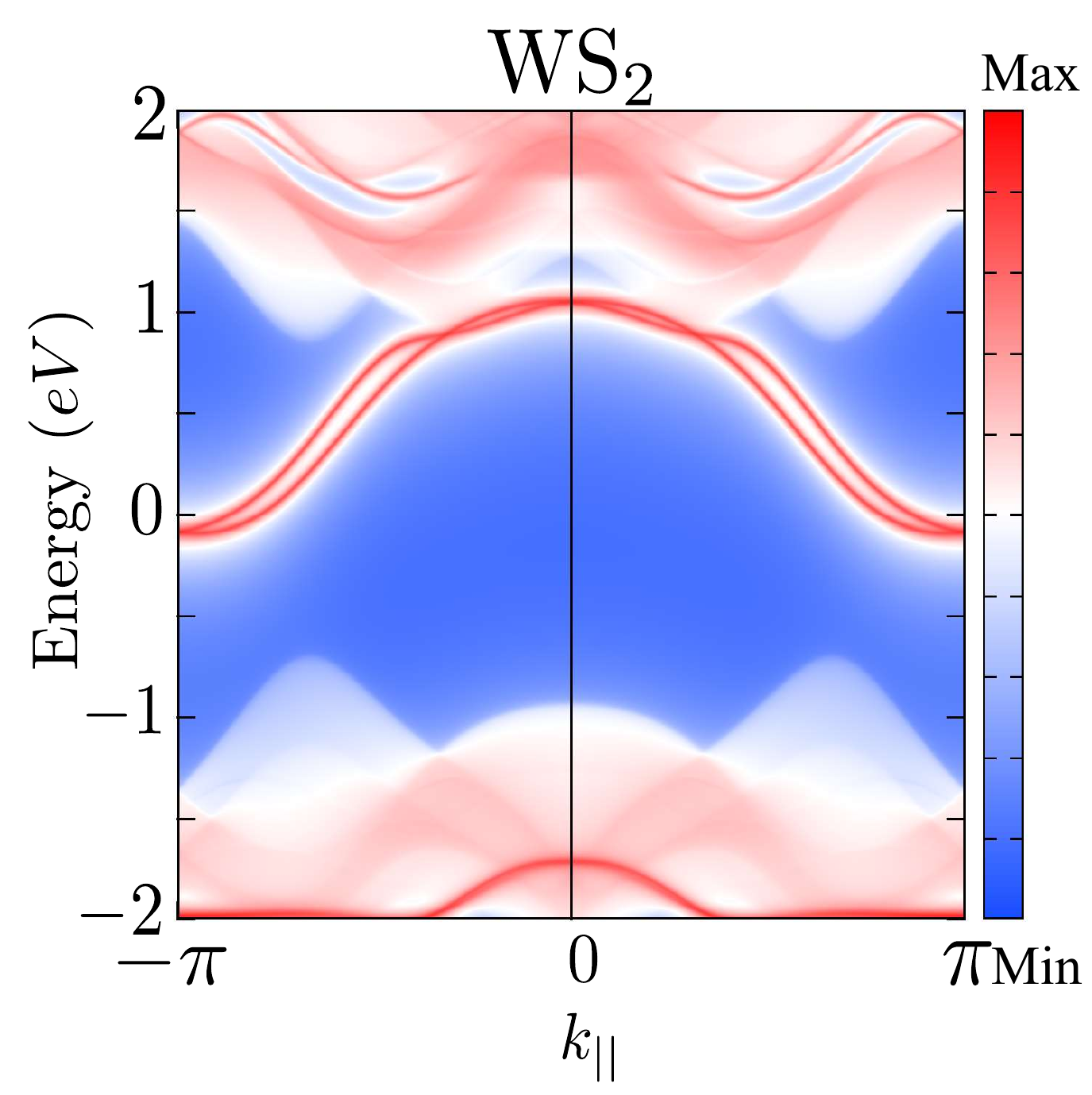}
\label{fig:}}
\subfigure[]{
\includegraphics[scale=0.3]{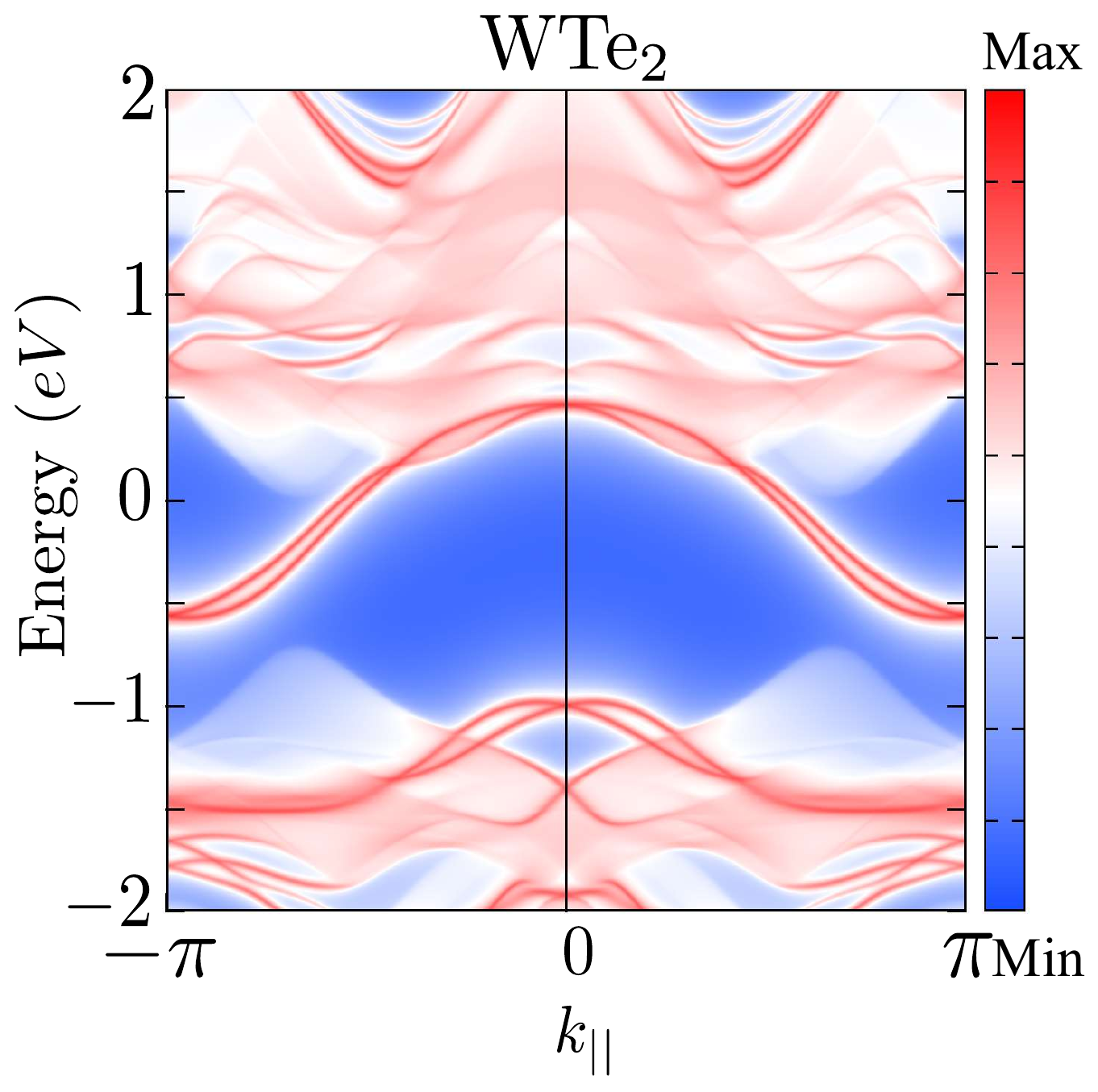}
\label{fig:}}
\caption{(a)-(d) Spin-hall conductivity for 1H-MX$_{2}$ TMDs admitting ground state spin Chern number, $|\mathcal{C}_{s,G}|=2$. In the absence of spin-conservation symmetry the spin-Hall conductivity is not quantized along a given principle axis. However, the non-zero result within the bulk gap is indicative of non-trivial bulk-topology. (e)-(h) Spectral density on zig-zag surface of 1H-MX$_{2}$ TMDs. While surface bound mid-gap states exist, indicating the existence of non-trivial bulk-topology, the spectra remains gapped, disallowing definitive assignment of the bulk invariant.}
\label{fig:MX2SurfaceSpinHall}
\end{figure*}

\par
In Tyner et. al\cite{tynerbismuthene}, it was demonstrated that in a two-dimensional insulator supporting a ground state spin Chern number, $\mathcal{C}_{s,G}=+2$, in the presence of Kramers-degeneracy ($\mathcal{PT}$-symmetry) and particle-hole symmetry, a magnetic flux tube serves as a spin-pump. At $\pi$-flux there exists four, zero-energy VBMs creating a ground-state degeneracy. Spin-charge separation was then demonstrated through direct calculation of induced charge at half-filling, and upon doping $N_{e}$ electrons where $N_{e}\in [-2,2]$. The results clearly show that the vortex acquired an induced charge given by $\delta Q=N_{e} \times e$, proving the insulator is a platform for spin-charge separation. 
\par
In many $\mathcal{T}$-invariant real materials, $\mathcal{P}$, as well as particle-hole symmetry, is not conserved. This is true in the 1H-MX$_{2}$ family of TMDs, where $\mathcal{P}$-symmetry is absent. In this case, the $2N$ VBMs for a two-dimensional insulator supporting $\mathcal{C}_{s,G}=|N|$ need not be degenerate at $\pi$-flux, creating a gap in the spectral-flow. To exemplify this situation, we employ a four-band tight-binding model on a square lattice, the Bloch Hamiltonian takes the form: 
\begin{multline}\label{eq:FuModel}
    H(\mathbf{k})=2t_{2}(\sin k_{x} \sin k_{y})\sigma_{0}\otimes \tau_{1}\\+t_{1}(\cos k_{x} - \cos k_{y})\sigma_{3}\otimes\tau_{2}\\+(t'_{1}+2t'_{2}(\cos k_{x}+ \cos k_{y})-2t)\sigma_{0}\otimes\tau_{3}\\+0.2t_{3}(\cos 2k_{x}-\cos 2k_{y})\sigma_{2}\otimes \tau_{2}+\\ 0.5t_{4}\left(\sin k_{x}\cos k_{y} +\sin k_{y}\cos k_{x} \right)\sigma_{1} \otimes \tau_{1}\\,
\end{multline}
 where $\sigma_{0,1,2,3}(\tau_{0,1,2,3})$ are the $2\times 2$ identity matrix and three Pauli matrices respectively, operating on the spin (orbital) indices. 
\par
We select the hopping parameters, $t_{1}=t_{3}=t_{4}=1, \; t_{2}=0.5, \; t'_{1}=2.5,\; t'_{2}=0.5$ and $t=1$, with the band structure shown in Fig. \eqref{fig:FuBands}. This model preserves $C_{4}$ rotational and time-reversal symmetry while breaking inversion symmetry. The time-reversal symmetry is explicitly implemented as, $\mathcal{T}=i(\sigma_{2}\otimes\tau_{0})\mathcal{K}$, where $\mathcal{K}$ indicates complex conjugation. Furthermore, both the WCC spectra and surface state spectra are gapped as seen in Fig. \eqref{fig:FuWCC} and Fig. \eqref{fig:FuSlab} respectively. 
\par 
We first establish the ground-state spin-Chern number in momentum space by performing an in-plane Wilson loop for occupied valence bands along the contour ABCD\cite{tyner2021quantized}. The area of the contour over which the Wilson loop is calculated is systematically increased from zero to be equivalent with the area of the first Brillouin zone. This is shown schematically in Fig. \eqref{fig:FuPWLSchem}. The in-plane Wilson loop is calculated as,
\begin{equation}
    W_{n,j}=\mathcal{P}\text{exp}\left[i\oint A_{j,n}(\mathbf{k})dk_{j}\right].
\end{equation}
Upon integration, we find $W_{n,j}=\text{exp}\left(i\theta_{n}\hat{n}\cdot \mathbf{\sigma}\right)$, where $\theta_{n}$ is the non-Abelian flux. Wannier center charges (WCCs), $\bar{j}$, follow as eigenvalues of $\text{Im(Ln(}W_{n,j}))$, therefore Wannier center charges are equivalent to $\pm |\theta|\text{mod}\pi$. By plotting the winding of WCCs we can identify the flux enclosed by the contour as a function of the area enclosed. The results, shown in Fig. \eqref{fig:FuPWLWCC}, demonstrate that the ground-state spin-Chern number, $\mathcal{C}_{s,G}=|2|$. 

\par 
We could have also computed the ground-state spin Chern number following the procedure outlined by Prodan\cite{Prodan2009}. This procedure requires defining the projected spin operator (PSO), $P(\mathbf{k})\hat{s}P(\mathbf{k})$, where $P(\mathbf{k})$ is the projector onto occupied bands and $\hat{s}$ is a chosen spin-quantization axis. In the absence of spin-orbit coupling the eigenvalues of the PSO are fixed as $\pm 1$, however since we have introduced spin-orbit coupling, the eigenvalues are no longer pinned at $\pm 1$. Nevertheless, a gap in the eigenvalue spectra remains for our given model when selecting $\hat{s}=s_{z}=\sigma_{3}\otimes \tau_{0}$, allowing for calculation of the spin-Chern number via spin-resolved Wilson loop as detailed in Lin et. al\cite{Lin2022Spin}. The results shown in Fig. \eqref{fig:FuPSP}, demonstrate the WCC spectra when performing the spin-resolved Wilson loop along the $\hat{x}$ axis as a function of transverse momenta $k_{y}$ for the band corresponding to negative eigenvalues of the PSO, $\theta^{-}_{x}$. The conclusion is in alignment with the earlier determination that the ground states supports $|\mathcal{C}_{s,G}|=2$. We emphasize though, that this method is extremely challenging to implement in DFT derived models where detailed information regarding our basis is not always easily accessible, particularly in an automated workflow. 
\par 
Having established the magnitude of the bulk invariant, we turn to insertion of a magnetic flux tube. The local density of states on the flux tube inserted at the origin as a function of flux strength in a system of size 60 $\times$ 60 unit cells is given in Fig. \eqref{fig:FuStateEnergy}. While four VBMs are visible within the mid-gap, they are not four-fold degenerate. As a result, we do not observe gapless spectral-flow connecting valence and conduction states upon tuning the strength of the vortex. 
\par
In order to demonstrate that in the absence of a ground-state degeneracy at $\pi$-flux, spin-charge separation remains unaffected, we directly calculate the induced charge as a function of filling following the procedure described in the main body. The results are visible in Fig. \eqref{fig:FuInduced}. We observe induced charge as a function of filling follows the relation $\delta Q = N_{e}\times e$ with $N_{e} \in [-2, 2]$. This conclusively illustrates that the non-degenerate VBMs originating from conduction/valence states are correlated and not acting as independent end states to spinful Su-Schieffer-Heeger chains. Moreover the results are identical to those presented by Tyner et. al\cite{tynerbismuthene} in the presence of gapless spectral-flow, allowing for the magnitude of the ground-state spin-Chern number to be determined unambiguously. 
\par
For completeness, we investigate the presence of induced spin on the inserted vortex at $\phi=\phi_{0}/2$ in a $20 \times 20$ lattice in Fig. \eqref{fig:InducedSpin}. As time-reversal symmetry is restored at this point, all states exist in Kramers pairs. This leaves an open question as to whether the two VBMs on one side of the Fermi energy are Kramers partners or whether the Kramers partner for a state below the Fermi energy exists above the Fermi energy. In order for the system to support spin-charge separation, VBM Kramers partners must exist on opposite sides of the Fermi energy. If both VBMs below the Fermi energy were to be Kramers partners, their spins would be opposite, leading to a vanishing induced spin at half-filling. By contrast, if at half-filling only a single Kramers partner is occupied, there will exist a finite induced spin on the vortex. To probe this, we calculate the real-space expectation value of $\hat{s}=s_{z}$ for the four VBMs at $\phi=\phi_{0}/2$. The results for each of the four states, in order of increasing energy are visible in Fig. \eqref{fig:state1sz}-\eqref{fig:stat4sz} respectively. We note that the two states lying below the Fermi-energy, \eqref{fig:state1sz} and \eqref{fig:state2sz}, support a positive, non-zero expectation value at the vortex core while the two states above the Fermi energy, \eqref{fig:state3sz} and \eqref{fig:stat4sz}, support negative non-zero expectation values at the vortex-core. We then compute the induced spin density on the vortex core at half-filling, adapting the procedure used for computation of induced charge, with the results shown in Fig. \eqref{fig:halffillsz}. This result is contrasted with induced charge density at half-filling, shown in Fig. \eqref{fig:halffillcharge}. 
\par 
Upon doping by two electrons, such that the vortex acquires a quantized induced spin as shown in Fig. \eqref{fig:FuInduced}, we find the induced charge and spin densities shown in Fig. \eqref{fig:2dopedcharge} and \eqref{fig:2dopedsz} respectively, demonstrating a vanishing induced spin. These images definitively show that, at half-filling the vortex acquires a finite induced spin and vanishing induced charge, while upon doping by $N_{e}=\mathcal{C}_{s,G}$ electrons, the vortex supports quantized induced charge and vanishing induced spin, verifying it as a platform for spin-charge separation. 

\begin{figure}[H]
    \centering
    \includegraphics[width=8cm]{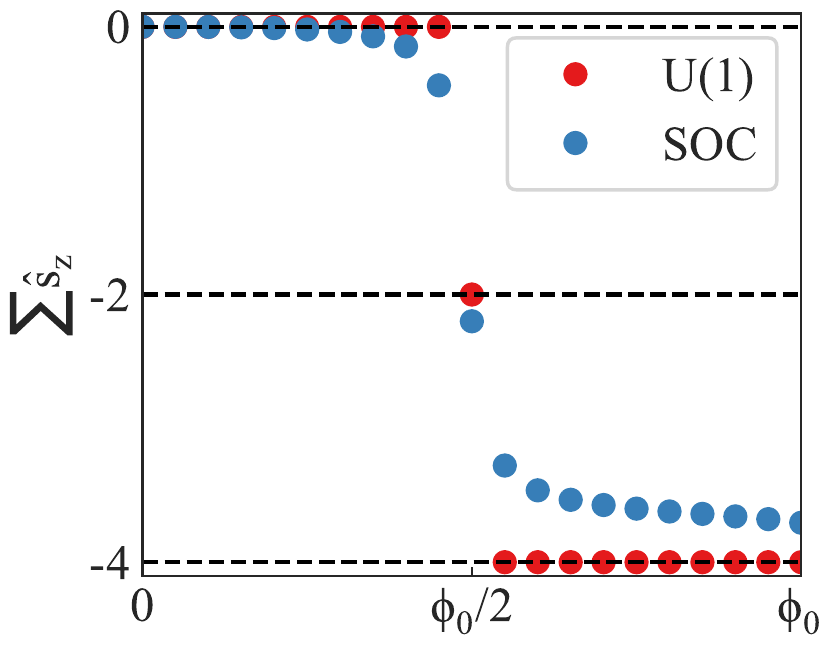}
    \caption{Spin-expectation value for half-filled system of $20\times 20$ lattice sites as a function of flux strength for inserted magnetic vortex in the presence of $U(1)$ spin-conservation symmetry (denoted $U(1)$) and generic spin-orbit-coupling (denoted SOC), fixing $t_{3,4}=0$ and $t_{3,4}=1$ respectively in eq. \eqref{eq:FuModel}. The results demonstrate that the flux tube continues to serve as a spin-pump in the absence of $U(1)$ symmetry.}
    \label{fig:SpinPump}
\end{figure}

\par
Finally, we address the role of the vortex as a spin-pump. To do so we compute a sum of the real-space expectation value, $\sum  \hat{s}_{z}=\sum_{i,j}\bra{\psi_{i,j}}s_{z}\ket{\psi_{i,j}}$, where $i$ is a sum over occupied states and j over all lattice sites, explicitly considering a $20 \times 20$ system. We begin by fixing, $t_{3,4}=0$, introducing $U(1)$ spin-conservation. The results in Fig. \eqref{fig:SpinPump} demonstrate that the half-filled spin expectation value transitions from zero to $-4$ as the flux strength is tuned past $\phi_{0}/2$ and spin is pumped away from the ground state; confirming the role of the vortex as a spin pump. The magnitude of spin being pumped is also in accordance with $2\times \mathcal{C}_{s,G}$ as expected. 
\par 
Restoring $t_{3,4}=1$, introducing spin-orbit coupling and breaking the spectral flow of VBMs, we continue to observe spin-pumping, as seen in Fig. \eqref{fig:SpinPump}. The spin expectation for the half-filled system in the presence of spin-orbit coupling no longer transitions between zero and the quantized value of -4. Spin-orbit coupling causes spin-quantization to become redistributed among the principal axes. Nevertheless, spin-pumping is conclusively observed proving the system is a generalized quantum spin-Hall insulator and that spectral flow is unnecessary for this classification.

\section*{Appendix B: Further details of 1H-MX$_{2}$ TMDs}\label{app:B}

\begin{figure*}
\centering
\subfigure[]{
\includegraphics[scale=0.3]{Figures/MoTe2bands.pdf}
\label{fig:}}
\subfigure[]{
\includegraphics[scale=0.3]{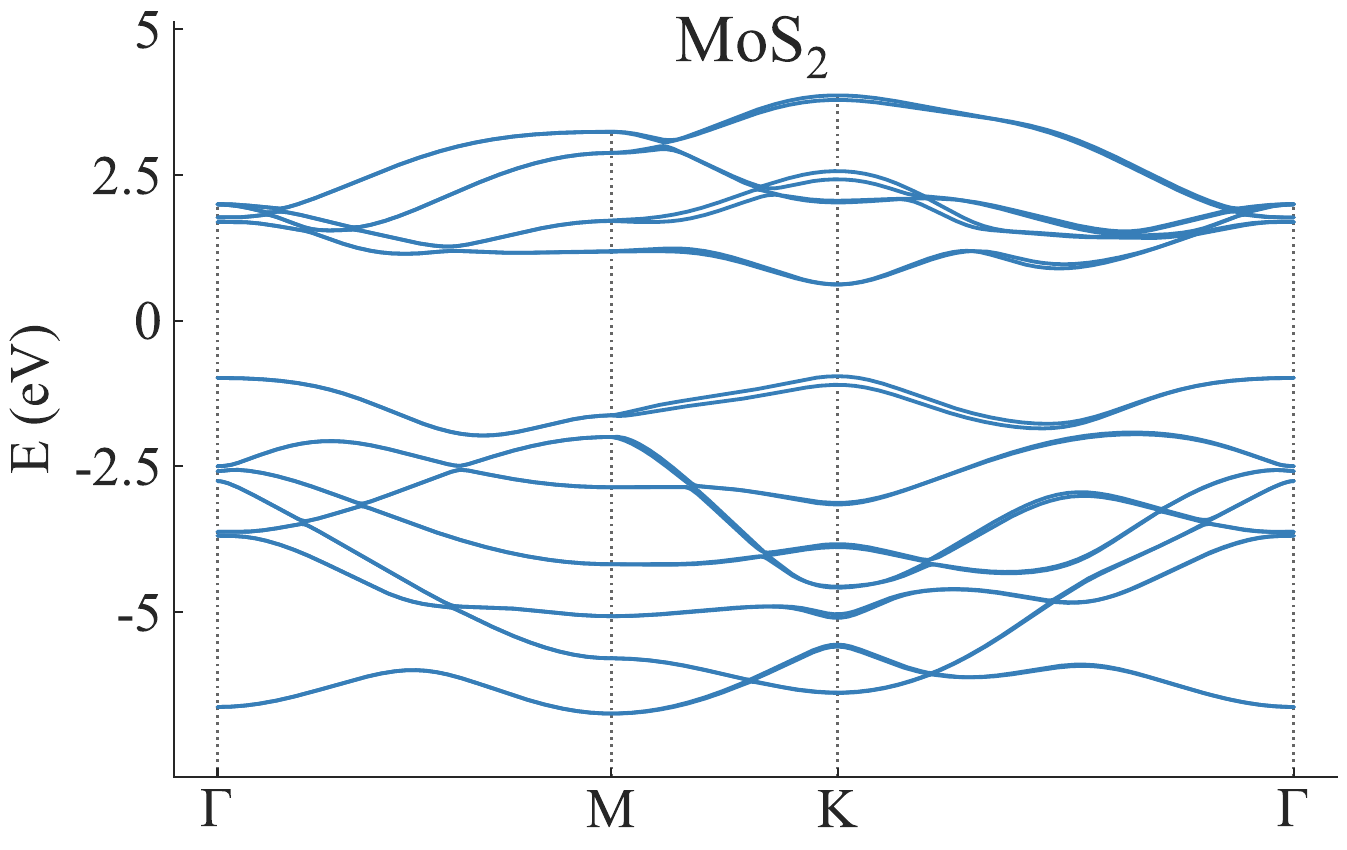}
\label{fig:}}
\subfigure[]{
\includegraphics[scale=0.3]{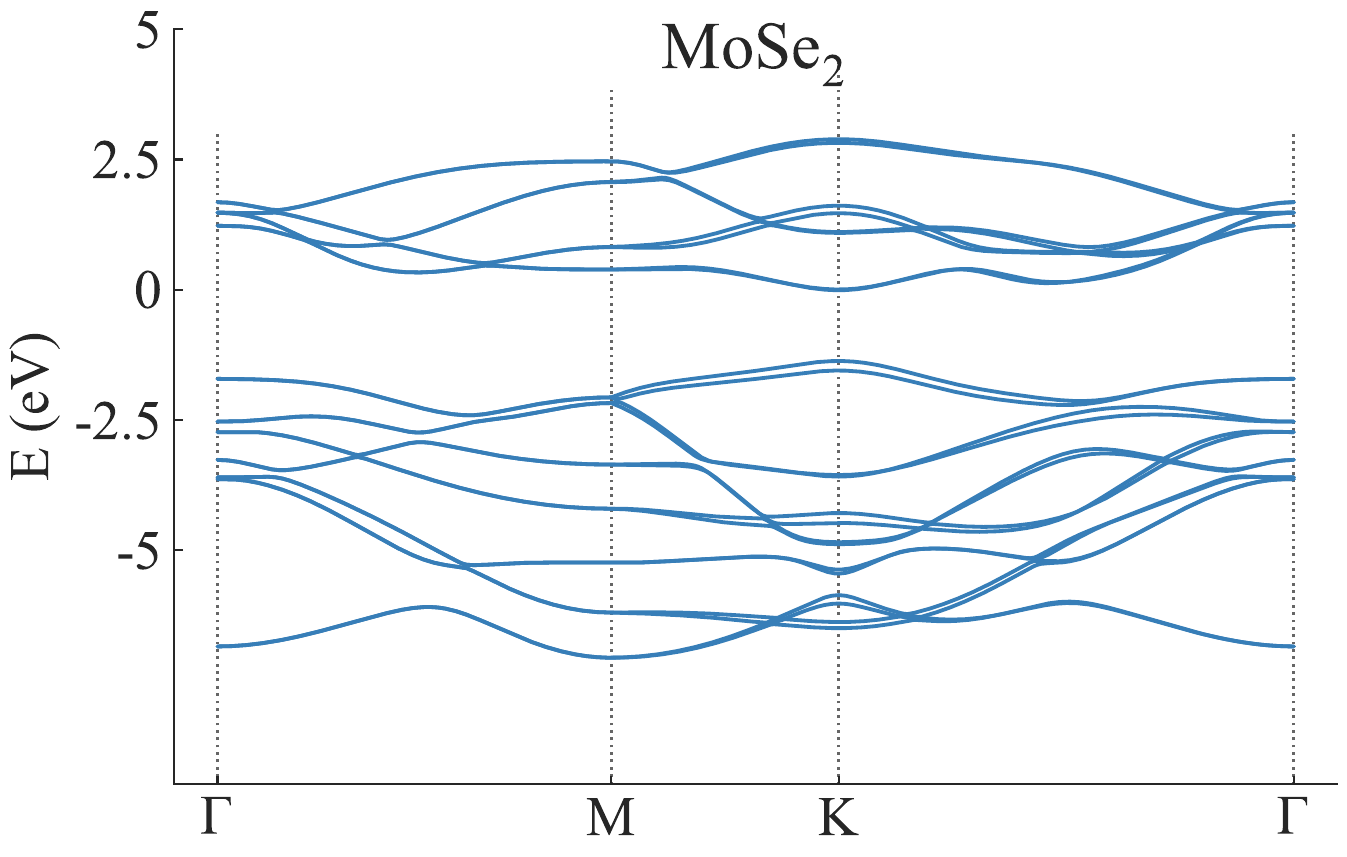}
\label{fig:}}
\subfigure[]{
\includegraphics[scale=0.3]{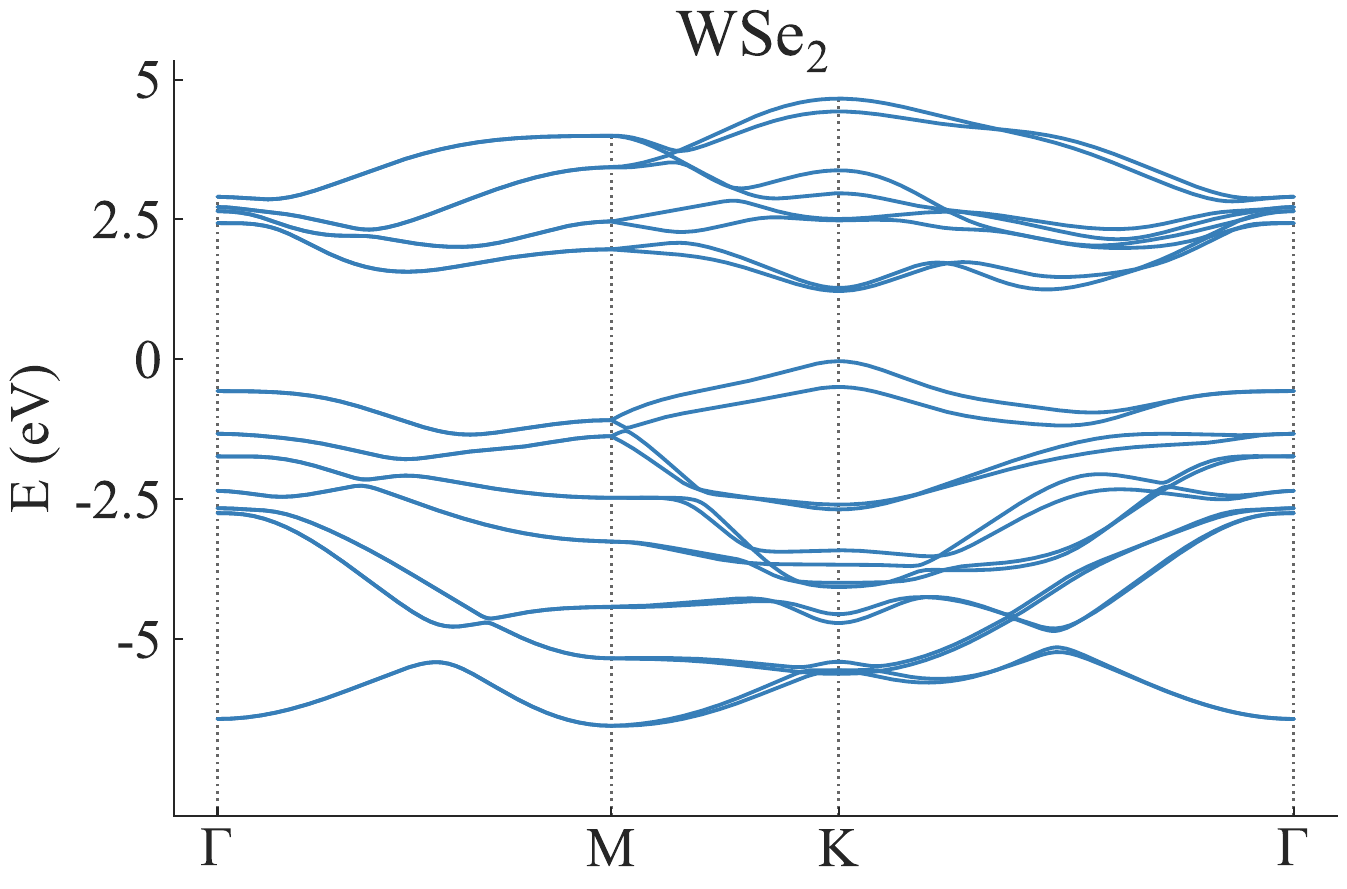}
\label{fig:}}
\subfigure[]{
\includegraphics[scale=0.3]{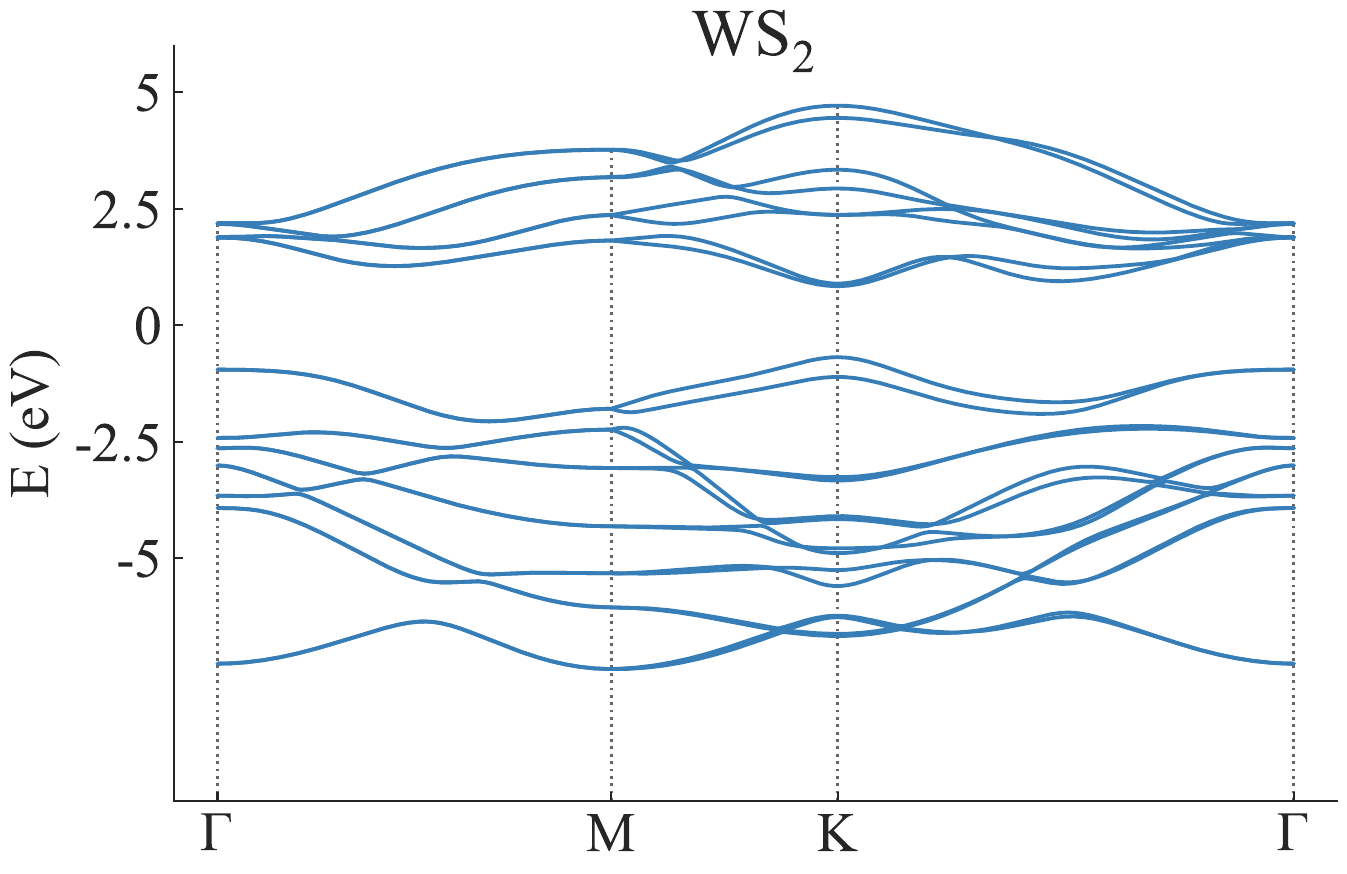}
\label{fig:}}
\subfigure[]{
\includegraphics[scale=0.3]{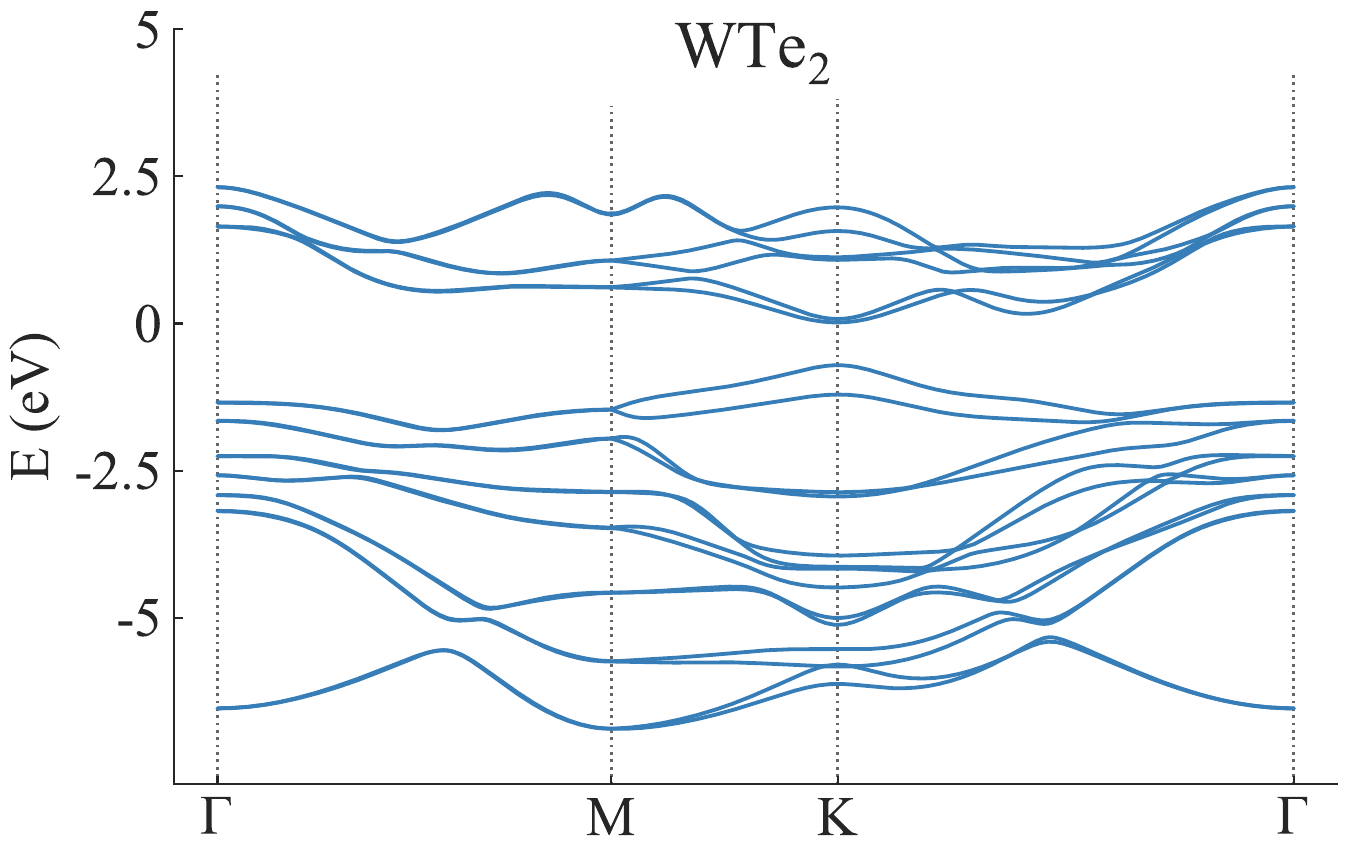}
\label{fig:}}
\subfigure[]{
\includegraphics[scale=0.3]{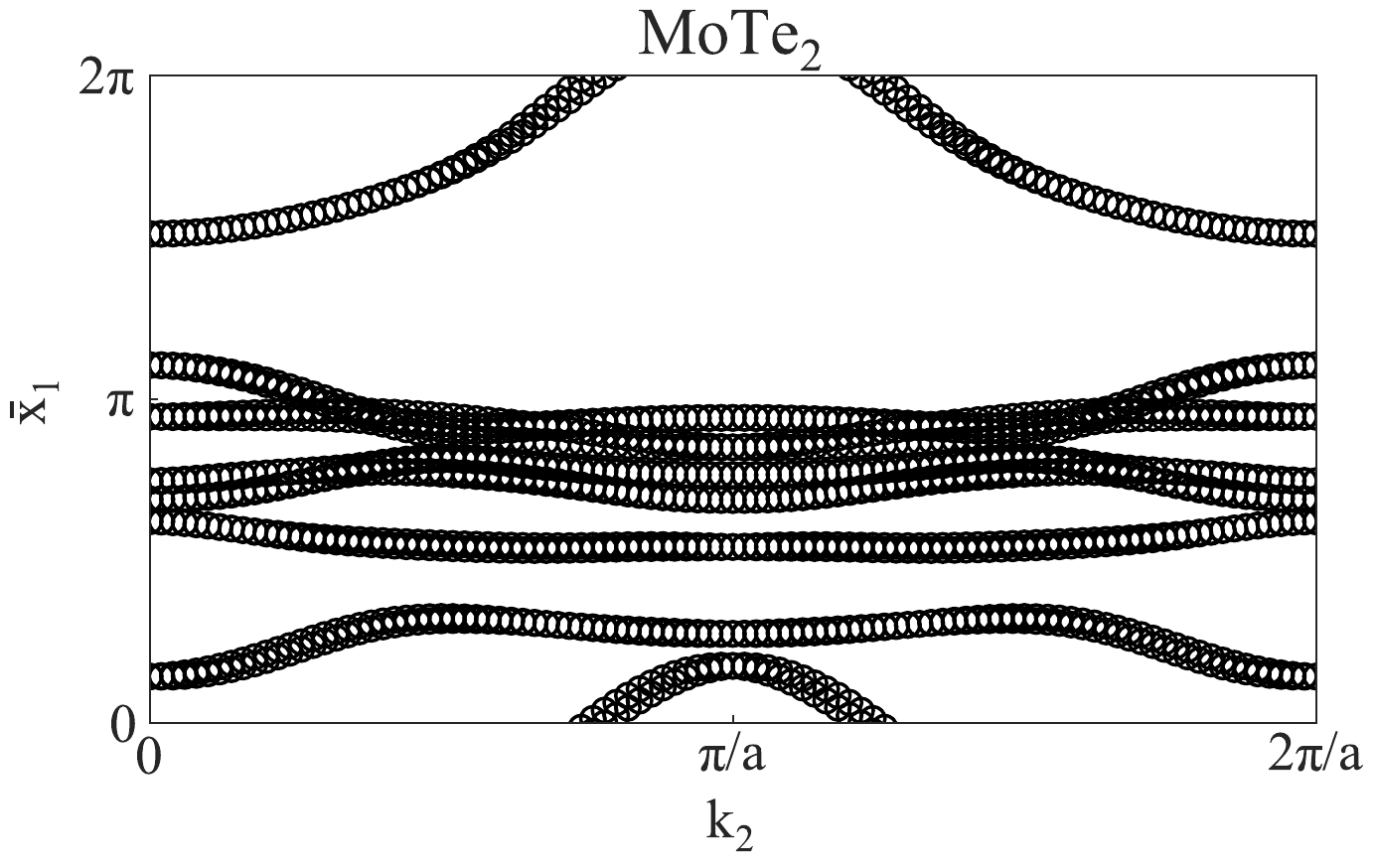}
\label{fig:}}
\subfigure[]{
\includegraphics[scale=0.3]{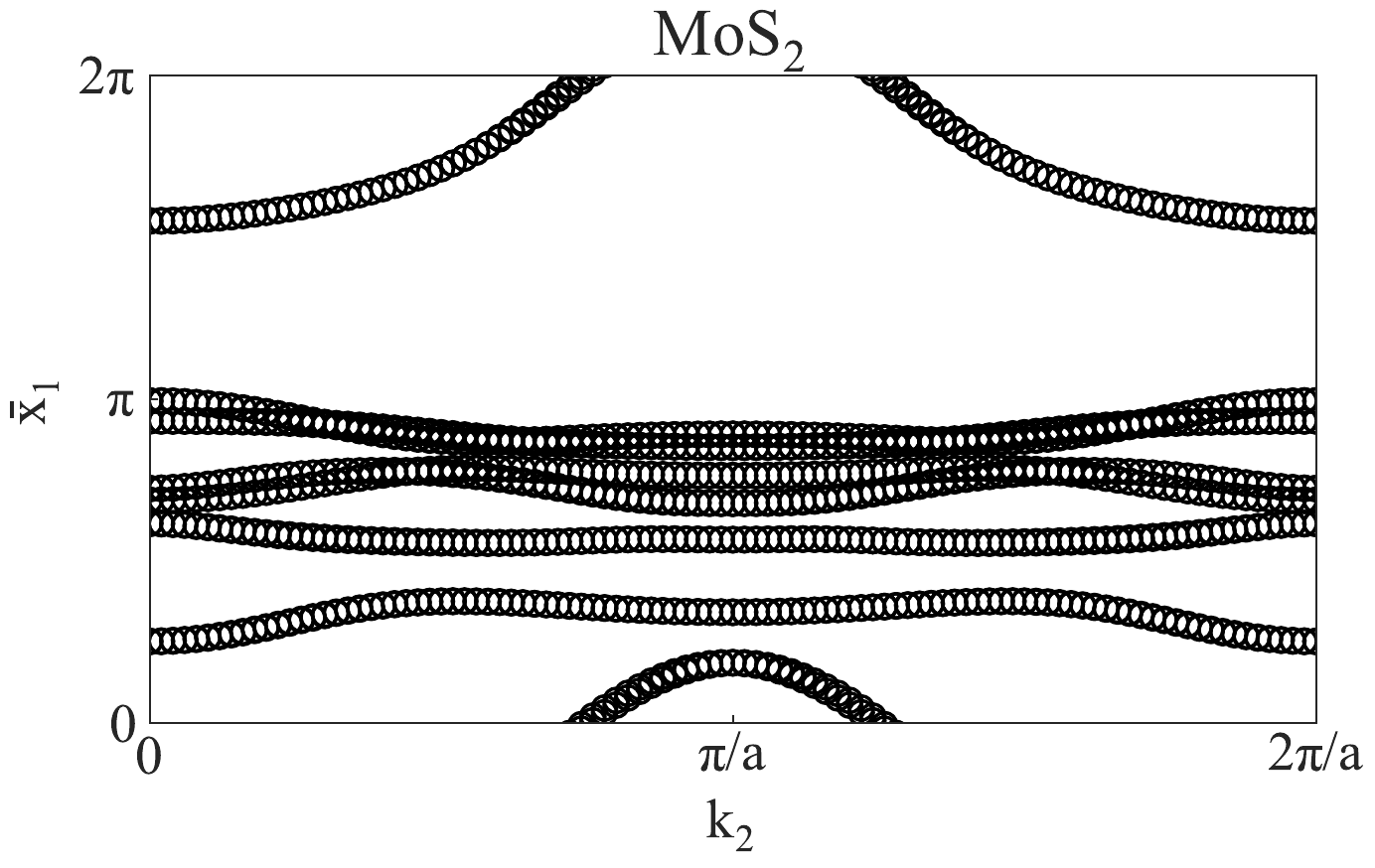}
\label{fig:}}
\subfigure[]{
\includegraphics[scale=0.3]{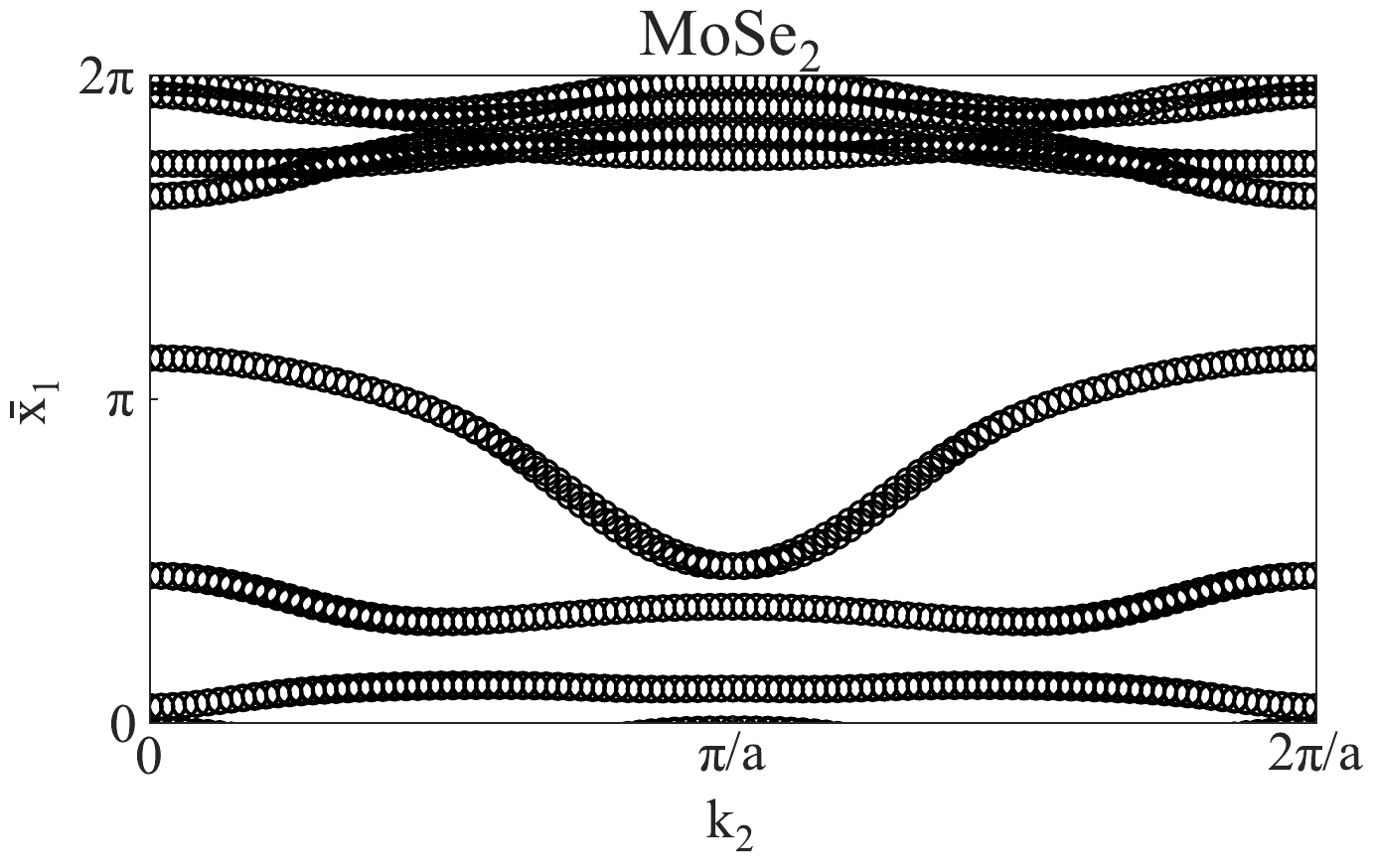}
\label{fig:}}
\subfigure[]{
\includegraphics[scale=0.3]{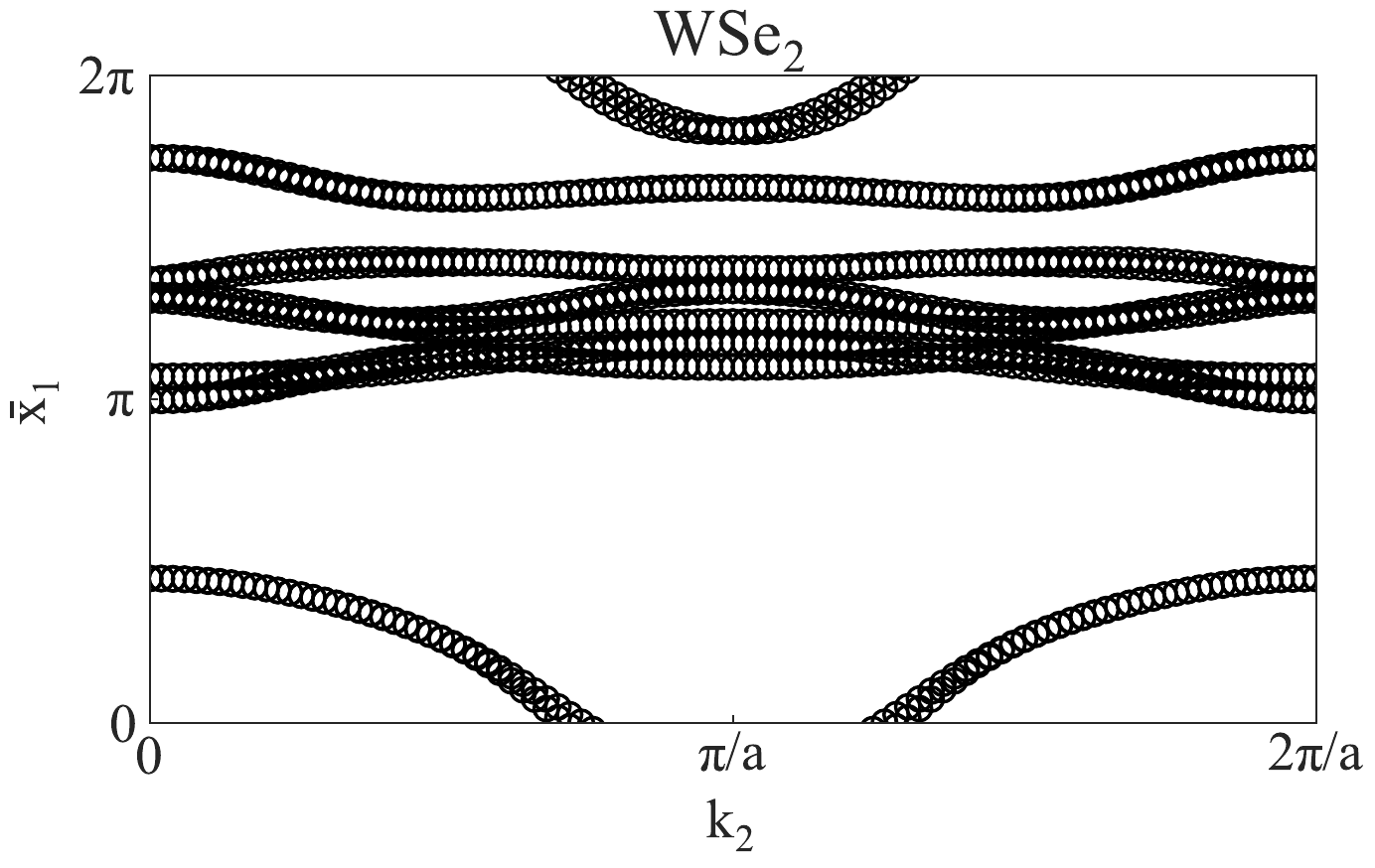}
\label{fig:}}
\subfigure[]{
\includegraphics[scale=0.3]{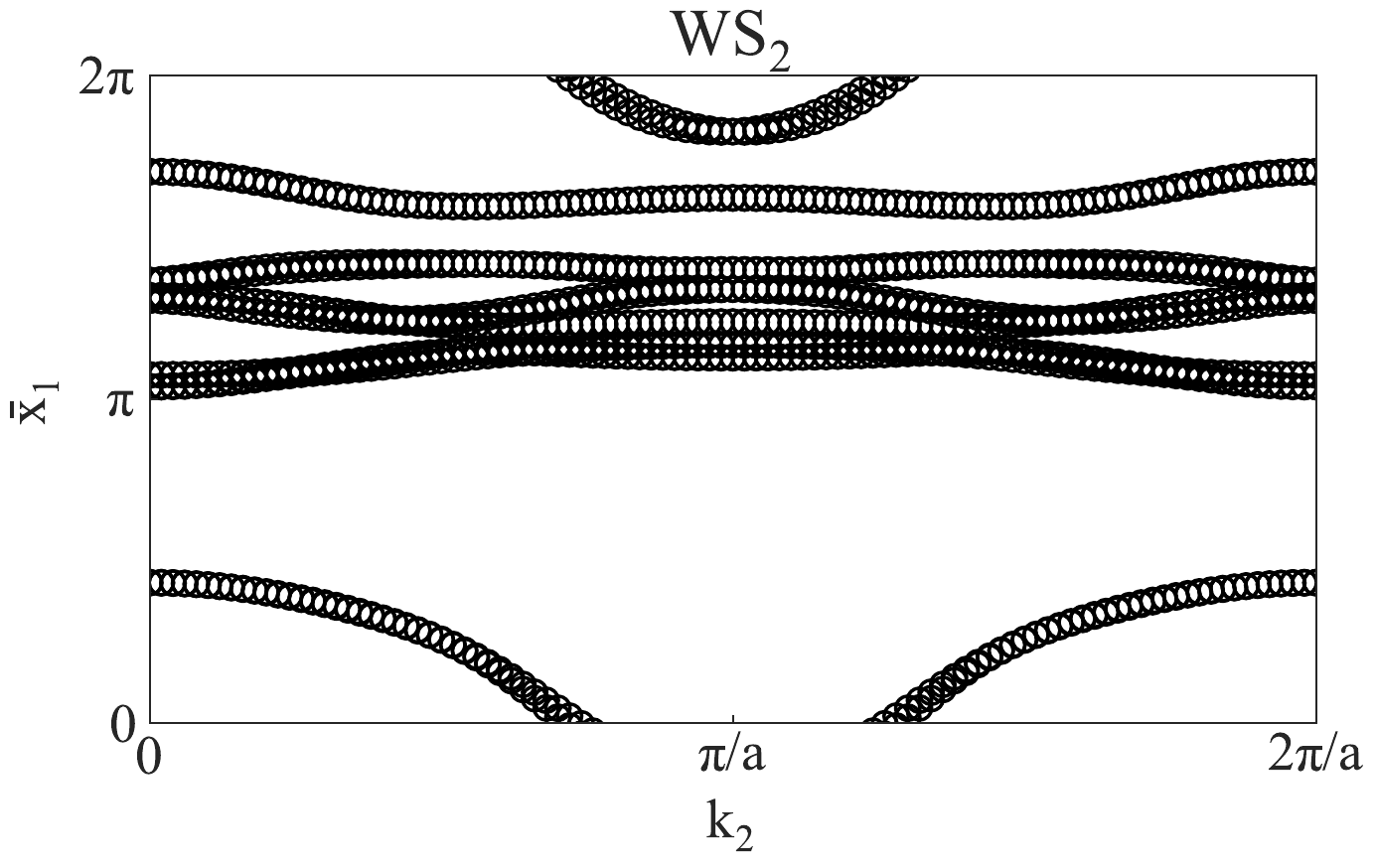}
\label{fig:}}
\subfigure[]{
\includegraphics[scale=0.3]{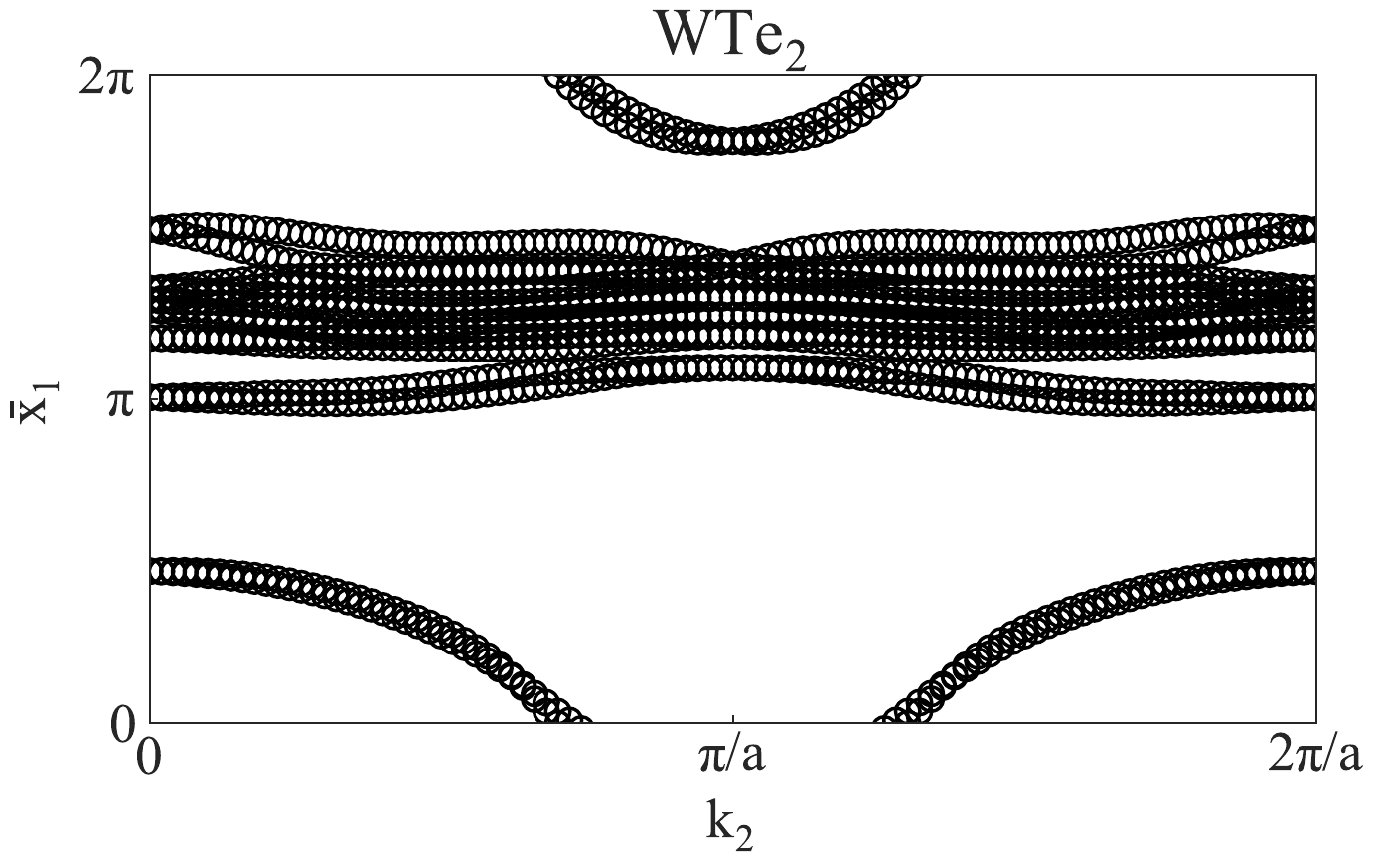}
\label{fig:}}
\caption{(a) Bulk band structure along high-symmetry path for 1H-MX$_{2}$ TMDs supporting $|\mathcal{C}_{s,G}|=2$. (b) Wannier center charge (WCC) spectra in the two-dimensional plane for 1H-MX$_{2}$ TMDs admitting ground state spin Chern number, $|\mathcal{C}_{s,G}|=2$. In each case the the spectra is gapped. As a result, no information about the topological nature of the ground-state can be determined.}
\label{fig:MX2BandsWCC}
\end{figure*}

\begin{figure*}
\centering
\subfigure[]{
\includegraphics[scale=0.23]{Figures/MoS2bands.pdf}
\label{fig:}}
\subfigure[]{
\includegraphics[scale=0.23]{Figures/MoS2_Induced_Charge.pdf}
\label{fig:}}
\subfigure[]{
\includegraphics[scale=0.23]{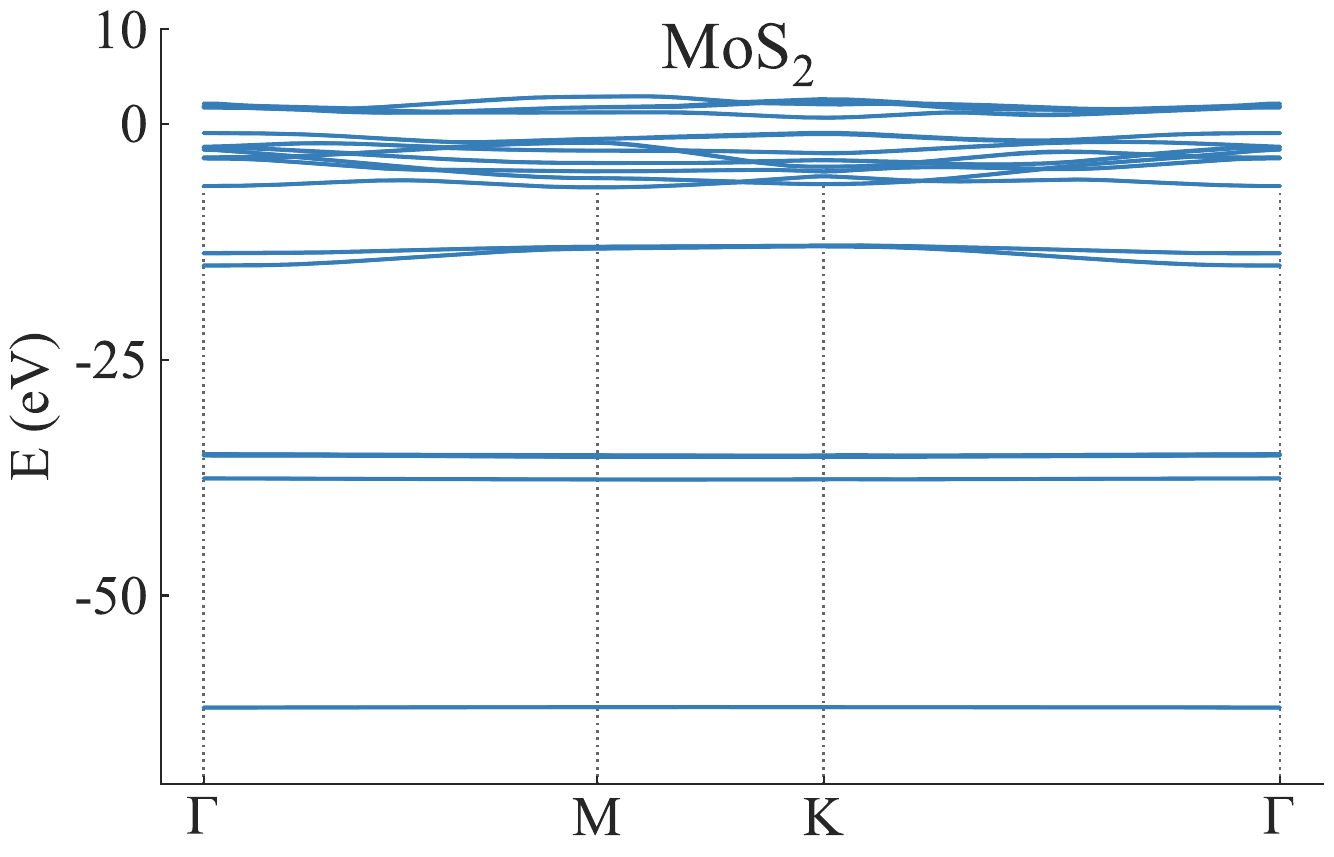}
\label{fig:}}
\subfigure[]{
\includegraphics[scale=0.23]{Figures/MoS2_Induced_Charge.pdf}
\label{fig:}}
\caption{The Wannier tight-binding band structure for MoS$_{2}$ considering (a) 14 and (c) 26 occupied valence bands. Induced charge on the vortex in the mid-gap as a function of doping the vortex bound modes for a slab of 30 $\times$ 30 unit cells remains unchanged regardless of whether we consider (b) 14 occupied bands or (d) 26 occupied bands, demonstrating robustness to the addition of trivial bands.}
\label{fig:Fragile}
\end{figure*}
\par 
Due to the complexity associated with calculation of bulk invariants in the absence of spin-conservation or inversion symmetry, previous studies have relied on alternative indicators of bulk topology to scan the available materials databases\cite{sodequist2022abundance,costa2021discovery}. Two common physical signatures examined in the literature are the surface state spectra and the spin-Hall conductivity. In Fig. \eqref{fig:MX2SurfaceSpinHall}, we display the surface state spectral density, calculated using the Wannier Tools software package\cite{WU2017,sancho1985highly}, in the vicinity of the bulk-gap, demonstrating the lack of gapless surface states and emphasizing that the insulators cannot be identified by the Fu-Kane $\mathbb{Z}_{2}$-index or calculation of WCCs along principal axes. 
\par 
Calculation of the spin-Hall conductivity as a method for detecting non-trivial bulk topology has also become commonplace due to the development of advanced software including the Wannier-Berri package\cite{destraz2020magnetism,tsirkin2021high}. We compute the spin Hall conductivity along principle spin-quantization axes and display the results in Fig. \eqref{fig:MX2SurfaceSpinHall}. It is clear in each case that the spin-Hall conductivity is non-vanishing in the bulk gap, a signature of non-trivial bulk topology. Although it is important to note that the conductivity is non-quantized in each case. This is to be expected given the lack of spin-conservation symmetry.

\section*{Appendix C: Momentum space analysis for 1H-MX$_{2}$ TMDs}\label{app:c}

\par 
The bulk band structure along the high-symmetry path for each member of the 1H-MX$_{2}$ TMD family is displayed in Fig. \eqref{fig:MX2BandsWCC}. We note that each band structure supports a bulk gap, $|\Delta E|\geq 1.1 \;eV$
\par 
Calculation of WCC spectra represents the most common tool for topological diagnosis involving the ground-state Berry connection\cite{yu2011equivalent,alexandradinata2014wilson,Z2pack}. The results of a WCC calculation along the $x_{1}$ direction as a function of transverse momenta $k_{2}$ are displayed in Fig. \eqref{fig:MX2BandsWCC}. These figures demonstrate that the WCC spectra is gapped in each case, in correspondence with the lack of non-trivial $\mathbb{Z}_{2}$ index and gapless surface states. The trivial classification due to WCCs provides further explanation as to why these materials have evaded topological classification prevously. 

\section*{Appendix D: Robustness of topological classification to inclusion of occupied bands}\label{AppendixD}
\par 
The subject of fragile topological phases has received significant attention in recent years\cite{FragilePo,song2020twisted,wieder2020strong}. Band structures which exhibit fragile topology are often classified as those in which the occupied bands exhibit a Wannier obstruction, however, the obstruction can be removed through addition of occupied topologically trivial bands. With the exception of $\beta$-bismuthene, we do not consider a two-dimensional insulator in which the occupied subspace considered in the Wannier tight-binding (WTB) model supports a Wannier obstruction when computing Wannier center charges along principle axes in this work. 
\par 
In this appendix, we focus on the question of whether inclusion of deeper-energy, occupied bands, can trivialize the observed behavior upon vortex insertion. A definitive answer can be given to this question. \emph{The inclusion of deeper lying, filled bands does not trivialize the observed topological behavior.} 
\par
This statement can be definitively made as the automated procedure providing the initial screening by constructing a WTB via the SCDM method considers bands far below the Fermi energy. As a result, we can directly compare the WTB generated in an automated fashion and the low-energy manually generated WTB for materials in the 1H-MX$_{2}$ class. We demonstrate this explicitly for MoS$_{2}$ in Fig. \eqref{fig:Fragile}. The automated WTB model includes 26 occupied bands. By contrast, the low-energy model constructed from d-orbitals of Mo and p-orbitals of S, admits only 14 occupied bands. Nevertheless, the results upon vortex insertion are identical as seen in Fig. \eqref{fig:Fragile}. Confirming that the inclusion of occupied bands, further from the Fermi energy, does not trivialize our observations. 

\section*{Appendix E: Results of automated materials database screening}\label{App:E}
A full list of materials screened, their respective band-gap and space group are given in Tab. \eqref{fig:fullres}. Materials displaying evidence of non-trivial topology through the presence of mid-gap vortex bound modes and quantized induced charge obeying the relation are high-lighted. For each of these systems, in Fig. (11) we display the bulk band structure following the path, $\Gamma-Y-S-\Gamma-X$, where we define $X=\mathbf{k}_{1}/2$, $Y=\mathbf{k}_{2}/2$ and $S=\mathbf{k}_{1}/2+\mathbf{k}_{2}/2$ setting $k_{1,2}$ to be the primitive reciprocal lattice vectors. We further display the Wannier center charge spectra $\bar{x}_{1}(k_{2})$, and the results of an induced charge calculation for a system of $30 \times 30$ unit cells. Each of these figures was generated using the automatically constructed Wannier tight-binding model.

\begin{figure}
    \centering
    \includegraphics[width=17cm]{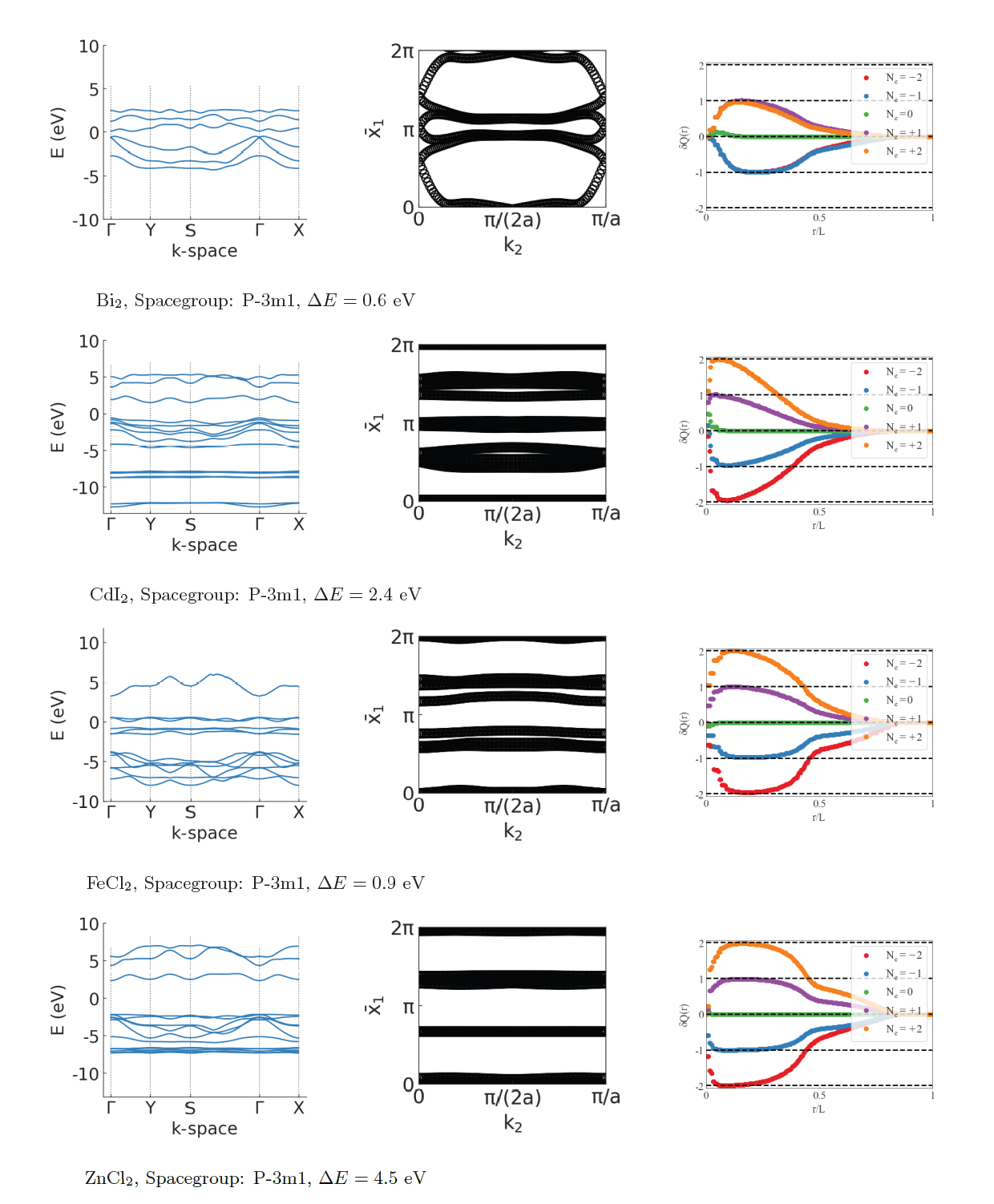}
    \label{fig:topmat}
\end{figure}
\begin{figure}
    \centering
    \includegraphics[width=16cm]{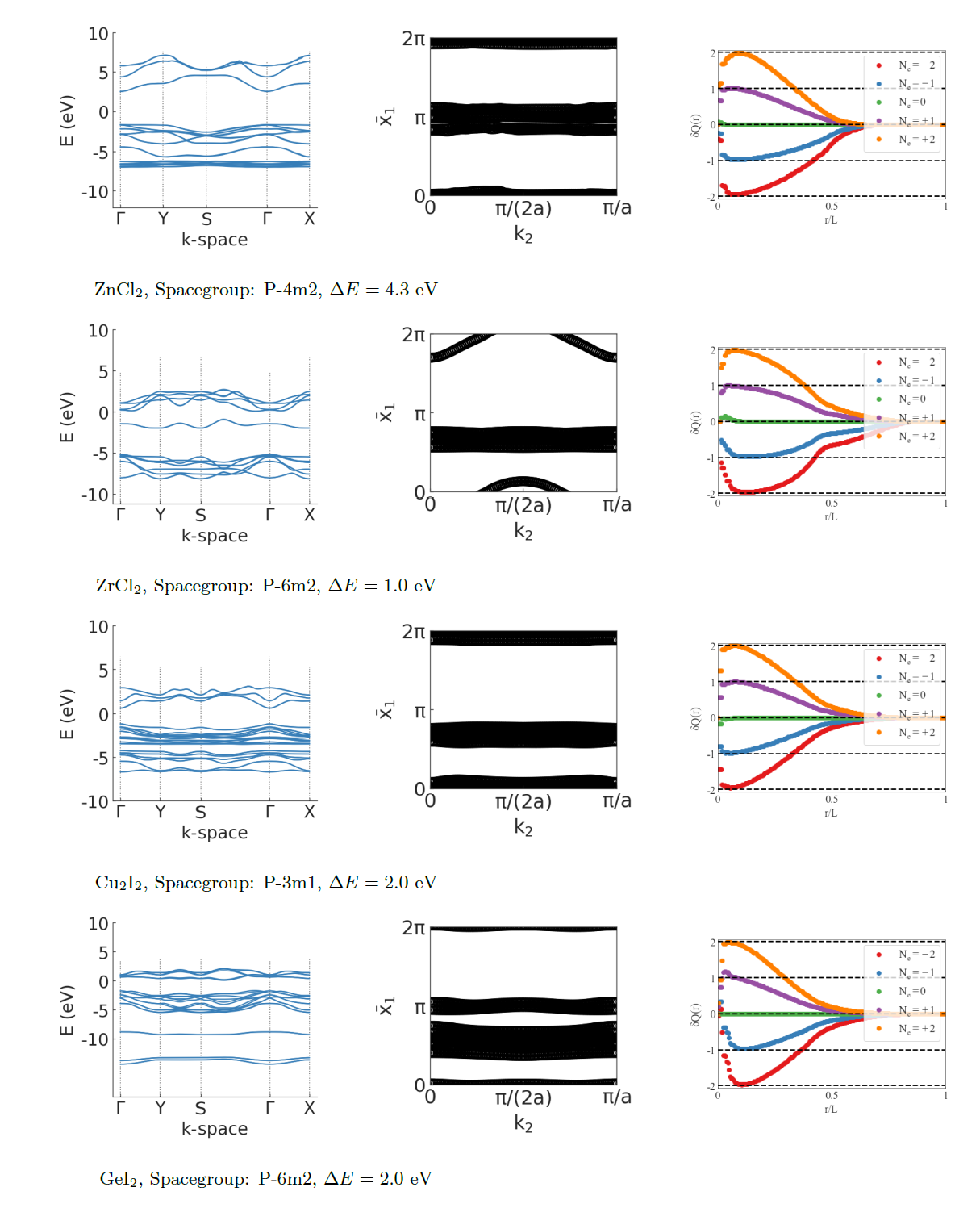}
    \label{fig:my_label}
\end{figure}
\begin{figure}
    \centering
    \includegraphics[width=16cm]{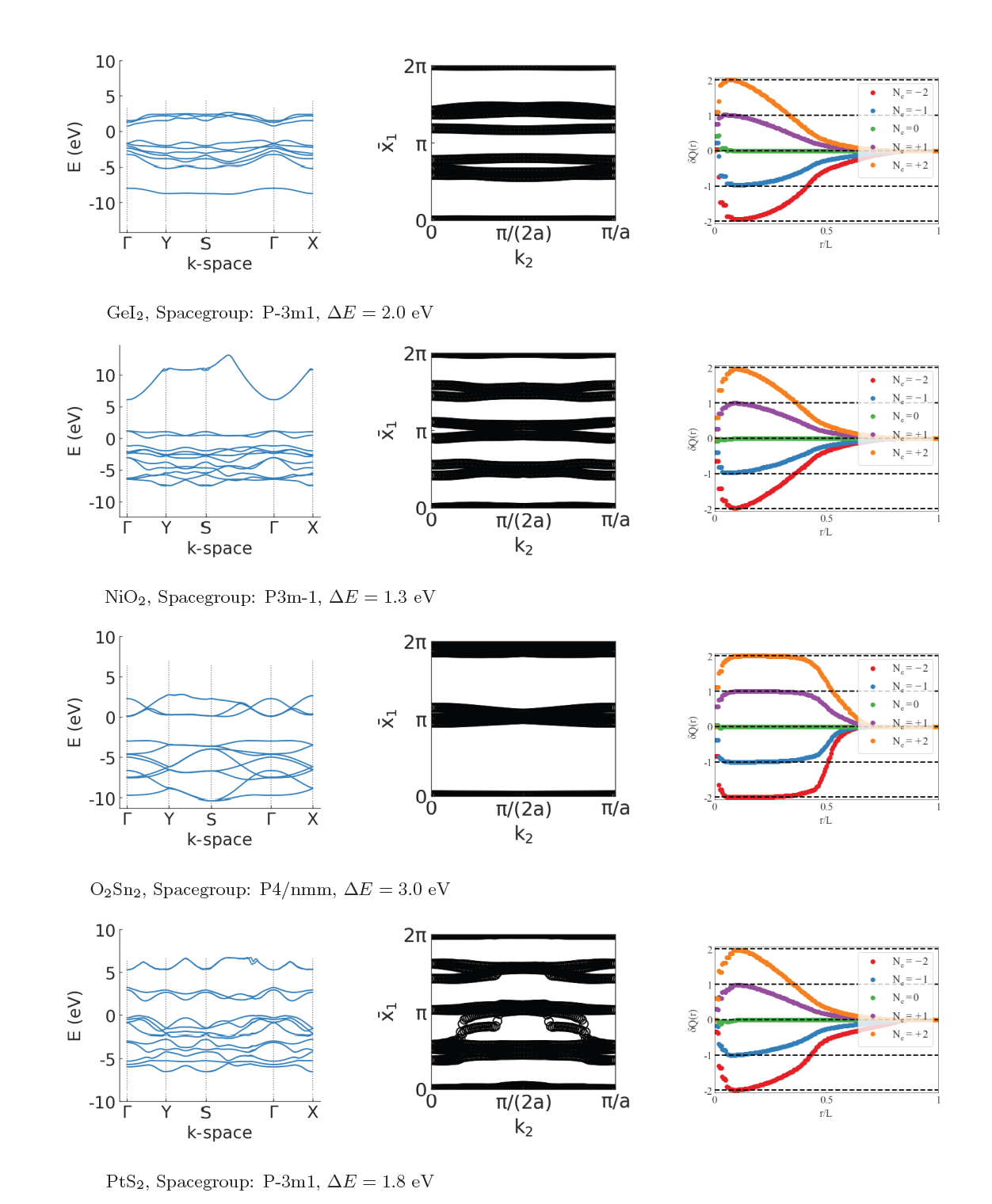}
    \label{fig:my_label}
\end{figure}
\begin{figure}
    \centering
    \includegraphics[width=16cm]{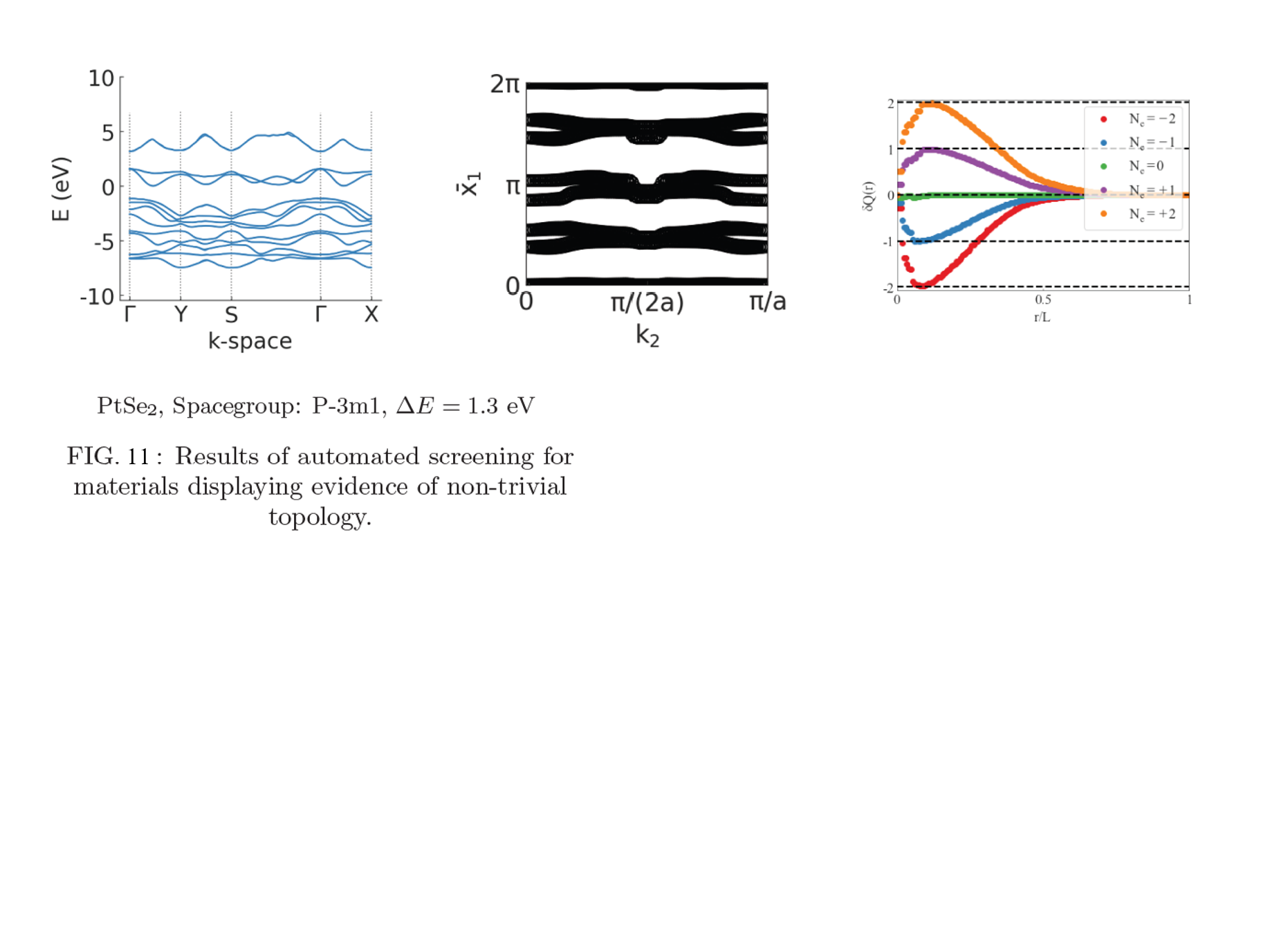}
\end{figure}

\begin{figure}
    \centering
    \includegraphics[width=20cm]{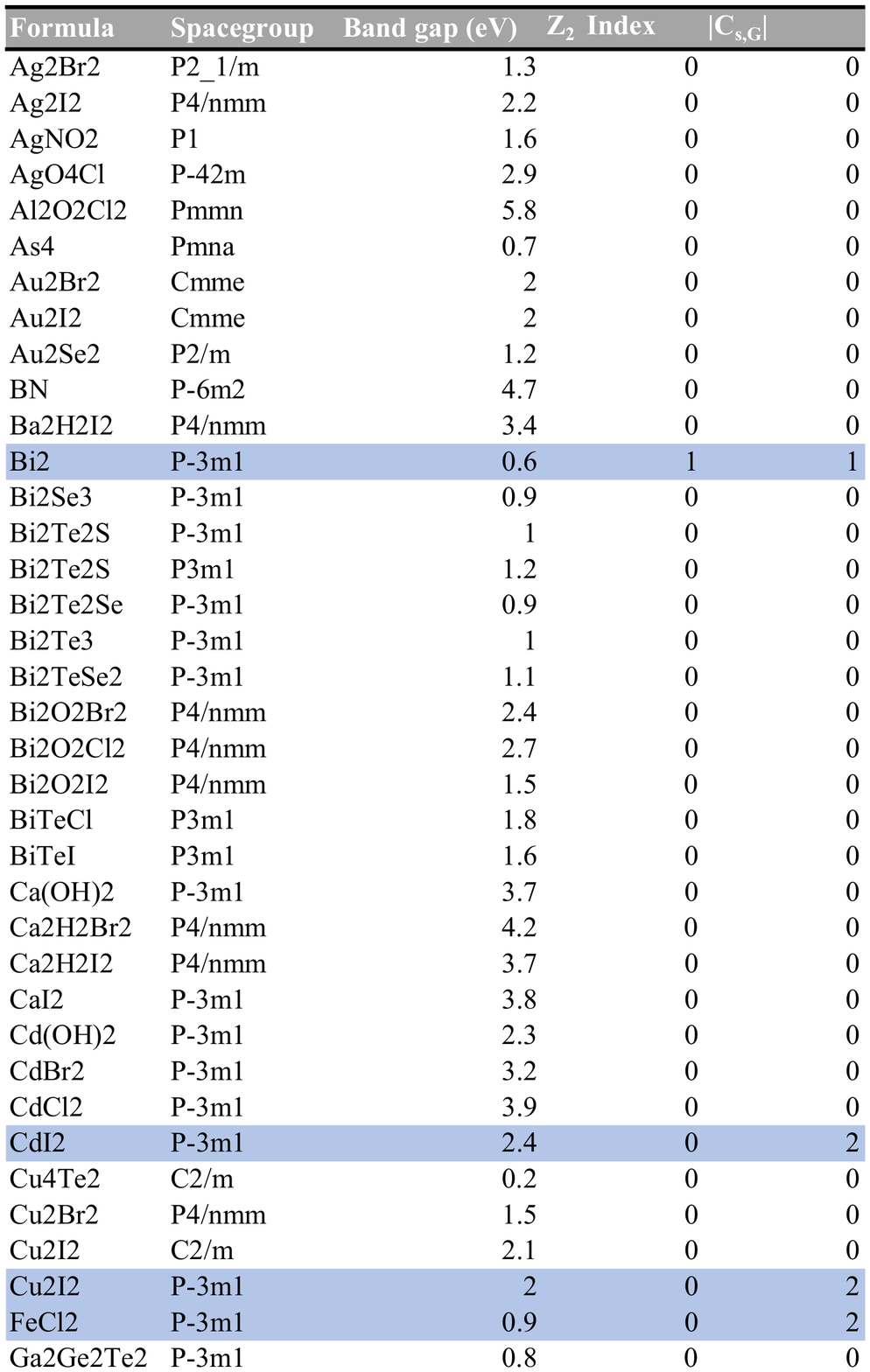}
\end{figure}
\begin{figure}
    \centering
    \includegraphics[width=20cm]{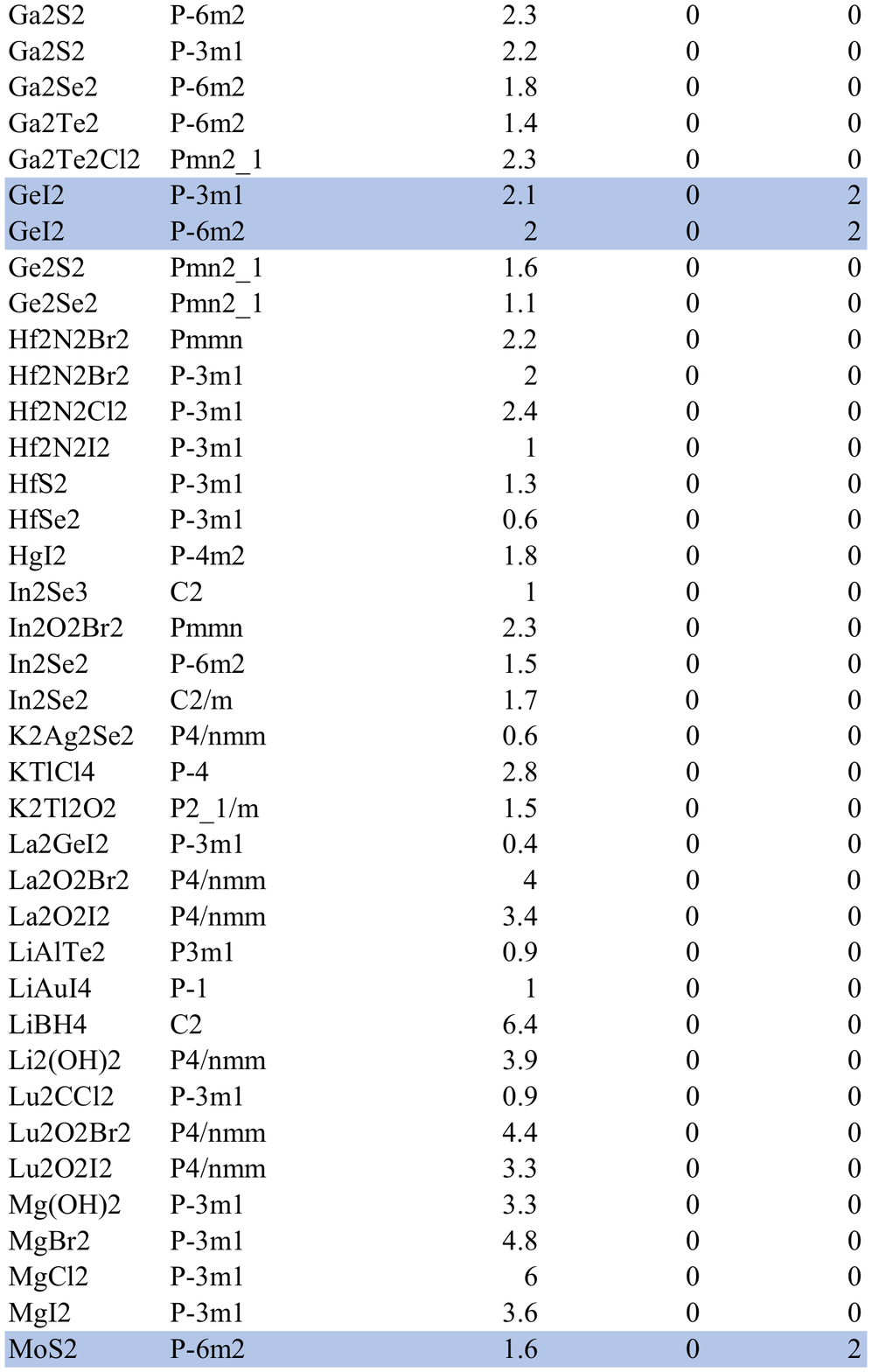}
    \label{fig:my_label}
\end{figure}
\begin{figure}
    \centering
    \includegraphics[width=20cm]{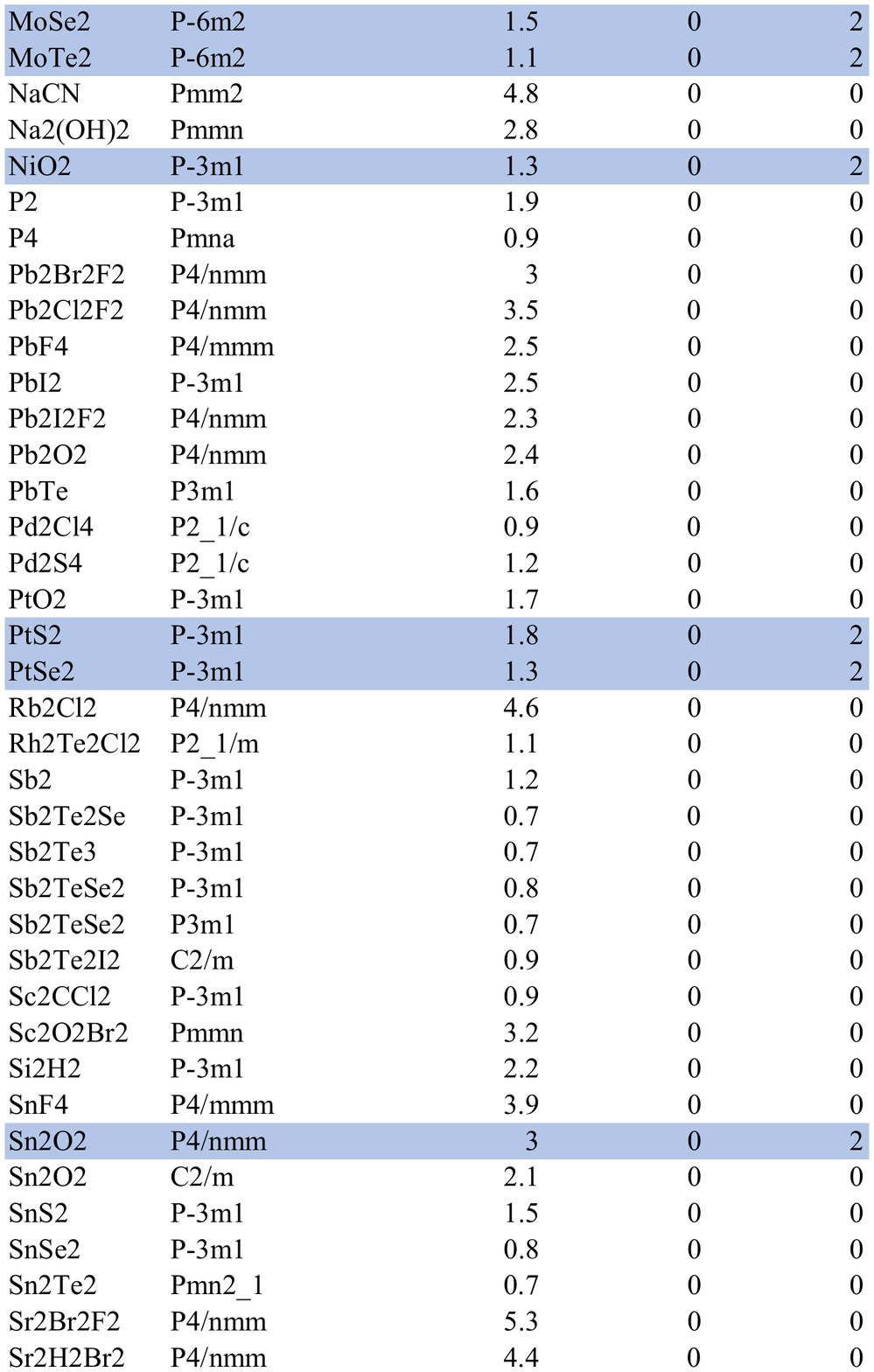}
    \label{fig:my_label}
\end{figure}
\begin{figure}
    \centering
    \includegraphics[width=20cm]{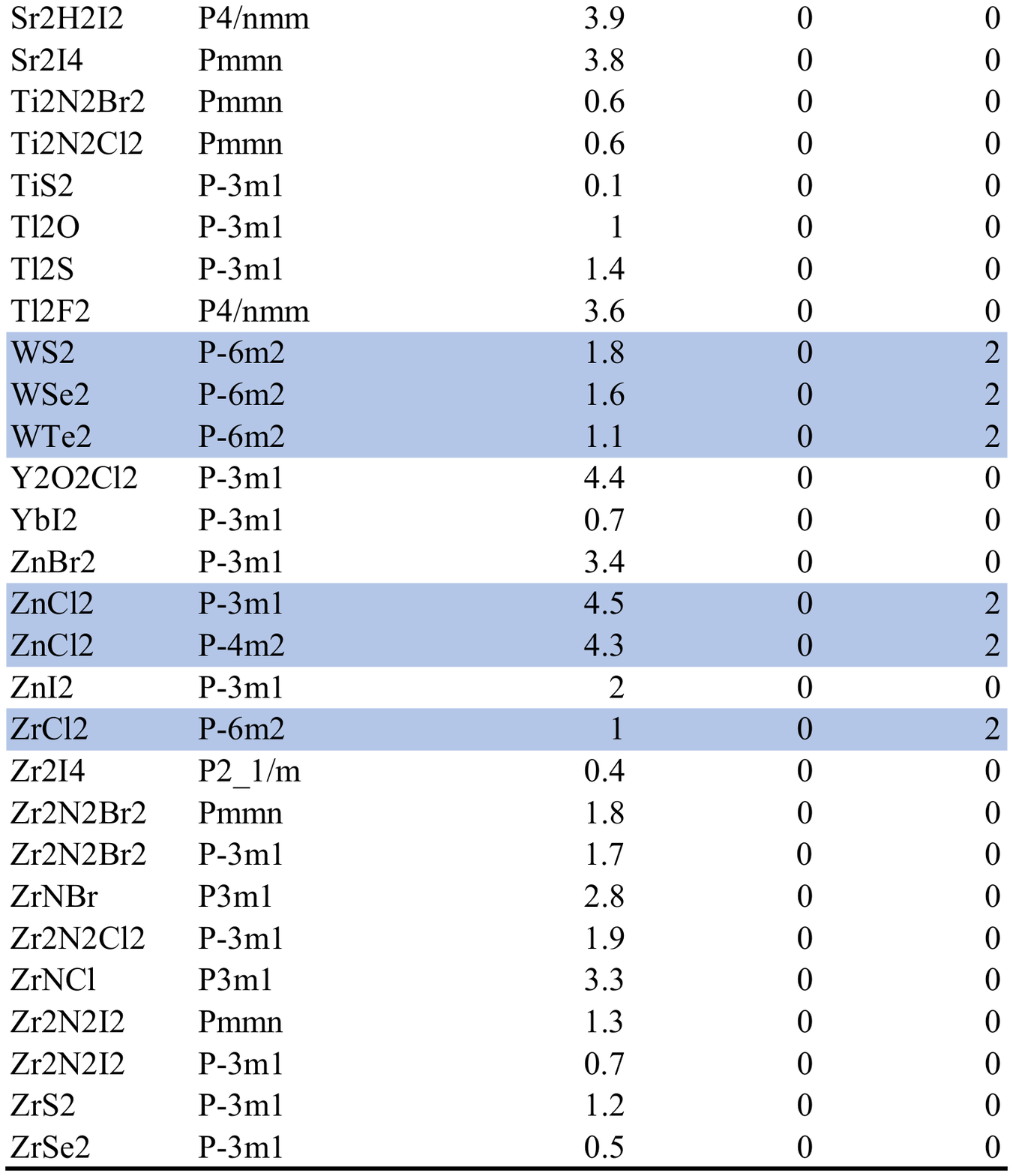}
    \caption{}
    \label{fig:fullres}
\end{figure}

\end{document}